\documentstyle[12pt,aaspp4]{article}
\def\n2hp{N$_2$H$^+$}
\def\nh3{NH$_3$}

\def\ga{\mathrel{\mathchoice {\vcenter{\offinterlineskip\halign{\hfil
$\displaystyle##$\hfil\cr>\cr\sim\cr}}}
{\vcenter{\offinterlineskip\halign{\hfil$\textstyle##$\hfil\cr
>\cr\sim\cr}}}
{\vcenter{\offinterlineskip\halign{\hfil$\scriptstyle##$\hfil\cr
>\cr\sim\cr}}}
{\vcenter{\offinterlineskip\halign{\hfil$\scriptscriptstyle##$\hfil\cr
>\cr\sim\cr}}}}}

\def\p {$\pm$}

\def\percc {$\hbox{{\rm cm}}^{-3}$}    
\def\MOLH {\hbox{${\rm H}_2$}}  
\def\AMM {\hbox{${\rm NH}_{3}$}} 
\def\THCO {\hbox{$^{13}{\rm CO}$}}   
\def\NTHP {\hbox{${\rm N}_2{\rm H}^+$}} 

\def\p{$\pm$}

\lefthead{Caselli et al.}
\righthead{Dense cores in dark clouds}

\received{2001 December 5}
\begin{document}

\title{Dense Cores in Dark Clouds. XIV. 
N$_2$H$^+$(1--0) maps of dense cloud cores}

\author{Paola Caselli}
\affil{Osservatorio Astrofisico di Arcetri, Largo E. Fermi 5, I-50125
Firenze, Italy}

\author{Priscilla J. Benson}
\affil{Whitin Observatory, Wellesley College, 106 Central Street, 
Wellesley, MA 02181--8286, USA}

\author{Philip C. Myers}
\affil{Harvard--Smithsonian Center for Astrophysics, 60 Garden Street, 
Cambridge, MA 02138, USA}

\and

\author{Mario Tafalla}
\affil{Observatorio Astron\'omico Nacional (IGN), Alfonso XII, 3, 
E-28014 Madrid, Spain}

\begin{abstract}
We present results of an extensive mapping survey of N$_2$H$^+$(1--0)
in about 60 low mass cloud cores already mapped in the NH$_3$(1,1) 
inversion transition line.  The survey has been carried out 
at the FCRAO antenna with an angular resolution of 54$^{\prime\prime}$,
about 1.5 times finer
than the previous ammonia observations made at the Haystack telescope. 
The comparison between 
N$_2$H$^+$ and NH$_3$ maps shows strong similarities in the 
size and morphology of the two molecular species indicating that
they are tracing the same material, especially in starless cores.  
Cores with stars typically have  map sizes about a factor of two 
smaller for \NTHP \ than for NH$_3$, indicating the presence of denser and more
centrally concentrated gas compared to starless cores. The mean aspect ratio is
$\sim$ 2.  Significant correlations are found between \AMM \ and \NTHP \ column 
densities and excitation temperatures in starless cores, but not in cores
with stars, suggesting a different chemical evolution of the two species.
Starless cores are less massive 
($<M_{\rm vir}>$ $\simeq$ 3 M$_{\odot}$)
than cores with stars ($<M_{\rm vir}>$ $\simeq$ 9 M$_{\odot}$).  
Velocity gradients 
range between 0.5 and 6 km/s/pc, similar to what has 
been found with \nh3 data, and the ratio $\beta$ of rotational kinetic
energy to gravitational energy have magnitudes between $\sim$ 10$^{-4}$ and 
0.07, indicating that rotation is not energetically dominant in the support
of the cores.  ``Local'' velocity gradients show significant variation
in both magnitude and direction, suggesting the presence of complex motions
not interpretable as simple solid body rotation.  Integrated intensity 
profiles of starless cores present a ``central flattening'' and are 
consistent with a spherically symmetric density law $n$ $\propto$ 
$r^{-\alpha}$ where $\alpha$ = 1.2 
for $r$ $<$ $r_{\rm break}$ and $\alpha$ = 2 for $r$ $>$ 
$r_{\rm break}$, where $r_{\rm break}$ $\sim$ 0.03 pc.  Cores with 
stars are better modelled with single density power laws with $\alpha$
$\geq$ 2, in agreement with observations of submillimeter 
continuum emission.  Line widths change 
across the core but we did not find a general trend: there are cores 
with significant positive as well as negative linear correlations between 
$\Delta v$ and the impact parameter $b$.  
The deviation in line width correlates with the mean 
line width, suggesting that the line of sight contains $\sim$ 10 coherence 
lengths.  The corresponding value of the coherence length, $\sim$ 0.01 pc, 
is similar to the expected cutoff wavelength for MHD waves.  This similarity
may account for the increased ``coherence'' of line widths on small scales.
Despite of the finer angular resolution, the majority of \NTHP \ and \AMM  \
maps show a similar ``simple'' structure, with single peaks and no elongation.

\end{abstract}


\section{Introduction}

Dense cores in dark clouds have been extensively studied  
through observations of the inversion transition lines of ammonia
(\cite{mb83}; \cite{bm89}, hereafter BM89) and other high density tracers, 
including CS (\cite{zef89}), C$_2$S (\cite{syo92}) and HC$_3$N 
(\cite{fm93}).  Molecular emission maps have improved our 
understanding of cloud structure and have given us insights on the initial 
conditions of the star forming process.  It is now well established 
that low mass cores as mapped in NH$_3$ lines are about 0.1 pc in size, 
have kinetic temperatures 
of about 10 K, and gas number density of $\sim$ 3$\times$10$^4$ cm$^{-3}$
(BM89).  Generally, cores have elongated maps 
with typical aspect ratio of 2 (\cite{mfg91}; \cite{r96}).  \cite{gbf93}
used ammonia maps to measure velocity gradients;  they found typical 
magnitudes between 0.3 and 4 km s$^{-1}$ pc$^{-1}$ and conclude 
that  rotation is not energetically dominant in the support of cores.  

The upgrade of the 37-m Haystack telescope \footnote{Radio 
Astronomy observations
at the Haystack Observatory of the Northeast Radio Observatory Corporation
are supported by a grant from the National Science Foundation.} at 3 mm
(\cite{bs94}) enabled high spatial (and spectral) resolution
observations of the molecular ion N$_2$H$^+$ towards those cores already 
studied in NH$_3$ (\cite{bcm98}, hereafter BCM98).  
The J = 1 $\rightarrow$ 0 rotational transition 
of diazenylium has been detected in most (94\%) of 
the cores indicating that this species is widespread and easy to detect. 
Moreover, a good correlation between N$_2$H$^+$ and NH$_3$ velocities and 
line widths indicates that the two species are probably tracing the same 
material.  N$_2$H$^+$ is known to be a selective tracer of quiescent gas 
(\cite{tt77}; \cite{wzw92}; \cite{b96}) and
 is particularly suitable for studying the structure and kinematics 
of cold star forming cores.  Being an ion, diazenylium is also important
to trace the ionized gas and to give information about the coupling between
ions and neutrals in star--forming dense cores (BCM98). We point out that 
N$_2$H$^+$, formed through the ion--molecule reaction
N$_2$ + H$_3^+$ and mainly destroyed by CO and electrons,  traces molecular 
nitrogen, N$_2$, which is a major repository of nitrogen.  
N$_2$H$^+$ is thought to be
a ``late depleter'' (\cite{bl97}; \cite{aoi01}; \cite{cwz01b}) and so is a good 
tracer of dense core gas.

In this paper we present N$_2$H$^+$ maps of 57 low mass cloud cores 
made at the FCRAO 14--m antenna equipped with the QUARRY receiver (\cite{egn92}). 
The list of objects is the same as in BCM98 and it includes
all cores already mapped in NH$_3$(1,1).  The angular resolution of the 
present observations (54$^{\prime\prime}$) is 1.5 times finer than the 
previous NH$_3$ study, allowing a more detailed analysis of the morphology 
and internal motions of dense cores. Technical details of the observations made at 
FCRAO are reported in Sect.~2. Results and discussion 
of this mapping program, including
integrated intensity maps, column density, size, and mass estimates, 
velocity gradients, intensity profiles, and the variation of line widths
across the cores,  are shown in Sect.~3. The main conclusions of this work 
are summarized in Sect.~5.  

\section{FCRAO Observations}

The $J$ = 1$\rightarrow$0 N$_2$H$^+$ observations were made in March 
and June 1995 and in March 1996 at the FCRAO 14 m telescope at New Salem, 
Massachusetts.  We used the 15 element QUARRY receiver (\cite{egn92}) 
and the autocorrelation spectrometer with a bandwidth of 20 MHz
over 1024 channels, giving a spectral resolution of $\sim$20 kHz or 
0.063 km s$^{-1}$.  The beam efficiency ($\eta_{\rm B}$) was 0.51 at 
93 GHz (\cite{pds97}) and the main beam width at half power (HPBW) 
was 54$^{\prime\prime}$.  The typical system temperature at 93 GHz was 
$\sim$500 K.  The data were acquired in frequency--switching mode with 
a throw of 8 MHz and calibrated using the ambient load vane method.  The 
rms pointing error, estimated by observing Venus and Mars, was $\sim$ 
5$^{\prime\prime}$.

\section{Results and discussion}

\subsection{Integrated intensity maps}

We have completed 13 Nyquist sampled maps (25$^{\prime\prime}$ spacing) and 
34 beam sampled maps (50$^{\prime\prime}$ spacing) to intensity levels below 
half maximum. An additional 
10 beam--sampled maps are not complete below the half maximum contour.  
Table~\ref{tpeak} reports in columns 2 and 3 the 
coordinates of the (0,0) map position.  These do not always correspond to 
the coordinates quoted in BM89 because of recentering
of the map to include all the emitting region.  The peak position 
is reported in columns 4 and 5 as offsets from the (0,0) coordinates. 
In column 6, the one--sigma level of the noise in the off-line 
channels (in antenna temperature units) is listed.  The integrated 
intensity of the emission at the map peak is in column 7; the error 
on the integrated intensity is $\sigma_{\rm I}$ = $\Delta T_{\rm A}^*$
$\times$ $\sqrt{N_{\rm ch}}$ $\times$ $\Delta v_{\rm res}$, where 
$N_{\rm ch}$ is the number of channels in the integrated intensity (listed 
in column 8),
and $\Delta v_{\rm res}$ (= 0.063 km s$^{-1}$) is the velocity resolution.  
Column 9 of 
Tab.~\ref{tpeak} indicates the map grid spacing: beam sampled maps 
(50$^{\prime\prime}$ spacing) are 
indicated with a ``B'', whereas Nyquist sampling 
(25$^{\prime\prime}$ spacing) is marked by an ``N''.
The size of the mapped area is given in column 10. The association with 
an IRAS source is indicated in column 11.

A sample of the \NTHP (1--0) data is shown in Fig.~\ref{fspectra}, where 
the averaged spectrum, i.e. the sum of all the spectra
inside the half maximum map contour, is presented for three starless cores 
(L1498, L1544, TMC--2) and three cores with stars (L1489, L1228, L1251E).
The line profiles dramatically change from quiescent starless cores, such as 
L1498, to L1251E, the core with the most complex velocity structure (see 
Sect.~\ref{sdv}).
Integrated intensity maps are shown in Figure~\ref{fmap}.  All the maps 
have the same angular scale and the contours are between 20\% and 
95\% of the peak, in steps of 0.15 $\times$ the peak intensity.  The 
mapped area is shown in the figure (the map type is reported in 
Tab.~\ref{tpeak}). IRAS sources associated with cores are marked with stars.
The criteria of association are described in \cite{jma99}.

\subsection{Column density}
\label{scolumn} 

The N$_2$H$^+$(1--0) line presents hyperfine structure (e.g. \cite{cmt95}; BCM98) 
and the hfs fitting program in CLASS 
(\cite{fgl89}), with the
hyperfine frequencies adopted from Caselli et al. 1995,  has been 
used to determine LSR velocities ($V_{\rm LSR}$), intrinsic line widths
($\Delta v$), total optical depths ($\tau_{\rm TOT}$), and  excitation 
temperatures ($T_{\rm ex}$).  These parameters are listed in 
Table~\ref{thfs} for the peak spectrum and the spectrum averaged inside
the half maximum contour.  Column 6 of Tab.~\ref{thfs} reports the 
N$_2$H$^+$ column density ($N_{\rm tot}$), which  has been calculated 
following the procedure described in \cite{cwz01b}\footnote{The approximated 
method used by BCM98, which assumes $J_{\nu}(T_{\rm ex})$ = $T_{\rm ex}$, 
with $J_{\nu}(T)$ the 
equivalent Rayleigh--Jeans temperature, underestimates the \NTHP \ column density 
by a factor of about four compared to the present calculation.}. 
In the case of optically thin emission, $T_{\rm ex}$ = 5 K has been assumed 
to compute $N_{\rm tot}$.  It is interesting to note that the peak column density 
averaged over the whole sample is (7$\pm$5)$\times$10$^{12}$ cm$^{-2}$, 
about three times smaller than the averaged column density in BCM98 
(when the same method is applied).  This discrepancy is probably due to 
the smaller beam of BCM98 observations (factor of 2) and suggests 
the presence of density structure in the studied cores (see 
Sect.~\ref{sprofile}).

BCM98 found good correlations between N$_2$H$^+$ and NH$_3$
velocities and line widths, but no correlation between column 
densities of the two species.  The authors claimed that this 
result may be due to the different spatial resolutions
in the N$_2$H$^+$ and NH$_3$ studies, and to the fact that 
N$_2$H$^+$ emission may have a different peak position.  
In this paper, the peak of the N$_2$H$^+$ emission has been 
determined and we can then compare the peak column densities 
reported in Tab.~\ref{tpeak} with those of NH$_3$.  The result is 
shown in Fig.~\ref{fcorr}a, where only cores with $N/\sigma_{N}$ $>$ 2 
have been included.  Indeed, the entire sample does not show a significant
correlation between $N({\rm N_2H^+})$ and $N({\rm NH_3})$. The scatter 
may be due to the significant errors associated with N$_2$H$^+$
column densities, caused by the large uncertainties on the total 
optical depth probably related to the presence of excitation
anomalies (e.g. Caselli et al. 1995).  However, it is interesting to note 
that for starless cores the correlation between \NTHP \ and \AMM \ column 
densities is significant (best--fit line in Fig.~\ref{fcorr}a):
\begin{eqnarray}
N_{\rm starless}({\rm N_2H^+}) \times 10^{12} & = & (2\pm2) + (0.6\pm0.3) 
N_{\rm starless}({\rm NH_3})
\times 10^{14} \,\, {\rm cm}^{-2},
\end{eqnarray}
where the quantities following $\pm$ are 1--sigma uncertainties and 
where the linear correlation coefficient is $cc$ = 0.52.    
The presence of a young stellar object probably affects the chemistry in a 
way that differentiation between the two species starts to be evident.  
Fig.~\ref{fcorr}a shows that in ``\AMM -rich starred'' cores,  
the $N(\NTHP )/N(\AMM )$ column density ratio is smaller than in the rest
of the sample. This may indicate that the \NTHP (1--0) line is more optically
thick and probably more affected by the lower density foreground material, so that
a correct measurement of $\tau $ is difficult to obtain without a correct radiative
transfer calculation.  

The N$_2$H$^+$ integrated intensity is plotted versus the ``equivalent''
NH$_3$ integrated intensity ($T_{\rm A} \times \Delta v$) in   
Fig.~\ref{fcorr}b. Although the associated uncertainties are significantly
smaller than in the case of column density (see Tab.~\ref{tpeak}), the 
dispersion is still large.  As in the previous case, starless 
cores show a more significant correlation:
\begin{eqnarray}
I_{\rm starless}({\rm N_2H^+}) & = & (0.4 \pm 0.3)+(2.0 \pm 0.7) 
(T_{\rm A} \times \Delta v)_{\rm starless,NH_3} \,\, {\rm K \, km \, s}^{-1}, 
\,\,\,\, cc = 0.67. 
\end{eqnarray}
In Fig.~\ref{fcorr}c, N$_2$H$^+$ and NH$_3$ excitation temperatures are 
plotted. Once again, the correlation is strong for starless cores:
\begin{eqnarray}
T_{\rm ex , starless}(\NTHP ) & = & (1.4 \pm 0.8) + (0.4 \pm 0.1) 
T_{\rm ex , starless} (\AMM ),
\end{eqnarray}
with $cc$ = 0.79 (dotted line in Fig.~\ref{fcorr}c), whereas no significant
correlation is present in cores with stars.  Typically, $T_{\rm ex}$(NH$_3$) 
$>$ $T_{\rm ex}$(N$_2$H$^+$), which suggests that the critical density
$n_{\rm cr}$(NH$_3$) $<$ $n_{\rm cr}$(N$_2$H$^+$).  In fact, $n_{\rm 
cr}$(N$_2$H$^+$) = 2$\times$10$^5$ cm$^{-3}$ (\cite{ubg97}),
about one order of magnitude larger than that of NH$_3$ ($2\times 10^4$ 
cm$^{-3}$; \cite{s89}). 

Correlations between velocities and linewidths of the N$_2$H$^+$ and 
NH$_3$ peak spectra are identical to those found in BCM98 and we will not 
further discuss them. This paper will concentrate on the structure 
and internal motions of individual cores.  

\subsection{Sizes and masses}
\label{ssize}

Sizes of the mapped cores are listed in Tab.~\ref{tsize}.
The 2D gaussian fitting routine in GRAPHIC (\cite{bdd01}) has been used to find
the position angle, the major and minor axis (see columns 2, 3, and
4, respectively).  The reported source sizes have been corrected for
beam size, i.e. we subtracted the gaussian beam size in quadrature 
for each dimension.  Aspect ratios (column 5) range between 1.1 and 6.4
with mean $\pm$ standard deviation 
2.0$\pm$0.9. The size $r$ in column 6 and 7 is the 
half power radius, given by the geometric mean of the semimajor and semiminor 
axis.  Note that $r$ = $R$/2, where $R$ is the size listed in Table~\ref{tgradient}
of BM89. The cloud distance is reported in column 8.  
Comparing the size of \n2hp and \nh3 cores with associated stars we find a good 
correlation ($cc$ = 0.8):
\begin{eqnarray}
r_{\rm star}({\rm NH_3}) & = & (-0.02\pm0.03) + (2.2\pm0.4) 
 r_{\rm star}({\rm N_2H^+}) \,\, {\rm pc},    
\end{eqnarray}
which indicates that the emission of the two molecules have similar 
morphology, despite the different beam sizes.  The factor of 2 difference
is probably due to differences in the critical density $n_{\rm cr}$ of the
two tracers since map sizes are already corrected for beam smoothing and the
resolution ratio is 1.5, less than the typical ratio of radii, 1.8.  Indeed, 
it is interesting to note that the 
above relation pertains to cores with stars, given that for starless cores we find:
\begin{eqnarray}
r_{\rm starless}({\rm NH_3}) & = & (-0.01\pm0.01) + (0.9\pm0.3) 
  r_{\rm starless}({\rm N_2H^+}) \,\, {\rm pc} \,\, (cc = 0.7)    
\end{eqnarray}
These relations suggest that N$_2$H$^+$(1--0) is tracing higher density material 
than NH$_3$(1,1), and that cores with stars are denser and more 
centrally condensed than starless cores.  On the other hand,
in starless cores, the two lines originate from the same regions.
Figure~\ref{fhisto} shows the distribution of N$_2$H$^+$ and NH$_3$ core radii in 
(i) the entire sample, (ii) cores with stars, and (iii) starless cores; 
the two tracers span different size ranges only for cores with stars.   
On average, starless cores have smaller sizes than cores
with stars. For 35 cores with stars, the \n2hp \, map radius $r$ has mean 
$\pm$ standard error of the mean (s.e.m.) 0.069$\pm$0.005 pc, while for 19 
starless cores,  $<r>_{\rm starless}$ = 0.054$\pm$0.005 pc.

The virial mass of an equivalent uniform density sphere: 
\begin{eqnarray}
M_{\rm vir}({\rm M_{\odot}}) & = & 210 \times r({\rm pc}) \times 
\Delta v_{\rm m}^2({\rm km^2 \, s^{-2}}),  
\label{evir}
\end{eqnarray}
is listed in column 3 of Tab.~\ref{tdensity}.  The corresponding number density 
$n_{\rm vir}$ is in column 2.  $\Delta v_{\rm m}$ in eqn.~\ref{evir} is the FWHM 
of the molecule of mean mass (2.33 amu, assuming gas with  90\% of H$_2$ 
and 10\% He):
\begin{eqnarray}
\Delta v_{\rm m}^2 & = & \Delta v^2 + 8 ln 2 \frac{k T}{m_{\rm H}}
\times \left( \frac{1}{2.33} - \frac{1}{m_{\rm N_2H^+}} \right),
\end{eqnarray}
where $\Delta v$ is the intrinsic linewidth of the N$_2$H$^+$
peak spectrum (see Tab.~\ref{thfs}), and $m_{\rm N_2H^+}$ (= 29 amu) is the
mass of the \n2hp molecule.  Column 4 reports the N$_2$H$^+$ fractional 
abundance: $X({\rm N_2H^+})$ = $N({\rm N_2H^+})/N({\rm H_2})$, where 
$N({\rm H_2})$ = (4/3) $\times$ ($n_{\rm vir}$/1.11) $\times$ $r$, and the
factor 1.11 takes into account the difference between $n_{\rm vir}$, the 
``virial'' number density of the molecule of mean mass, and
$n_{\rm vir}({\rm H_2})$.  The N$_2$H$^+$ column density used for estimating
$X(\rm N_2H^+)$ is the peak or the average column density, depending 
on the associated errors.  The average value is 
$<X({\rm N_2H^+})>$ = (3$\pm$1)$\times$10$^{-10}$, close to that
 found in BCM98.

Tab.~\ref{tdensity} also gives
in column 5 the volume density $n_{\rm ex}$,  calculated from the 
($n_{\rm ex}/n_{\rm cr}^{\prime}$) ratio (see eqn.~\ref{encr} below), 
where $n_{\rm cr}^{\prime}$ is the critical density (corrected for
trapping) of the N$_2$H$^+$(1--0) line; 
and in column 6 the ``excitation'' mass of a uniform core ($M_{\rm ex}$ = (4/3) 
$\pi m n_{\rm ex} r^3$).  
The quantity $n_{\rm ex}/n_{\rm cr}^{\prime}$ has been obtained by using 
the expression for two--level statistical equilibrium (\cite{g92}):
\begin{eqnarray}
\frac{n_{\rm ex}}{n_{\rm cr}^{\prime}} & = & 
\frac{\tilde{T}_{\rm ex} - \tilde{T}_{\rm cb}}
{\frac{h \nu}{k} \left( 1 - \frac{\tilde{T}_{\rm ex}}{\tilde{T}_{\rm kin}}
\right)}
\label{encr}
\end{eqnarray}
where $\nu$ is the transition frequency, $h$ and $k$ are the 
constants of Planck and Boltzmann, respectively, 
\begin{eqnarray}
n_{\rm cr}^{\prime} & = & n_{\rm cr} \times \frac{1 - exp (- \tau)}
	{ \tau} \label{encrp}, 
\end{eqnarray}
and
\begin{eqnarray}
\tilde{T} & = & \left( \frac{h \nu}{k} \right) 
\left( 1 - exp \left( - \frac{h \nu}{k T} \right) \right)^{-1} .
\end{eqnarray}
In eqn.~\ref{encrp}, 
$\tau$ is the optical depth of a ``typical'' hyperfine component ($\sim$ 
$\tau_{\rm TOT}$/9). 
$T_{\rm ex}$, $T_{\rm cb}$, and $T_{\rm kin}$ in eqn.~\ref{encr} 
are the excitation, the cosmic 
background, and the kinetic temperatures, respectively. We assumed 
$T_{\rm cb}$ = 2.7 K, $T_{\rm kin}$ = 10 K (see BM89), whereas $T_{\rm ex}$
is the excitation temperature of the averaged spectrum (see Tab.~\ref{thfs}).
The error on $n_{\rm ex}$ is obtained by propagating the error associated
with $T_{\rm ex}$ and $\tau$ into eqn.~\ref{encr} (see Appendix~\ref{aerror}).

In Fig.~\ref{fvir} $M_{\rm ex}$ is plotted as a function of $M_{\rm vir}$ for
the whole sample (with the exception of cores with 
(i) $M_{\rm ex}/\sigma_{M_{\rm ex}}$ or $M_{\rm vir}/\sigma_{M_{\rm vir}}$ $<$ 2, 
(ii) assumed excitation temperature, and (iii) deconvolved size less than
the beam size, as in the case of B335).  
In average, starless cores are less massive than 
cores with stars: $<M_{\rm vir}>_{\rm star}$ = 9$\pm$3 M$_{\odot}$, 
$<M_{\rm vir}>_{\rm starless}$ = 3.3$\pm$0.4 M$_{\odot}$.
  However, there is not significant difference in the 
$M_{\rm ex}/M_{\rm vir}$ ratio between the two classes of cores
(in average $<M_{\rm ex}/M_{\rm vir}>_{\rm star}$ = 1.4$\pm$0.3, and
$<M_{\rm ex}/M_{\rm vir}>_{\rm starless}$ = 1.3$\pm$0.3). 
We note that the assumption of a uniform density sphere to estimate the
virial mass is a very crude one.  In Sect.~\ref{sprofile} we will show that
spheroidal cores are consistent with density profiles of the form 
$n(r)$ $\propto$ $r^{-\alpha}$, with $<\alpha>$ $\sim$ 2, although most of the
starless cores present ``central flattening'', at impact parameters $b$ $\leq$
5000 AU. If cores were approximated with singular isothermal spheres, our virial 
mass estimates should be reduced by a factor of 1.6.

\subsection{Velocity Gradients}

Following \cite{gbf93} (hereafter GBF93), a least squares fitting of a 
velocity gradient has been performed in all the cores with at least 9 
positions with a good determination of 
the LSR velocity $V_{\rm LSR}$.  In Table~\ref{tgradient}, the magnitude 
of the velocity gradient ${\cal G}$ and its direction ($\theta_{\cal G}$,
the direction
of increasing velocity, measured east of north) are reported in column 3
and 4, respectively; the number of velocity points used in the fit is in 
column 2; the product
between ${\cal G}$ and the core size $r$, or the typical velocity shift across
the map, is in column 5; the ratio $\beta$ of 
rotational kinetic energy to gravitational energy (see eqn. (6) in 
GBF93) is shown in column 6.  

For thirteen of the cores in Tab.~\ref{tgradient}, the same quantities have 
been calculated by GBF93 using NH$_3$ maps (see their Table 2).  
By comparing the magnitude of the velocity gradient calculated
from N$_2$H$^+$(1--0)  and NH$_3$(1,1) data (${\cal G}_{\rm N_2H^+}$ and 
${\cal G}_{\rm NH_3}$, respectively) we found no correlation and the majority 
of the cores (with the exception of L1495 and L1251E) have  
${\cal G}_{\rm N_2H^+}$ $>$ ${\cal G}_{\rm NH_3}$, although the 
scatter is large (the average of the 
${\cal G}_{\rm N_2H^+}/{\cal G}_{\rm NH_3}$ ratio $\pm$ standard deviation is 
 1.6$\pm$1.0).  On the other hand, the correlation between  
$\Theta_{\cal G} ({\rm N_2H^+})$ and $\Theta_{\cal G} ({\rm NH_3})$
is significant ({\it cc} = 0.7).  The slightly larger magnitude of the 
velocity gradient in most of N$_2$H$^+$ cores probably reflects the 
finer spatial resolution of these observations compared to that for 
NH$_3$.  In fact, a larger beam will tend to smooth out 
$V_{\rm LSR}$ variations across the map. We note, however, that this also
implies broader \AMM \ lines, which are not observed (see end of 
Sect.~\ref{scolumn} and BCM98).  
Finally,  $\beta$ is similar in N$_2$H$^+$ and NH$_3$ cores: 
in those cores which are in common in the two samples, the average value 
of $\beta$ is $\sim$ 0.02 from N$_2$H$^+$ data, and $\sim$ 0.03 from NH$_3$ data. 

Tab.~\ref{tgradient} also lists the average value and its standard 
deviation (columns 
7 and 8, respectively) of the fit residuals $V_{\rm LSR}(i)$ - 
$V_{\rm fit}(i)$ across the core, where $V_{\rm fit}(i)$ is the LSR 
velocity at position ($\alpha(i)$, $\beta(i)$) determined by the least square
fit of a linear
velocity gradient.  These quantities are useful to describe the more 
complex motions in the core and will be discussed in section
3.6.

For 12 cores it has also been possible to determine the magnitude
and the direction of ``local'' velocity gradients, i.e. variations
of $V_{\rm LSR}$ in portions of a cloud core.  
The selected cores have at least 12 observed positions where the 
determination of $V_{\rm LSR}$ is possible via hfs fitting (see Sect.~3.2).  
In each of these
cores, the least square fitting routine to determine ${\cal G}$ and 
$\Theta_{\cal G}$ has been successively applied to adjacent grids
of 3$\times$3 points spaced by $\leq 35^{\prime\prime}$ or 
$\leq 71^{\prime\prime}$ (depending on the map type; see Tab.~\ref{tpeak}, 
column 8).  Incomplete grids of at least 7 points have also been included
in the computation of ``local'' velocity gradients.
The results are shown in Fig.~\ref{fgradient}, for the 12 selected cores, together 
with the ``global'' velocity gradient.  The arrows point toward increasing
$V_{\rm LSR}$. From the figure it is evident that many cores present 
internal variations of the magnitude and direction of the velocity gradient 
(see, in particular, L1228 and L1251E), which illustrates the presence of 
complex motions not interpretable as simple solid body rotation of the
whole core.  

The velocity gradient maps of Fig.~\ref{fgradient} show a wide range of structure
which departs from the uniform rotation model used by GBF93
to analyse the velocities of the corresponding NH$_3$ maps which generally had
fewer data points than do the present N$_2$H$^+$ maps.  Of the 9 maps in 
Fig.~\ref{fgradient} having at least 5 velocity gradient vectors, only 
2 -- L1512 and L1221 -- have velocity gradient maps which are nearly uniform
in both magnitude and direction, indicating simple uniform rotation.  The rest
show patterns which are fairly uniform in magnitude but not direction (L183, 
L1544), uniform in direction but not magnitude (L483), or which have significant
variation in both magnitude and direction (L1228, L1251E, L43, and L1498). The
most complex pattern is in L1251E, which shows distinct reversal of gradient
direction between the East and West parts of the core.  The departures from 
uniform rotation are most pronounced in the cores having associated YSOs and 
outflows (L483, L43, L1228, and L1251E).  But substantial departures are
also evident in the starless cores L183, L1544, and L1498 -- which show 
evidence of contracting motion in CS(2--1) (\cite{tmm98}; \cite{lmt99}, 2000),
although HCO$^+$(3--2) lines  show no infall asymmetry toward L1498 
(\cite{ge00}).
These results are consistent with a picture where simple
uniform core rotation is rare, where more turbulent motions are relatively
common, and where such turbulence is associated with cores with YSOs and outflows
or with starless cores having evidence of inward flows.  An interpretation of 
dense core velocity gradients in terms of turbulent motions was recently
presented by \cite{bb00}.

We note that cores with complex velocity gradient maps should be poorly fit 
by a simple model of uniform rotation, and inspection of the ``normalized'' 
standard deviation of the fit residuals 
$s_{<V_{\rm LSR} - V_{\rm fit}>}/({\cal G} \times r)$, as well as 
$s_{<V_{\rm LSR}-V_{\rm fit}>}$ (see Tab.~\ref{tgradient}), 
bears out this expectation.  The most uniform pattern of gradient vectors, for 
L1512 (Fig.~\ref{fgradient}), corresponds to one of the smallest ``normalized'' 
fit residuals, 0.3, while the most complex pattern,
for L1251E, has one of the largest 
$s_{<V_{\rm LSR}-V_{\rm fit}>}/({\cal G} \times r$), 2.4 (and the largest
$s_{<V_{\rm LSR}-V_{\rm fit}>}$, 0.36 km s$^{-1}$). The ``normalized'' standard
deviation relative to L1498 lies between L1512 and L1544.  This suggests 
that L1498 is probably in an evolutionary stage later than the ``static'' 
core L1512 and earlier than the collapsing L1544. In fact, L1498 has been 
described as an extremely quiescent core (\cite{lww95}, \cite{wlg97}) with evidence 
of slow contraction, outer envelope growth and strong chemical differentiations
(\cite{klv96}; \cite{tmc01}).  Large values 
of $s_{<V_{\rm LSR}-V_{\rm fit}>}/{\cal G} \times r$ ($\sim$ 1) are also present 
in TMC--2, an infall candidate with CS asymmetry (\cite{lmt00}).

\subsection{Integrated intensity profiles}
\label{sprofile}

The ``standard model'' of isolated star formation (\cite{sal87}) states 
that cores lose 
magnetic support by ambipolar diffusion until they become so 
concentrated that the central regions are gravitationally unstable
and start to collapse.  The collapse of the central region deprives 
the above layers of pressure support and causes them to also fall towards
the center.  In this way, gravitational collapse propagates from the 
inside out and continues until the core runs out of mass or a powerful
wind from the central star-disk system reverses the collapse and disperses
the core.  

Some critical parameters of the ``standard model'' are unconstrained.
The two most crucial unknowns are the radial dependence of density and 
turbulent velocity (alternatively of magnetic field), which together 
determine how gravitational collapse
occurs.  For example, a change in the density power law from $r^{-2}$
(isothermal sphere, \cite{s77}) to $r^{-1}$ (logotropic sphere, 
\cite{mp96}), changes completely the core star forming 
properties (\cite{mp97}), and our observations are not
fine enough at this point to rule out any of these options.  In fact, 
our ignorance of  these basic core properties constitutes the most
serious limitation in our understanding of how stars form in isolated 
dense cores. 

From submillimeter continuum dust emission, \cite{wsh94}
found that the radial density profiles of pre--stellar cores are significantly
different from the singular isothermal sphere, and qualitatively consistent
with models of magnetically--supported cores undergoing ambipolar diffusion
(e.g. \cite{cm95}).  Typically, the radial density 
profile inferred assuming a constant dust temperature is as flat as 
$\rho(r)$ $\propto$ $r^{-\alpha}$, with $\alpha$ $\sim$ 0.4--1.2, depending on the
core shape, at radii less than $\sim$ 4000 AU, and approaches      
$\rho(r)$ $\propto$ $r^{-2}$ only between $\sim$ 4000 AU and $\sim$
15000 AU (\cite{awm96}; \cite{wma99}; \cite{all01}).  However, recent model 
calculations of the 
dust temperature in pre--stellar cores (\cite{zwg01}; \cite{ers01})
predict a temperature gradient, with a drop from $\sim$ 14 K at the edges
to $\sim$ 7 K at the centers, which implies more peaked density 
distributions than in the isothermal case.   
On the other hand, millimeter
continuum observations of circumstellar envelopes of low--mass protostars are 
in good agreement with the standard protostellar model of Shu et al. (1987),
with power--law density gradients such as $\rho(r)$ $\propto$ $r^{-2}$ or 
$r^{-1.5}$ (\cite{ma00}).

N$_2$H$^+$(1--0) maps can be used to investigate the column density 
structure of dense cores and make comparisons with results from
submillimeter maps.  We already noted (see Sect.~3.2) that the 
uncertainty associated with N$_2$H$^+$ column density is quite large,
especially for low sensitivity spectra (such as those away from the 
map peak), because of the difficulty in 
determining an accurate value of the total optical depth of the 
J=1--0 transition.  Therefore, instead of using N$_2$H$^+$ column density 
profiles, we made plots of the N$_2$H$^+$(1--0) intensity integrated 
below the seven hyperfine components as a function of impact parameter. 
The use of integrated intensity may be dangerous in those cases
where the optical depth is large (when the column density is no longer 
simply proportional to the integrated intensity), but we will see that 
our  conclusions are not affected by this problem. 

The integrated intensity profiles of spheroidal cores (with aspect ratio
$\leq$ 2, see Tab.~\ref{tsize}) have been fitted using two models.
Model 1 consists of a spherically symmetric cloud model with a density 
profile $\rho(r)$ $\propto$ 
$r^{-\alpha}$.  The resultant intensity profile vs. impact parameter
$b$ ($I$ $\propto$ $b^{-p}$, with $p$ = $\alpha-1$), has been convolved
with a 2D Gaussian, with FWHM equal to the FCRAO beam (54$^{\prime\prime}$).  
For each core, we change the value of 
$p$ from 0.5 to 2.0, in steps of 0.1, and find the best $\chi^2$ 
convolved profile and the corresponding $\alpha$ value. 
Fig.~\ref{fprofiles} (thin curves) and 
Tab.~\ref{tprofile} (columns 2 and 3) show the results of this procedure.  

Model 2 considers a spherically symmetric cloud with a radial density 
profile inferred from submillimeter continuum observations of 
spheroidal cores, with $\rho(r)$ $\propto$ $r^{-1.2}$ at $r < r_{\rm break}$, 
and $\rho(r)$ $\propto$ $r^{-2}$ at $r > r_{\rm break}$ (see e.g. 
~\cite{awb00}). The two profiles have been convolved with the 
FCRAO beam and joined at $r_{\rm break}$. We run several models with
different values of impact parameter at the break ($b_{\rm break}$
from 10$^{\prime\prime}$ to 100$^{\prime\prime}$, in steps of 
5$^{\prime\prime}$). From the $\chi^2$ minimization, we determined
$b_{\rm break}$ for each core (see column 4 of Tab.~\ref{tprofile}).
Model 2 profiles are shown in Fig.~\ref{fprofiles} by the dashed 
curves.

The comparison between the $\chi^2$ values of Model 1 and 2 
(columns 3 and 5 of Tab.~\ref{tprofile}) allows one to find
the appropriate model density profile for each object.  It is 
interesting to note that most (6 out of 9) of the cores with stars are best
fitted with single power laws.  With the exception of Per 6, L1495, and 
L43, all cores with stars have $b_{\rm break}$ = 10$^{\prime\prime}$  
(the minimum value in Model 2), which is equivalent to having a
single slope.  Starless cores show a different behaviour: 6 out of 8 
objects have intensity profiles consistent with central flattening
at impact parameters less than $\sim$ 5000 AU.  This is in good agreement
with results from submillimeter continuum observations.  
These results are interpreted by \cite{wma99}
as indicating that the cores are probably magnetically--supported and evolving 
through ambipolar diffusion to star formation (e.g. \cite{ls89}; 
\cite{bm95}).  However, caution must be used 
with the above models because, as pointed out by \cite{awm96} and
\cite{awb00}, they require fairly strong magnetic fields on parsec scales 
($\sim$ 100 $\mu$m), difficult to reconcile with available Zeeman 
measurements (e.g. \cite{c99}).  Higher spatial resolution observations
are needed to make quantitative conclusions on the column and volume
density structure of star forming cores and better constrain theory.

In cores with stars, $\alpha$ is typically greater than 2 
(Tab.~\ref{tprofile}).  This is probably 
due to an excitation temperature increase at the center caused by the presence
of a sufficiently luminous protostar.  A similar result (greater map
``spikiness'' of cores with stars over that of starless cores) was found by
\cite{moh94} using H$^{13}$CO$^+$ data.  From Fig.~\ref{fprofiles} we also 
note that the model integrated intensity tends to overestimate data points
at large impact parameters.  This reflects the fast drop in 
$T_{\rm ex}$ caused by the density drop and it is well reproduced by 
Monte Carlo simulations. It is not related to the ``sharp 
edges'' found in isolated prestellar cores at radii $>$ 15000 AU
(\cite{bap00}).

To check the effects of optical depth in the observed shallow profiles, 
in those
cores with $\tau_{\rm TOT}$ \footnote{$\tau_{\rm TOT}$ is the sum of the peak 
optical depths of the seven hyperfine components} 
$>$ 10 (L1498, L1489, and L483)  we made 
integrated intensity profiles by using the area below the thinnest hyperfine 
component of N$_2$H$^+$(1--0) having the least optical depth 
(F$_1$,F = 1,0 $\rightarrow$ 1,1; \cite{cmt95}).  
In Fig.~\ref{fprofiles}, the dotted curves indicate these profiles, 
normalized so that the peak integrated intesity of the ${\rm F_1, F}$ = 
${1,0}$ $\rightarrow$ ${1,1}$
component is the same as the peak integrated intensity of all the
components.  The dotted curves in the L1498, L1489, and L483 plots
 (Fig.~\ref{fprofiles}) closely follow the data points, suggesting that
optical depth effects are not affecting our conclusions.  

In summary, the radial density profiles of starless cores 
in Fig.~\ref{fprofiles} present a consistent
picture of ``central flattening'' where the shape of the profile is 
shallower at small radii than at large radii.   The N$_2$H$^+$ 
integrated intensity profile is modelled by a spherically
symmetric density law $n \sim r^{-\alpha}$ where $\alpha$ = 1.2
for $r < r_{\rm break}$ and $\alpha$ = 2.0 for $r > r_{\rm break}$ $\sim$ 
0.03 pc. Cores with stars  are better modelled 
with single power laws $n \propto r^{-\alpha}$ with $\alpha$ $\geq$ 2.
These results are the first to show central flattening 
in a significant sample of molecular line maps.  In contrast to dust continuum maps,
these maps offer the prospect of relating core density structure to core 
turbulence structure, in more detailed studies to be made in the future. 
The agreement between dust and N$_2$H$^+$ profiles suggests that the 
molecular gas traced by the N$_2$H$^+$ line is not significantly depleted
in relation to the dust, unlike CO and CS (\cite{cmt99}; \cite{cwz01a},b; 
\cite{bcl01}; \cite{tmc01}).

\subsection{Variation of $\Delta v$ across the cores}
\label{sdv}

\cite{gbw98} described a 
physical picture of star--forming dense cores and their 
environs where the cores are ``velocity coherent'' regions
of nearly constant line width.  
From NH$_3$ maps, \cite{bg98} found that
within the interiors of dense cores, the line widths are roughly constant 
and appear to increase at the map edges. Although many theories of low mass 
star formation begin with 
an isothermal sphere having no turbulence 
(e.g. \cite{s77}), the line width inside the coherent 
cores is not purely thermal.  A clearly measurable turbulent 
component remains even in these ``coherent'' regions (\cite{fm92};
\cite{camy95}).  
The ``transition to coherence'' may occur because of a decrease in the 
magnetic field's ability to control gas motions in regions of very low 
ionization (\cite{m91}; \cite{m97}; \cite{gbw98}; \cite{m98}).

We have made a similar study with our \n2hp maps in those cores
having a sufficient 
number of high sensitivity spectra across the map
to allow us to see systematic variations in line width, if they are present.  
In Fig.~\ref{fdvmap}, integrated intensity contour maps
of 9 cores selected to have at least 9 spectra with 
$\Delta v$/$\sigma_{\rm \Delta v}$ $\geq$ 3 and $I/\sigma_{\rm I}$ $\geq$ 5 are 
superposed on grey scale maps of line width (light grey indicates narrow lines).
From the figure, it is evident that starless cores 
show a spotty pattern of low line width at central positions
 inside the integrated intensity half--maximum 
contour, whereas more internal structure is present in cores with stars. 
In fact, L43 (\cite{bpi98}),  L483 (\cite{tmm00}, \cite{fw00}, 
\cite{ppo00}) and L1228 (\cite{tm97})
are associated with well known protostellar outflows, which may contribute
to the broadening of N$_2$H$^+$ line widths.  However, in starless cores,
the broadest lines are often located at the edges of the map.  
The positions indicated by dots on the maps in Fig.~\ref{fdvmap} are those 
where a good ($\Delta v$/$\sigma_{\Delta v}$ $\geq$ 3) estimate of the intrinsic 
line width via hfs fitting was possible.  

Another way of looking at core velocity structure is to consider plots similar 
to those shown by \cite{gbw98}, where the intrinsic line width
is reported as a function of the antenna temperature, which in turn is used 
as a measure of the distance from the peak of the map.  Instead of using
antenna temperature we consider the impact parameter $b$, already
introduced in Sect.~3.5.  In practice, at each map point $i$ we 
 associate an ``effective radius'' $b(i)$ by counting the number of positions
$N_{\rm mp}$ with integrated intensity $I$ equal or larger than $I(i)$, so 
that $b$ = $\sqrt{(a \times N_{\rm mp})/\pi}$, where $a$ is the area 
of the map pixel.  Thus we can consider in more quantitative fashion how the
line width varies with effective map radius.  We point out that 
this way of looking at core coherence is equivalent to that 
described in \cite{gbw98} (we tried both methods for three cores
and found essentially the same results), but this approach shows the dependence
on size scale explicitly.

We first examine how core line widths vary with effective radius by fitting 
a simple linear relation between $\Delta v$ and $b$.  
Table~\ref{tdeltav} (columns 2, 3, and 
4) lists the intercept $p$, the slope $q$, and the correlation coefficient 
$cc$ of the linear least squares fit to the $\Delta v$ -- $b$ data for each 
core (only those cores with at least 9 positions having $\Delta v/\sigma_{\rm 
\Delta v} \geq$ 3 and $I/\sigma_{\rm I}$ $\geq$ 5 are included in the table).  
From Tab.~\ref{tdeltav}, three classes 
of cores are recognized: a) cores with a ``significant'' ($cc$ $\geq$ 0.4)
positive $\Delta v$ -- $b$ correlation (L1498, L1495, L1524, L1400K, L260); b) 
cores with no significant correlation; and c) 
cores with a significant negative correlation (PER4, B5, TMC--1C2, L1174).  
Starless cores and cores with stars are found in all three
classes.  Fig.~\ref{fdvb}  shows some example of the three classes.  
Evidently the slight increase of single--tracer 
line width with effective 
map radius discussed by \cite{gbw98} is significant in a few cores, 
but these are not representative of the 22 cores in Tab.~\ref{tdeltav}.

We next consider how the variation in line width relates to the mean line
width, for all the usable spectra in the map.
Columns 5 and 6 of Tab.~\ref{tdeltav}, report for each core 
the average linewidth $<\Delta v>$
= $(1/M) \times \sum_{i=1}^{M} \Delta v(i)$, 
with $M$ = number of positions in the 
map where $\Delta v$ has been calculated), and
the corresponding standard deviation of the sample population  
$s_{<\Delta v>}$ ($\equiv$ $(1/\sqrt{M-1}) \times \sqrt{\sum_{i=1}^M (\Delta 
v_i - < \Delta v> )^2}$ ). Note that $<\Delta v>$ is 
different from $\Delta v$ of the average spectrum given in Tab.~\ref{thfs}.
The corresponding errors are reported in Appendix~\ref{aerror}.

These data can be compared with a simple statistical model of ``cells''
or ``zones'' along the line of sight to estimate the ``coherence
length'' or length over which the motions are correlated along
the line of sight (\cite{kd87}). We assume that   
each cell moves as a coherent unit, with a velocity along the line
of sight that follows a Gaussian probability distribution with 
the dispersion $\sigma$ equal to the nonthermal component of the
overall velocity distribution that we see in the line profile
($\sigma^2$ = $\sigma_{\rm NT}^2$ = $\Delta v ^2/(8 ln2)$ - 
$kT/m_{\rm obs}$, where $k$ is the Boltzmann constant, $T$ is the kinetic 
temperature, and $m_{\rm obs}$ is the mass of the observed molecule).  Then
from basic statistics we can write relations for the rms of the line
width and the rms of the mean velocity:
\begin{eqnarray}
\sigma_{<\Delta v_{\rm NT}>} & = & \frac{<\Delta v_{\rm NT}>}{\sqrt{N}}
\label{esigma1} \\
\sigma_{<v>} & = & \frac{\sqrt{<\Delta v_{\rm NT}^2>}}{\sqrt{8 ln 2 N}}
\label{esigma2}
\end{eqnarray}
where $N$ is the typical number of cells along the line of sight.  
Figure~\ref{fsigmadv} shows $\sigma_{<\Delta v_{\rm NT}>}$  versus 
$<\Delta v_{\rm NT}>$.  In the figure, thin lines represent 
eqn.(~\ref{esigma1})
for different $N$ values.   A linear least square fit to the data, taking into 
account the errors on $\sigma_{\Delta v_{\rm NT}}$, gives 
$N$ = 10, with a linear correlation coefficient $cc$ = 0.9.  

Relation (\ref{esigma2}) is investigated in Fig.~\ref{fdvvlsr}, where the 
dispersion of the average velocity gradient fit residuals 
$<V_{\rm LSR} - V_{\rm fit}>$ is plotted versus 
the rms of the nonthermal line width $\sqrt{<\Delta v_{\rm NT}^2>}$.  
$V_{\rm fit}$ is
the velocity predicted by fitting a first--order gradient, so that 
$V_{\rm LSR} - V_{\rm fit}$ can be used to analyse higher order structure 
in the velocity field.  In this case, a linear least square fit to the 
data gives $N$ = 13, with $cc$ = 0.5.  
Only cores with at least 9 data 
points available to estimate means and standard deviations have been included 
in Fig.~\ref{fsigmadv} and \ref{fdvvlsr}.  
L1251E has been excluded from 
the two figures because of the complex velocity structure clearly 
seen in Fig.~\ref{fgradient}, suggestive of two adjacent cores rotating
in almost opposite directions.  

In each of Fig.~\ref{fsigmadv} and \ref{fdvvlsr}, there is a clear tendency 
for the ``dispersion'' quantity on the $y$--axis to correlate with the ``mean''
quantity on the $x$--axis.
From this simple ``cell'' model we conclude that $N$ $\simeq$ 10.
Given that the length along the line of sight to which our observations
are sensitive is comparable to the map diameter, typically 0.1 pc, 
then the ``coherence length'' deduced
from our data is about 0.01 pc. This size scale is comparable 
to the ``cutoff''  wavelength below which Alfv\'en waves cannot 
propagate because the neutral--ion collision frequency in the neutral
medium becomes comparable to or less than the wave frequency
(\cite{mz95}; \cite{gbw98}):
\begin{eqnarray}
\lambda_{\rm cut} ({\rm pc}) & = & 0.007 \frac{(B/10 \mu {\rm G})}
   	{(x_i/5 \times 10^{-8})(n/10^5 {\rm cm^{-3}})^{3/2}},
\end{eqnarray}
where $B$ is the magnetic field strength, $x_i$ is the ionization 
fraction ($n_i/n$), and $n$ is the volume density of the molecule of 
mean mass (2.33 amu).  The choice of $B$ $\sim$ 10 $\mu$G is based on the 
OH Zeeman measurements in L1544 (a core in our sample) of \cite{ct00}. At the
typical densities of the observed cores ($\sim$ 10$^5$ \percc , see 
Tab.~\ref{tsummary}), the ionization fraction is about 5$\times$10$^{-8}$, if the
standard relation between $x_i$ and $n(\MOLH )$ is used ($x_i$ = 1.3 $\times$ 
10$^{-5}$ $n(\MOLH )^{-0.5}$, \cite{m89}\footnote{We note that \cite{cwz01b} have 
recently found a different $x_i$--$n(\MOLH )$ relation for the L1544 core: $x_i$ = 
5.2$\times$10$^{-6}$ $n(\MOLH )^{-0.56}$, which, at $n(\MOLH )$ = 10$^5$ \percc ,
 implies an electron fraction of 8$\times$10$^{-9}$, six times lower than that 
deduced from the ``standard'' relation.  With this new value of $x_i$, the 
``typical'' cutoff wavelength is about 0.001 pc.}). 
Therefore, the ``transition to coherence'' 
may arise from insufficient wave coupling on size scales 
of $\sim$ 0.01 pc, as (i) proposed by \cite{m91}, (ii) elaborated by 
Myers (1997, 1998) as a reason why \AMM \ line widths are
nearly thermal and narrower than their surrounding 
\THCO \ gas (as demonstrated by \cite{fm92} and \cite{camy95}), and (iii) 
suggested by  \cite{gbw98} as a reason why core line widths are ``coherent''
-- nearly constant in the interior of \AMM \ maps.
In general, the positive correlations shown in Fig.~\ref{fsigmadv}
and \ref{fdvvlsr} strongly suggest the presence of internal motions,  
not ascribed to simple solid-body rotation and probably due to 
turbulence,  which broaden lines and contribute to line 
width and LSR velocity dispersion across the cores. 
 
To further quantify the ``coherence'' of core line widths, we ask whether 
the ``local'' dispersion of line widths at radius near $b$, 
$s_{\Delta v}$, increases with $b$ in N$_2$H$^+$ cores.
To answer this question, the standard deviation of the average
$\Delta v$ inside all the contour levels between 20\% and 100\% of the 
integrated intensity peak has been calculated (in steps determined 
by the following requirements: i) consecutive levels are associated
with different impact parameters; ii) each level contains at least 
5 positions).  As with previous analysis, we only considered 
those positions where $\Delta v/\sigma_{\Delta v} \geq 3$ and 
$I/\sigma_{\rm I} \geq 5$.  We then performed linear least square 
fits to the $s_{\Delta v}$--$b$ 
data in each core and the results are listed in columns 7, 8, and 9
 of Tab.~\ref{tdeltav}. Note that all the cores (with the exception of 
L1495, L1536, and L1512) show a strong positive correlation between the
two quantities, indicating that dispersion is indeed 
increasing with projected radius.  We tested whether the general increase
in $s_{\Delta v}$ with increasing $b$ could be due to the corresponding 
decrease in signal--to--noise (S/N) ratio with increasing $b$.  We raised the
threshold $(I/\sigma_{\rm I})_{\rm min}$ from 5 to 10 and repeated the 
linear lest--square fits as in Tab.~\ref{tdeltav} with smaller, lower--noise
data sets.  For the 12 surviving cores with enough data points (at least 9), 
the 
tendency for $s_{\Delta v}$ to increase with $b$ is evident in 5 starless
cores -- L1544, L183, TMC--1CS, TMC--1NH3, TMC--2 -- and 5 cores with stars
-- L1489, L43, L1228, L1221, L1251E.  The remaining two cores show a 
negative (L1512) and a null (L483) $s_{\Delta v} - b$ correlation.

This analysis of 22 low--mass \n2hp \, cores indicates that the brightest part 
of a core map has nearly constant line width, typically 0.3--0.4 km s$^{-1}$,
while the surrounding positions with fainter emission have line widths with 
much greater variation.  This confirms the conclusions of \cite{bg98} based on
4 NH$_3$ core maps.  However, our data show no significant increase in 
\n2hp line width with effective map radius for the typical core, in contrast 
to the result of \cite{bg98}.

\subsection{Map structure}
	
The N$_2$H$^+$ maps presented in Fig.~\ref{fmap} are generally similar in 
shape and orientation to the NH$_3$ maps of BM89 and \cite{lmg94}.
This similarity is easily understood, since N$_2$H$^+$ and NH$_3$ 
are closely related chemically (N$_2$ is the precursor molecule for both 
species).  In cores with stars, the factor of 
$\sim$ 2 smaller size of the N$_2$H$^+$ maps compared to its NH$_3$ 
counterpart probably reflects both the higher
critical density ($\sim$ 2$\times$10$^5$ cm$^{-3}$ for N$_2$H$^+$ compared to 
$\sim$ 3$\times$10$^4$ cm$^{-3}$ for NH$_3$) and the finer angular 
resolution ($\sim$ 54$^{\prime\prime}$ instead of $\sim$ 83$^{\prime\prime}$)
and sampling (25$^{\prime\prime}$ or 50$^{\prime\prime}$ instead of 60$^{\prime
\prime}$) used in the N$_2$H$^+$ observations.

To quantify the complexity of the projected map structure, we have
counted the number of map "peaks" and "elongations" in each of the 59 \n2hp \,
maps in this paper and in each of the 47 NH$_3$ maps in BM89 and in \cite{lmg94}.
Here a "peak" is a local maximum of integrated intensity which exceeds its
surrounding valley by at least 3-sigma, and an "elongation" is an extension
from the position of a peak by at least one beam, which does not terminate
in a new peak. We find that the \n2hp \, and NH$_3$ maps have similar proportions
of simple and complex structure.  The fraction of "simple" maps with single
peaks and no elongation is 0.64 for NH$_3$ and 0.70 for \n2hp .  The fraction of
maps with at least two peaks is 0.12 for NH$_3$ and 0.07 for \n2hp .  Because of
small number statistics, these NH$_3$ and \n2hp \, fractions are not significantly
different.  One might expect that  more peaks per map would be seen in the
finer-resolution \n2hp \,  observations than in the coarser-resolution NH$_3$
observations, but this is not the case.  We interpret this result as
arising from the fact that the \n2hp \, maps have both finer resolution and
smaller spatial extent, each by about the same factor of ~1.5. Thus some
N$_2$H$^+$ 
maps show two peaks where the NH$_3$ map shows one, but other \n2hp \, maps
do not extend far enough to sense the second peak seen in the NH3 map.
Although 
relatively few N$_2$H$^+$ maps have more than one local maximum (14 of 57), 
we note that most cases with double peaks have peak--to--peak separation of 
only 1--2 FWHM beam diameters.  Therefore these core maps are nearly as 
``clumpy'' as the map resolution allows. 

\subsection{Overview of core parameters}

An overview of the parameters determined in previous sections is presented
in Table~\ref{tsummary}, for cores with stars and starless cores separately.
Although the dispersion in large, cores associated with young stellar 
objects are typically more turbulent (non--thermal line widths are 
about 1.5 times larger),  have larger sizes (factor of $\sim$ 1.4), and
are more massive (factor of $\sim$ 2), than starless cores.  Other quantities 
($T_{\rm ex}$, $N_{\rm TOT}$(\n2hp ), aspect ratio, ${\cal G}$, $\beta$,
have very similar
values in the two classes of cores, suggesting that there is not a definite
separation between them, at least from the analysis of \n2hp (1--0) lines at 
the present angular resolution. 

\section{Conclusions}

We have mapped 57 low mass cores in the rotational transition $J$ = 1$\rightarrow$0
of the molecular ion \n2hp, using the FCRAO antenna.  This extensive mapping
survey has allowed us to study physical properties of dense cores with an 
angular resolution about 1.5 times finer than previous ammonia maps from BM89.
The main conclusions of this work are summarized below.

\noindent
1. The excitation temperature of the \n2hp (1-0) line is typically $\sim$ 5 K, 
indicating that \n2hp \, lines are subthermally excited.  The peak \n2hp 
column density averaged over the whole sample is $N$(\n2hp ) $\sim$ 
7$\times$10$^{12}$ 
cm$^{-2}$, about two orders of magnitude less than $N$(\nh3 ) (BM89). 
There is  a positive correlation between \n2hp \, and \nh3 \, column densities 
and excitation temperatures in starless cores, whereas in cores with stars the 
scatter is large and no significant correlations are found. Although this may 
partially be due to the difficulty in estimating the \NTHP (1--0) total optical 
depth, the lack of correlation in cores with stars suggests a different 
chemical evolution of \AMM \ and \NTHP.  However, the good  
correlations between LSR velocities and line widths in the entire sample, 
indicates that the two tracers generally originate from the same regions.   

\noindent
2. The mean aspect ratios of the mapped sources is $\sim$ 2.  Starless cores
have about the same linear sizes than those found with \nh3 (1,1) maps.  On the
other hand,  cores with stars have $r$(\nh3 ) $\sim$ $2\times r$(\n2hp ).  This 
gives evidence that cores associated with young stellar objects are more centrally
concentrated than starless cores and that \n2hp (1--0) traces denser gas than
\nh3 (1,1).  

\noindent
3. In average, starless cores have virial and ``excitation'' masses 
$M_{\rm vir}$ $\sim$ $M_{\rm ex}$ $\sim$ 3 M$_{\odot}$, and are less massive 
than cores with stars ($M_{\rm vir}$ $\sim$ $M_{\rm ex}$ $\sim$ 8 M$_{\odot}$).
The $M_{\rm ex}$/$M_{\rm vir}$ ratio averaged over the whole sample is $\sim$ 1.

\noindent
4. Typical values of velocity gradient magnitudes are $\sim$ 2 km/s/pc, both in
cores with stars and starless cores.  If the gradient represents rotation, the
ratio $\beta$ of rotational energy to
gravitational energy ranges between $\sim$ 10$^{-4}$ and 0.07 so that 
rotation is not significant in the support of the core.  Maps of ``local'' 
velocity gradients 
reveal the presence of complex internal motions that deviate strongly from a 
simple model of solid body rotation of the whole core. 

\noindent
5. Six out of nine spheroidal starless cores present central flattening in the
integrated intensity profile.  This is consistent with a spherically symmetric
density law $n(r)$ $\sim$ $r^{-\alpha}$ where $\alpha$ = 1.2 for 
$r$ $<$ $r_{\rm break}$ and $\alpha$ = 2 for $r$ $>$ $r_{\rm break}$,
with $r_{\rm break}$ $\sim$ 0.03 pc.  Cores with stars are better modelled 
with single power law density profiles with $\alpha$ $\ga$ 2. These results 
are in qualitative agreement with submillimeter 
continuum observations, suggesting that 
\n2hp \, is not significantly depleted inside dense cores
(unlike CO and CS).

\noindent
6. Most \n2hp cores are ``coherent'' in having more uniform line widths in their
bright interior than in their faint periphery, as seen earlier in NH$_3$ 
observations.  The fluctuations in line width also increase significantly with 
mean line width.  However this sample shows no significant tendency  for the 
line widths themselves 
to increase with map radius -- a few cores have positive and 
negative trends, while most have no significant trend.
For 20 of 26 cores, the standard deviation of the 
average line width, $s_{\Delta v}$, increases with $b$, indicating that 
core line widths vary more with increasing radius 
as previously found by \cite{bg98} and 
\cite{gbw98} using ammonia data.  Yet, line widths $\Delta v$ positively 
correlate with the impact parameter $b$ in only 5 sources (L1498, L1495, 
L1524, L1400K, and L260).  Four sources present negative 
correlations (PER4, B5, TMC--1C2, L1174).  The remaining 17 cores do not 
show a significant $\Delta v - b$ correlation.  

\noindent
7. The ``coherence length'' deduced from our data is about 0.01 pc, comparable
to the cutoff wavelength below which Alfv\'en waves cannot propagate.  Thus,
the ``transition to coherence'' may arise from a decay of turbulence in the 
innermost parts of the cores,  due to insufficient wave coupling, on 
size scales of $\sim$ 0.01 pc. 

\noindent
8. Although \NTHP \ maps have finer angular resolution than \AMM \ maps, 
they do not show a more complex structure.  The majority (70\%) of the cores 
in our sample have ``simple'' \NTHP \ maps, with single peaks and no elongation.   
Most cases with double--peaks have peak--to--peak separation of only 1--2 
FWHM beam diameters.

\acknowledgments
 P.C. acknowledges support from ASI (grants ARS--96--66, ARS--98--116, and
ARS--78--1999) and from MURST (project ``Dust and Molecules in Astrophysical 
Environments).  PJB acknowledges support from NSF AST 9417359 and The Bunting 
Institute at Radcliff College. P.C.M. acknowledges support from NASA Origins 
of Solar Systems Grant NAG5--6266. 

\appendix

\section{Error expressions}
\label{aerror}

The error on the volume density $n_{\rm ex}$ (see Sect.~\ref{ssize}) is found by 
propagating the error in eqn.(\ref{encr}):

\scriptsize
\begin{eqnarray}
\sigma_{n_{\rm ex}} & = & \left[ \left( n_{\rm cr} \sigma_{T_{\rm ex}} 
\frac{k}{h \nu} \frac{\tilde{T}_{\rm kin}(\tilde{T}_{\rm kin} -
\tilde{T}_{\rm cb})}{(\tilde{T}_{\rm kin} - \tilde{T}_{\rm ex})^2}
\frac{\tilde{T}_{\rm ex}^2 exp(-hv/(k T_{\rm ex}))}{T_{\rm ex}^2}
\frac{1 - e^{-3 \tau}}{3 \tau} \right)^2 + 
   \left( \sigma_{\rm \tau} \frac{n_{\rm ex} 3 \tau}{1 -
  e^{-3 \tau}} \frac{e^{-3 \tau} (3 \tau + 1) - 1}{\tau} \right)^2
  \right]^{1/2}
\end{eqnarray} 

\normalsize
The errors on the average line width $<\Delta v>$ and the corresponding 
sample standard deviation $s_{\Delta v}$ have been calculated with the 
following expressions (from the propagation of error):

\begin{eqnarray}
\sigma_{<\Delta v>} & = & \frac{\sqrt{\sum_i \sigma_{\Delta v(i)}^2}}{N}, \\
\sigma_{s_{<\Delta v>}} & = & \frac{ \sqrt{ \sum_i \sigma_{\Delta v(i)}^2  
(\Delta v(i) - <\Delta v>)^2 }}{s_{<\Delta v>} (N - 1)}. 
\end{eqnarray}

\clearpage

\begin{center}
\begin{deluxetable}{lccrrcccccc}
\tablewidth{0pc}
\footnotesize
\tablecaption{N$_2$H$^+$(1-0) map results: peak position ($\Delta \alpha$, 
  $\Delta \delta$) and integrated intensity ($I$)}
\tablehead{
\colhead{Core}           & \colhead{RA(1950)\tablenotemark{a}}      &
\colhead{Dec(1950)\tablenotemark{a}}      & \colhead{$\Delta \alpha$}  &
\colhead{$\Delta \delta$} & \colhead{$\Delta T_{\rm A,rms}^*$} 
 & \colhead{$I\pm \sigma_I$\tablenotemark{b}}  & 
\colhead{N$_{\rm ch}$\tablenotemark{i}} &
\colhead{Map\tablenotemark{c}}            & \colhead{Area\tablenotemark{d}}&
\colhead{IRAS\tablenotemark{h}} 
\nl
 & & & \colhead{($\prime$)} & \colhead{($\prime$)} & \colhead{(K)} &
  \colhead{(K km s$^{-1}$)} & & & \colhead{(arcmin$^2$)}& }
\startdata
PER4-A& 03$^{\rm h}$26$^{\rm m}$31$^{\rm s}$.7& 
31$^{\circ}$17$^{\prime}$13$^{\prime\prime}$ & 
-2.49& 0.84& 0.16& 1.64$\pm$0.09& 91& B& 85.7& n \nl
PER4-B& & & -4.18& 0.00& 0.19& 1.56$\pm$0.11& & & & n\nl
PER4-C& & & -3.32& -1.70& 0.22& 1.05$\pm$0.13& & & & y\nl
PER5& 03 26 45.5& 31 28 48& 0.00& 0.00& 0.17& 2.00$\pm$0.10& 76& B& 21.4& y \nl
PER6& 03 27 10.3& 30 12 34& 0.00& 0.84& 0.16& 1.67$\pm$0.10& 93& B& 21.4& y \nl
PER7& 03 29 39.5& 30 49 50& 0.00& 0.00& 0.17& 1.45$\pm$0.11& 100& B& 21.4& y \nl
PER9\tablenotemark{e}& 03 30 10.4& 31 10 14& -0.88& 0.04& 0.19& 
  1.17$\pm$0.11& 84& B& 21.4& y \nl
B5&   03 44 32.7& 32 44 30& 0.00& -1.69& 0.19& 1.99$\pm$0.12& 95& B& 42.8& y \nl
L1389& 04 00 38.0& 56 47 59& 0.00& 0.00& 0.17& 1.11$\pm$0.10& 84& B& 21.4& y \nl
L1489& 04 01 45.0& 26 10 33& 0.42& 0.00& 0.16& 2.11$\pm$0.10& 98& N& 21.4& y \nl
L1498& 04 07 50.0& 25 02 13& 0.42& -0.42& 0.16& 1.20$\pm$0.10& 88& N& 21.4& n \nl
L1495& 04 11 02.7& 28 00 43& 0.00& 0.84& 0.19& 1.75$\pm$0.11& 77& B& 21.4& y \nl
L1400G& 04 21 12.1& 54 12 20& \nodata& \nodata& 0.18& \nodata& \nodata& B& 21.4& n \nl
B217& 04 24 48.1& 26 11 38& -0.88& 0.04& 0.19& 2.03$\pm$0.11& 85& B& 64.3& n \nl
L1524& 04 26 22.3& 24 28 36& 0.00& 0.00& 0.18& 1.51$\pm$0.10& 81& B& 42.8& y \nl
L1400K& 04 26 51.0& 54 45 27& 0.00& 0.00& 0.12& 0.66$\pm$0.06& 69& B& 21.4& n \nl
TMC-2A& 04 28 54.0& 24 26 27& 0.00& 0.00& 0.16& 1.87$\pm$0.10& 100& B& 21.4& n \nl
TMC-2& 04 29 41.2& 24 20 09& 0.86& -0.86& 0.20& 1.64$\pm$0.12& 90& B& 128.5& n \nl
L1536& 04 30 33.2& 22 37 50& -1.69& -1.69& 0.17& 1.30$\pm$0.10& 89& B& 85.7& n \nl
L1534\tablenotemark{e}& 04 36 42.3& 25 34 16& -3.32& 1.77& 0.19& 
  1.62$\pm$0.11& 90& B& 42.8& y \nl
L1527& 04 36 49.3& 25 57 16& 0.00& 0.00& 0.17& 1.25$\pm$0.10& 91& B& 21.4& y \nl
TMC-1NH$_3$\tablenotemark{e}& 04 38 19.0& 25 42 30& -0.86& 0.04& 0.14& 
  2.34$\pm$0.09& 107& B& 21.4& n \nl
TMC-1C2\tablenotemark{e}& 04 38 25.5& 25 56 00& 1.69& -1.69& 0.17& 
  1.34$\pm$0.10& 80& B& 21.4& n \nl
TMC-1C\tablenotemark{e}& 04 38 34.5& 25 55 00& 0.00& -0.86& 0.18& 
  1.34$\pm$0.10& 74& B& 21.4& n \nl
TMC-1CS\tablenotemark{e}& 04 38 38.9& 25 35 00& -0.86& 2.61& 0.13& 
  1.53$\pm$0.08& 90& B& 21.4& n \nl
L1517B& 04 52 07.2& 30 33 18& 0.00& 0.00& 0.19& 1.05$\pm$0.10& 74& B& 21.4& n \nl
L1512& 05 00 54.4& 32 39 00& -0.45& 0.04& 0.16& 1.06$\pm$0.09& 75& N& 21.4& n \nl
L1544& 05 01 14.0& 25 07 00& 0.00& -0.42& 0.12& 1.75$\pm$0.09& 158& N& 21.4& n \nl
L1582A& 05 29 11.9& 12 28 20& 0.00& 0.00& 0.10& 0.68$\pm$0.07& 104& B& 21.4& y \nl
B35& 05 41 45.3& 09 07 40& 0.00& 0.00& 0.14& 1.48$\pm$0.09& 90& B& 21.4& y \nl
L134A\tablenotemark{e}& 15 50 58.1& -04 26 36& 0.00& 0.00& 0.14& 
  0.68$\pm$0.08& 79& B& 21.4& n \nl
L183\tablenotemark{e}& 15 51 35.7& -02 40 54& -0.88& -2.41& 0.21& 
  1.95$\pm$0.11& 73& B& 85.7& n \nl
L1681B-A& 16 24 41.1& -24 35 32& -0.88& -0.86& 0.26& 1.81$\pm$0.16& 98& B& 
  42.8& n \nl
L1681B-B& & & -3.32& 0.87& 0.25& 1.98$\pm$0.15& & & & n \nl
L1696A& 16 25 30.0& -24 12 32& 0.00& 0.84& 0.27& 1.60$\pm$0.18& 109& B& 21.4& n \nl
L43& 16 31 42.1& -15 40 50& 0.42& 0.42& 0.19& 3.57$\pm$0.13& 109& N& 21.4& y \nl
 L260\tablenotemark{e}& 16 44 22.3& -09 30 02& 0.00& 0.00& 0.16& 
  0.74$\pm$0.08& 65& B& 21.4& y \nl
L158& 16 44 33.7& -13 54 03& 0.00& 0.00& 0.24& 0.97$\pm$0.14& 85& B& 21.4& y \nl
L234A& 16 45 21.0& -10 46 33& 0.86& 0.04& 0.22& 0.66$\pm$0.08& 37& B& 21.4& n \nl
L234E\tablenotemark{e}& 16 45 23.0& -10 51 43& 0.00& -0.86& 0.25& 
  0.69$\pm$0.13& 74& B& 21.4& n \nl
L63& 16 47 21.0& -18 01 00& 0.00& 0.00& 0.29& 1.75$\pm$0.17& 85& B& 21.4& n \nl
B68& 17 19 36.0& -23 47 13& 0.00& 0.00& 0.20& 0.64$\pm$0.10& 64& B& 21.4& n \nl
L483& 18 14 50.5& -04 40 49& 0.00& 0.00& 0.15& 4.32$\pm$0.10& 109& N& 21.4& y \nl
B133& 19 03 25.3& -06 57 20& 0.00& 0.00& 0.19& 0.56$\pm$0.10& 67& B& 21.4& y \nl
L778& 19 24 26.4& 23 52 37& 0.00& -0.81& 0.16& 1.72$\pm$0.10& 97& B& 21.4& y \nl
B335& 19 34 35.3& 07 27 34& 0.00& 0.00& 0.14& 1.70$\pm$0.09& 106& N& 21.4& y \nl
L1152& 20 35 19.6& 67 42 13& 0.00& 0.00& 0.20& 1.40$\pm$0.11& 79& B& 21.4& y \nl
L1155C& 20 43 00.0& 67 41 47& 0.00& 0.00& 0.11& 0.72$\pm$0.06& 81& B& 21.4& n \nl
L1082C& 20 50 19.5& 60 07 15& 0.42& 0.00& 0.19& 1.39$\pm$0.11& 83& N& 21.4& y \nl
L1082A\tablenotemark{f}& 20 52 20.7& 60 03 14& 0.00& 0.00& 0.12& 
  0.67$\pm$0.07& 82& B& 21.4& y \nl
L1228& 20 58 11.0& 77 24 00& 0.41& 0.00& 0.16& 3.85$\pm$0.11& 117& N& 21.4& y \nl
L1174& 20 59 46.3& 68 01 04& -0.44& -0.04& 0.20& 2.23$\pm$0.15& 131& N& 21.4& y \nl
BERN48& 21 00 20.0& 78 11 00& 0.41& 0.42& 0.23& 1.93$\pm$0.14& 103& N& 21.4& y \nl
L1172A& 21 01 45.0& 67 42 13& 0.00& 0.84& 0.19& 1.58$\pm$0.12& 98& B& 21.4& y \nl
B361& 21 10 35.0& 47 12 01& 0.00& 0.84& 0.17& 1.17$\pm$0.11& 112& B& 21.4& y \nl
L1031C& 21 44 35.6& 47 04 20& \nodata& \nodata& 0.14& \nodata& \nodata& B& 21.4& y \nl
L1031B& 21 45 32.0& 47 18 13& 0.00& 0.00& 0.14& 2.31$\pm$0.12& 186& B& 21.4& y \nl
L1221& 22 26 37.1& 68 45 37& 0.00& 0.00& 0.22& 4.14$\pm$0.15& 125& N& 21.4& y \nl
L1251A\tablenotemark{g}& 22 29 34.1& 74 58 51& 0.00& 0.00& 0.19& 
  1.39$\pm$0.12& 95& B& 21.4& y \nl
L1251C& 22 34 37.5& 75 02 32& 0.88& 0.80& 0.18& 1.91$\pm$0.12& 111& B& 21.4& y \nl
L1251E& 22 38 36.4& 74 55 50& -3.38& -0.42& 0.17& 4.03$\pm$0.15& 197& N& 42.8& y \nl
L1262& 23 23 32.2& 74 01 45& 0.86& -0.04& 0.15& 1.64$\pm$0.10& 111& B& 21.4& y \nl
\tablenotetext{a}{Coordinates of the (0,0) map position; they do not always
correspond to the coordinates quoted in BM89.}
\tablenotetext{b}{Integrated intensity at the peak position;  
$\sigma_I$ =  $\Delta T_{\rm A,rms}^* \times \sqrt{N_{\rm ch}} \times 
\Delta v_{\rm res}$, where $\Delta T_{\rm A,rms}^*$ [K] is the 1 $\sigma$
level of the noise in the off-line channels, $N_{\rm ch}$ is the number of 
channels in the integrated intensity, and $\Delta v_{\rm res}$ [km s$^{-1}$] 
is the velocity resolution (=0.063 km s$^{-1}$).}
\tablenotetext{c}{Type of map: B $\equiv$ beam sampling; N $\equiv$ Nyquist 
sampling.}
\footnotesize
\tablenotetext{d}{Mapped area.}
\tablenotetext{e}{The half maximum contour extends over the mapped area.}
\tablenotetext{f}{A second component is present in the West direction; it 
has not been included in the integrated intensity estimate.}
\tablenotetext{g}{There is another peak in the mapped area which belongs 
to another core (whose half maximum contour extends outside the mapped area).}
\tablenotetext{h}{IRAS association, from Jijina et al. 1999.}
\tablenotetext{i}{Number of channels in integrated area.}
\enddata
\label{tpeak}
\end{deluxetable}

\end{center}

\clearpage

\begin{center}
\begin{deluxetable}{lccccc}
\small
\tablewidth{16cm}
\tablecaption{N$_2$H$^+$(1-0) map results: multicomponent (hfs) fit to 
 the peak and averaged\tablenotemark{a} spectrum.}
\tablehead{
\colhead{Core}           & \colhead{$V_{\rm LSR}$}       &
\colhead{$\Delta v$}      & \colhead{$\tau_{\rm TOT}$\tablenotemark{b}}  &
\colhead{$T_{\rm ex}$\tablenotemark{c}} & 
 \colhead{N$_{\rm tot}$$\times$10$^{-12}$} \nl
 & \colhead{(km s$^{-1}$)} & \colhead{(km s$^{-1}$)}&  & \colhead{(K)} &
 \colhead{(cm$^{-2}$)}}
\startdata
PER4-A\tablenotemark{d}& 7.60\p0.02& 0.58\p0.06& 0.1 & 5.0& 4.7\p0.6 \nl  
  &      7.60\p0.02& 0.47\p0.05& 4\p2& 5\p1 & 5\p3 \nl
PER4-B& 7.61\p0.01& 0.36\p0.03& 0.1& 5.0 & 4.5\p0.5 \nl
   &     7.565\p0.007& 0.36\p0.01& 0.1& 5.0& 3.5\p0.1 \nl
PER4-C& 7.38\p0.02& 0.33\p0.04& 0.1& 5.0& 3.0\p0.4 \nl
   &	7.41\p0.02& 0.38\p0.04& 0.1& 5.0& 2.5\p0.4 \nl
PER5&   8.221\p0.008& 0.33\p0.02& 0.1& 5.0 & 5.5\p0.5 \nl 
   & 	8.203\p0.006& 0.29\p0.02& 5\p2& 5\p1 & 5\p2\nl
PER6&   5.87\p0.01& 0.44\p0.03& 0.1& 5.0& 4.8\p0.4 \nl	
   &    5.880\p0.008& 0.38\p0.02& 6\p2& 4.1\p0.5 & 6\p2 \nl
PER7&   6.81\p0.02& 0.33\p0.04& 15\p7& 3.8\p0.6 & 11\p5 \nl	
   &	6.82\p0.02& 0.45\p0.05& 5\p3& 3.9\p0.8 & 5\p3 \nl
PER9&   6.91\p0.02& 0.41\p0.04& 0.1& 5.0& 3.4\p0.4  \nl	
   &	6.82\p0.01& 0.42\p0.03& 0.1& 5.0& 2.9\p0.2 \nl
B5&    10.31\p0.02& 0.47\p0.03& 0.1& 5.0& 6.0\p0.5  \nl	
   &   10.25\p0.01& 0.39\p0.02& 4\p2& 4.7\p0.9 & 5\p2 \nl
L1389& -4.65\p0.03& 0.49\p0.05& 0.1& 5.0& 3.2\p0.4 \nl	
   &   -4.60\p0.01& 0.33\p0.04& 7\p3& 3.9\p0.7 & 5\p3 \nl
L1489&  6.80\p0.01& 0.28\p0.02& 17\p7& 4.3\p0.8 & 13\p5 \nl	
   &	6.783\p0.004& 0.277\p0.009& 7\p1& 4.7\p0.4 & 6\p1 \nl
L1498&  7.833\p0.007& 0.18\p0.02& 18\p9& 4.1\p0.8 & 8\p4 \nl	
   &	7.840\p0.002& 0.191\p0.005& 11\p1& 3.9\p0.2 & 4.7\p0.6 \nl
L1495&  6.832\p0.007& 0.22\p0.02& 4\p3& 7\p3 & 5\p3 \nl	
   &	6.824\p0.004& 0.27\p0.01& 4\p1& 6\p1 & 4\p1 \nl
L1400G\tablenotemark{e}& \nodata& \nodata& \nodata& \nodata& \nodata \nl
B217&   7.02\p0.01& 0.34\p0.03& 4\p3& 6\p3& 6\p4 \nl	
   &	7.001\p0.005& 0.33\p0.01& 4\p1& 5.1\p0.8& 5\p2 \nl
L1524&  6.36\p0.01& 0.26\p0.03& 15\p8& 4.1\p0.9& 10\p5 \nl	
   &	6.339\p0.008& 0.35\p0.02& 9\p2& 3.7\p0.3& 7\p2  \nl
L1400K& 3.28\p0.01& 0.19\p0.02& 11\p6& 3.6\p0.6& 4\p2 \nl	
   &	3.300\p0.006& 0.24\p0.01& 6\p2& 3.6\p0.3 & 3\p1\nl
TMC-2A& 5.917\p0.006& 0.22\p0.02& 16\p4& 4.7\p0.7& 11\p3 \nl	
   &	5.935\p0.006& 0.27\p0.01& 9\p2& 4.2\p0.4& 7\p2 \nl
TMC-2&  6.26\p0.02& 0.40\p0.03& 0.1& 5.0& 4.9\p0.5 \nl	
   &	6.197\p0.005& 0.36\p0.01& 5\p1& 4.3\p0.4& 5\p1 \nl
L1536&  5.53\p0.01& 0.30\p0.02& 0.1& 5.0& 3.9\p0.4 \nl	
   &	5.649\p0.003& 0.263\p0.009& 6\p1& 4.2\p0.3& 3.9\p0.8 \nl
L1534&  6.40\p0.01& 0.36\p0.02& 0.1& 5.0& 4.6\p0.4\nl	
   &	6.304\p0.006& 0.39\p0.01& 3\p1& 5\p1& 4\p2 \nl
L1527&  5.90\p0.01& 0.29\p0.03& 13\p7& 3.9\p0.7& 8\p5 \nl
   &	5.922\p0.008& 0.34\p0.02& 9\p2& 3.6\p0.3& 6\p2 \nl
TMC-1NH$_3$& 5.956\p0.009& 0.36\p0.02& 11\p3& 4.6\p0.5& 11\p3 \nl 
   &	     5.955\p0.004& 0.39\p0.01& 5.6\p0.8& 4.6\p0.3& 6.2\p0.9  \nl
TMC-1C2& 5.27\p0.02& 0.29\p0.03& 14\p6& 3.9\p0.7& 9\p4 \nl	
   &	 5.219\p0.006& 0.29\p0.01& 8\p2& 3.8\p0.3& 5\p1  \nl
TMC-1C& 5.27\p0.02& 0.27\p0.04& 18\p10& 3.8\p0.8& 11\p6 \nl	
   &	5.257\p0.004& 0.25\p0.01& 12\p2& 3.6\p0.2& 6\p1 \nl
TMC-1CS& 5.89\p0.01& 0.38\p0.03& 8\p3& 4.3\p0.8& 8\p3 \nl	
   &	 5.853\p0.005& 0.41\p0.01& 4\p1& 4.3\p0.4& 5\p1 \nl
L1517B& 5.80\p0.01& 0.27\p0.02& 0.1& 5.0& 3.0\p0.3 \nl	
   &	5.830\p0.008& 0.22\p0.02& 9\p4& 4.0\p0.6& 5\p2 \nl
L1512& 7.108\p0.007& 0.18\p0.02& 5\p3& 5\p2& 3\p2 \nl	
   &   7.088\p0.002& 0.195\p0.006& 7\p1& 4.2\p0.3& 3.6\p0.6 \nl
L1544& 7.169\p0.008& 0.31\p0.01& 10\p2& 4.5\p0.5& 9\p2 \nl	
   &   7.162\p0.003& 0.307\p0.006& 8.1\p0.9& 4.0\p0.2& 6.0\p0.7 \nl
L1582A\tablenotemark{f}& 10.20\p0.02& 0.43\p0.05& 0.1& 5.0& 1.9\p0.3 \nl
   &	\nodata      & \nodata& \nodata& \nodata& \nodata \nl
B35& 11.69\p0.04& 0.61\p0.08& 9\p4& 3.6\p0.5& 12\p6 \nl	
   & 11.86\p0.04& 0.89\p0.09& 0.1& 5.0& 4.2\p0.5 \nl
L134A& 2.74\p0.01& 0.29\p0.03& 0.1& 5.0& 2.1\p0.3 \nl	
   &   2.764\p0.009& 0.34\p0.02& 0.1& 5.0& 1.4\p0.1 \nl
L183& 2.44\p0.01& 0.25\p0.02& 22\p7& 4.3\p0.7& 14\p5 \nl	
   &  2.422\p0.004& 0.303\p0.009& 8\p1& 4.1\p0.2& 6.1\p0.9 \nl
L1681B-A& 3.62\p0.02& 0.40\p0.03& 0.1& 5.0& 5.5\p0.6 \nl	
   &	  3.69\p0.02& 0.45\p0.04& 0.1& 5.0& 3.8\p0.4 \nl 
L1681B-B& 4.12\p0.01& 0.36\p0.03& 0.1& 5.0& 5.7\p0.7 \nl	 
   &	4.13\p0.01& 0.39\p0.04& 0.1& 5.0& 3.9\p0.5 \nl
L1696A& 3.37\p0.01& 0.25\p0.03& 0.1& 5.0& 3.9\p0.6 \nl	
   &	3.40\p0.01& 0.30\p0.02& 0.1& 5.0& 2.9\p0.3 \nl
L43& 0.678\p0.009& 0.36\p0.02& 6\p2& 7\p1& 11\p4 \nl	
   & 0.692\p0.003& 0.396\p0.007& 5.9\p0.5& 5.4\p0.3& 8.6\p0.8 \nl
L260& 3.503\p0.008& 0.20\p0.02& 0.1& 5.0& 2.2\p0.3 \nl	
   &  3.480\p0.005& 0.19\p0.01& 11\p3& 3.4\p0.2& 4\p1 \nl
L158\tablenotemark{f}& 3.91\p0.01& 0.23\p0.04& 0.1& 5.0& 2.7\p0.5 \nl	
   &	\nodata& \nodata   & \nodata & \nodata& \nodata  \nl
L234A& 2.95\p0.01& 0.21\p0.02& 0.1& 5.0& 2.3\p0.3 \nl
   &	2.936\p0.007& 0.23\p0.02& 0.1& 5.0& 1.9\p0.2 \nl
L234E\tablenotemark{g}& \nodata & \nodata& \nodata& \nodata& \nodata \nl
   &	3.04\p0.02& 0.30\p0.04& 12\p7& 3.2\p0.3& 7\p4 \nl
L63& 5.78\p0.01& 0.21\p0.03& 13\p6& 5\p1& 8\p4 \nl
   & 5.780\p0.007& 0.27\p0.02& 9\p2& 4.3\p0.5& 6\p2 \nl
B68& 3.35\p0.02& 0.27\p0.04& 0.1& 5.0& 1.9\p0.4\nl
   & 3.38\p0.02& 0.33\p0.05& 0.1& 5.0& 1.4\p0.3\nl
L483& 5.53\p0.01& 0.59\p0.03& 16\p3& 4.6\p0.4& 27\p5 \nl 
   &  5.432\p0.004& 0.495\p0.008& 11.6\p0.7& 4.5\p0.1& 16\p1 \nl
B133\tablenotemark{h}& 12.1\p0.2& 0.9\p0.3& 0.1& 5.0& 2\p1 \nl
   &	\nodata& \nodata& \nodata& \nodata& \nodata \nl
L778& 9.98 \p0.02 & 0.44\p0.04& 6\p3& 4.4\p0.9& 7\p4\nl
   &  9.944\p0.009& 0.40\p0.02& 5\p2& 4.4\p0.7& 5\p2  \nl
B335& 8.36\p0.01& 0.39\p0.03& 6\p3& 4.5\p0.9& 6\p3 \nl
   &  8.345\p0.008& 0.40\p0.02& 7\p2& 4.0\p0.4& 6\p2 \nl
L1152& 2.70\p0.02& 0.46\p0.04& 0.1& 5.0& 4.2\p0.6 \nl
   &   2.60\p0.02& 0.48\p0.03& 0.1& 5.0& 2.8\p0.2 \nl
L1155C& 2.69\p0.02& 0.36\p0.04& 7\p4& 3.6\p0.5& 5\p3 \nl
   &	2.70\p0.01& 0.33\p0.03& 6\p3& 3.5\p0.4& 4\p2 \nl
L1082C& -2.53\p0.02& 0.42\p0.04& 0.1& 5.0& 3.8\p0.5 \nl
   &	-2.55\p0.01& 0.33\p0.03& 16\p6 & 3.4\p0.3& 10\p4 \nl
L1082A& -2.13\p0.02& 0.35\p0.04& 0.1& 5.0& 1.9\p0.3  \nl
   &	-2.19\p0.02& 0.46\p0.04& 0.1& 5.0& 1.7\p0.2\nl
L1228& -8.06\p0.02& 0.61\p0.04& 9\p2& 4.9\p0.6& 18\p4 \nl
   &   -8.041\p0.008& 0.68\p0.02& 5.9\p0.8& 4.5\p0.3& 11\p2 \nl
L1174& 2.67\p0.06& 1.2\p0.2& 0.1& 5.0& 7\p1\nl
   &   2.66\p0.02& 1.05\p0.06& 0.1& 5.0& 4.8\p0.4\nl
BERN48& -7.35\p0.03& 0.54\p0.06& 0.1& 5.0& 5.4\p0.8\nl
   &	-7.34\p0.01& 0.46\p0.04& 4\p2& 5\p1& 5\p3 \nl
L1172A& 2.91\p0.04& 0.54\p0.09& 8\p5& 3.8\p0.7& 10\p6 \nl
   &	2.88\p0.02& 0.50\p0.04& 8\p3& 3.7\p0.4& 9\p3 \nl
B361& 2.78\p0.05& 0.70\p0.09& 0.1& 5.0& 3.5\p0.6 \nl
   &  2.64\p0.04& 0.9\p0.2& 5\p3& 3.3\p0.3& 9\p5 \nl
L1031C\tablenotemark{e}& \nodata& \nodata& \nodata& \nodata& \nodata  \nl
L1031B& 4.18\p0.05& 1.6\p0.2& 0.1& 5.0& 7.0\p0.9 \nl 
   &	3.91\p0.03& 1.1\p0.1& 0.1& 5.0& 4.6\p0.6 \nl
L1221& -4.42\p0.02& 0.67\p0.05& 6\p2& 6\p1& 15\p5 \nl
   &   -4.430\p0.009& 0.69\p0.02& 6.9\p0.9& 4.5\p0.3& 13\p2 \nl
L1251A& -3.93\p0.02& 0.45\p0.05& 0.1& 5.0& 4.1\p0.6 \nl
   &    -3.93\p0.01& 0.36\p0.04& 8\p4& 3.8\p0.6& 6\p3 \nl
L1251C& -4.71\p0.01& 0.29\p0.03& 12\p5& 4.3\p0.9& 9\p4 \nl
   &	-4.75\p0.02& 0.55\p0.06& 4\p3& 3.9\p0.9& 6\p4 \nl
L1251E& -3.93\p0.03& 1.35\p0.08& 0.1& 5.0& 11.8\p0.8 \nl
   &	-3.872\p0.009& 1.50\p0.02& 0.1\p0.1& 5.0& 7.8\p0.2 \nl
L1262& 4.11\p0.01& 0.39\p0.05& 18\p10& 3.8\p0.8& 15\p9 \nl
   &	4.06\p0.01& 0.43\p0.02& 8\p2& 3.8\p0.3& 8\p2 \nl
\tablenotetext{a}{The averaged spectrum is obtained by adding together 
all the spectra inside the half maximum contour.}
\tablenotetext{b}{$\tau_{\rm TOT}$ is the sum of the peak optical depth
of the seven hyperfine components.}
\tablenotetext{c}{Excitation temperature calculated by assuming a main 
beam efficiency $\eta_{\rm B}$=0.51.  If $\tau_{\rm TOT}$ $<$ 1, $T_{\rm ex}$
= 5 K has been assumed (see text).}
\tablenotetext{d}{The first row refers to the peak spectrum, whereas the 
second row refers to the averaged spectrum.}
\tablenotetext{e}{No detection.}
\tablenotetext{f}{Compact source: only one spectrum inside the half 
maximum contour.}
\tablenotetext{g}{Individual spectra cannot be hfs-fitted because 
of low S/N.}
\tablenotetext{h}{Only the peak spectrum can be hfs-fitted because 
of low S/N in the other map spectra.}
\enddata
\label{thfs}
\end{deluxetable}
\end{center}

\clearpage

\begin{center}
\begin{deluxetable}{lrcccccc}
\footnotesize
\tablewidth{16cm}
\tablecaption{Angular and linear size of N$_2$H$^+$(1--0) cores}
\tablehead{
\colhead{Core\tablenotemark{a}}& \colhead{PA\tablenotemark{b}}&
\colhead{Major\tablenotemark{c}}& \colhead{Minor}& \colhead{Aspect}&
\colhead{$r$\tablenotemark{d}}& \colhead{$r$}& \colhead{D\tablenotemark{e}}
   \nl
 & \colhead{(deg)} & \colhead{(arcmin)}&  \colhead{(arcmin)} & \colhead{Ratio} 
 & \colhead{(arcmin)}& \colhead{(pc)} & \colhead{(pc)}  } 
 \startdata  
PER4-A\tablenotemark{f}&    45& 3.4& 0.5& 6.4& 0.7& 0.069& 350 \nl
PER4-B\tablenotemark{f}&    98& 3.9& 0.8& 4.6& 0.9& 0.092& 350 \nl
PER4-C&    22& 2.0& 1.3& 1.6& 0.8& 0.08& 350 \nl
PER5&      59& 2.2& 1.3& 1.7& 0.8& 0.09& 350 \nl
PER6&      64& 3.1& 1.5& 2.0& 1.1& 0.11& 350 \nl
PER7&      40& 1.7& 1.0& 1.6& 0.7& 0.07& 350 \nl
PER9&      71& 1.8& 1.1& 1.7& 0.7& 0.07& 350 \nl
B5&\tablenotemark{f}        98& 3.9& 0.9& 4.3& 0.9& 0.095& 350 \nl
L1389\tablenotemark{f}&     22& 1.2& 0.5& 2.5& 0.4& 0.065& 600 \nl
L1489&    135& 1.6& 1.1& 1.4& 0.7& 0.027& 140 \nl
L1498&    121& 2.6& 2.0& 1.4& 1.1& 0.046& 140 \nl
L1495&    115& 2.9& 1.9& 1.5& 1.2& 0.048& 140 \nl 
B217&      58& 2.7& 1.6& 1.7& 1.0& 0.042& 140 \nl
L1524&     71& 5.3& 1.8& 2.9& 1.6& 0.064& 140 \nl
L1400K&    35& 2.9& 1.8& 1.6& 1.2& 0.057& 170 \nl 
TMC-2A&    89& 2.0& 1.4& 1.4& 0.8& 0.034& 140 \nl
TMC-2&     53& 5.3& 3.8& 1.4& 2.2& 0.091& 140 \nl
L1536&    162& 5.5& 3.2& 1.7& 2.1& 0.085& 140 \nl 
L1534&    147& 5.7& 2.0& 2.8& 1.7& 0.069& 140 \nl
L1527&     19& 2.5& 1.9& 1.3& 1.1& 0.044& 140 \nl  
L1517B&   165& 1.6& 1.4& 1.1& 0.7& 0.030& 140 \nl 
L1512&    114& 2.3& 1.6& 1.5& 1.0& 0.039& 140 \nl
L1544&    125& 2.0& 1.2& 1.7& 0.8& 0.032& 140 \nl 
L1582A\tablenotemark{f}&    19& 1.2& 0.5& 2.5& 0.4& 0.046& 400 \nl
B35&\tablenotemark{f}        3& 1.8& 0.9& 2.1& 0.6& 0.072& 400 \nl
L134A&    121& 3.3& 2.0& 1.7& 1.3& 0.060& 160 \nl
L1681B-A& 146& 2.3& 1.0& 2.2& 0.8& 0.036& 160 \nl
L1681B-B& 162& 2.4& 1.1& 2.2& 0.8& 0.039& 160 \nl
L1696A&    34& 2.3& 1.3& 1.8& 0.9& 0.040& 160 \nl
L43&      172& 3.0& 1.6& 1.9& 1.1& 0.050& 160 \nl
L260&      38& 3.7& 2.1& 1.8& 1.4& 0.065& 160 \nl
L158&\tablenotemark{f}     112& 0.9& 0.8& 1.2& 0.4& 0.020& 160 \nl
L234A&     45& 3.6& 1.2& 3.0& 1.0& 0.048& 160 \nl
L63&       94& 3.7& 1.9& 1.9& 1.3& 0.062& 160 \nl
B68&       97& 2.1& 1.1& 1.9& 0.8& 0.044& 200 \nl
L483&      47& 1.5& 1.2& 1.2& 0.7& 0.040& 200 \nl
B133&      30& 1.3& 1.0& 1.3& 0.6& 0.101& 600 \nl
L778&     148& 2.0& 1.6& 1.2& 0.9& 0.109& 420 \nl
B335\tablenotemark{f}&      78& 0.7& 0.3& 2.1& 0.2& 0.018& 250 \nl
L1152&\tablenotemark{f}     46& 3.4& 0.9& 3.6& 0.9& 0.115& 440 \nl 
L1155C&    80& 1.7& 1.1& 1.6& 0.7& 0.088& 440 \nl
L1082C&\tablenotemark{f}    45& 1.2& 0.8& 1.6& 0.5& 0.062& 440 \nl
L1082A&    81& 1.6& 1.2& 1.3& 0.7& 0.092& 440 \nl
L1228&    95& 1.5& 1.0& 1.5& 0.6& 0.055& 300 \nl
L1174&    165& 1.3& 1.0& 1.3& 0.6& 0.072& 440 \nl
BERN48\tablenotemark{f}&     8& 1.3& 0.5& 2.9& 0.4& 0.023& 200 \nl
L1172A&    28& 1.7& 1.5& 1.2& 0.8& 0.100& 440 \nl
B361&     156& 2.7& 1.6& 1.7& 1.1& 0.108& 350 \nl
L1031B&   133& 1.5& 1.0& 1.5& 0.6& 0.164& 900 \nl   
L1221&    126& 1.8& 1.0& 1.8& 0.7& 0.039& 200 \nl 
L1251A&    58& 2.7& 1.4& 2.0& 1.0& 0.056& 200 \nl 
L1251C&    45& 1.8& 1.3& 1.4& 0.7& 0.043& 200 \nl
L1251E&     1& 5.2& 1.8& 2.9& 1.5& 0.088& 200 \nl
L1262&    155& 2.1& 1.4& 1.5& 0.9& 0.050& 200 \nl
\tablenotetext{a}{Six cores do not appear in this table because their 
half maximum contours extend beyond the mapped area: TMC--1NH$_3$, 
TMC--1C2, TMC--1C, TMC--1CS, L183, and L234E. In the cases of  
L1534, L134A, and L260 it was possible to fit the core with a 
2D Gaussian because only a small fraction of the half maximum contour
lies beyond the mapped area.}
\tablenotetext{b}{The position angle PA is defined as the angle in a 
clockwise direction from the positive right ascension axis.}
\tablenotetext{c}{The major and minor axes have been corrected for 
beam size.}
\tablenotetext{d}{$r$ is the half-power radius, 0.5 times the 
geometric mean of the major and minor axis.  Note that $r$ = $R$/2, 
where $R$ is the size listed in Tab.~\ref{tgradient}  of BM89.}
\tablenotetext{e}{Distance references are listed in BM89,
Ladd et al. (1994), GBF93, Jijina et al. (1999).}
\tablenotetext{f}{These cores have deconvolved sizes similar to 
the beam size, so their small sizes are less certain than the sizes 
of the larger sources because of beam subtraction.}
\enddata
\label{tsize}
\end{deluxetable}
\end{center}

\clearpage

\begin{center}
\begin{deluxetable}{lccccc}
\footnotesize
\tablewidth{0pc}
\tablecaption{Volume Density and Mass}
\tablehead{
\colhead{Core}& \colhead{$n_{\rm vir}$\tablenotemark{a}}& 
\colhead{$M_{\rm vir}$\tablenotemark{a}}&
\colhead{X(N$_2$H$^+$)\tablenotemark{b}}& 
\colhead{$n_{\rm ex}$\tablenotemark{c}} &
\colhead{$M_{\rm ex}$\tablenotemark{c}} \nl
 & \colhead{(10$^{4}$ cm$^{-3}$)}& \colhead{(M$_{\odot}$)} &
 \colhead{($10^{-10}$)}&  
 \colhead{(10$^{4}$ cm$^{-3}$)}& \colhead{(M$_{\odot}$)}} 
 \startdata
PER4-A&   10\p1&        8\p1&      1.9\p0.4&       15\p11&    15\p11 \nl      
PER4-B&   3.2\p0.3&    6.0\p0.5&   4.1\p0.6&       23&        49     \nl
PER4-C&   3.9\p0.4&    4.9\p0.5&   2.6\p0.5&       23&        32     \nl
PER5&     3.2\p0.2&    5.3\p0.2&   5.0\p0.5&       18\p8&     30\p14 \nl
PER6&     2.8\p0.2&    8.6\p0.6&   4.3\p0.5&       8\p3&      28\p10 \nl
PER7&     5.6\p0.5&    4.1\p0.4&   8\p4&           4\p2&      4\p1   \nl
PER9&     6.0\p0.6&    5.3\p0.5&   2.2\p0.3&       23&        23     \nl
B5&       3.9\p0.3&    8.1\p0.6&   4.4\p0.5&       15\p8&     35\p19 \nl
L1389&    9\p1&        5.7\p0.7&   1.5\p0.3&       7\p4&      6\p4 \nl
L1489&    32\p1&       1.47\p0.06& 4\p2&           13\p3&     0.7\p0.1 \nl
L1498&    8.8\p0.3&    2.08\p0.07& 5\p3&           5.7\p0.7&  1.4\p0.2 \nl
L1495&    8.8\p0.3&    2.33\p0.09& 3\p2&           28\p14&    8\p4 \nl
B217&     15\p1&       2.6\p0.2&   3\p2&           20\p8&     4\p2 \nl
L1524&    5.4\p0.3&    3.3\p0.2&   7\p4&           5\p1&      3.1\p0.8 \nl
L1400K&   5.9\p0.2&    2.61\p0.09& 3\p2&           5\p2&      2.3\p0.7 \nl
TMC-2A&   17.7\p0.7&   1.64\p0.06& 5\p1&           8\p2&      0.8\p0.2 \nl
TMC-2&    3.6\p0.3&    6.5\p0.5&   4.0\p0.5&       11\p3&     20\p5 \nl
L1536&    3.3\p0.1&    4.9\p0.2&   3.8\p0.4&       10\p2&     14\p3 \nl
L1534&    5.7\p0.3&    4.5\p0.2&   3.2\p0.3&       23\p13&    19\p10 \nl
L1527&    12.0\p0.8&   2.5\p0.2&   4\p2&           4\p1&      0.9\p0.3 \nl
TMC-1NH3&  \nodata&     \nodata&   \nodata&        13\p2&     \nodata \nl
TMC-1C2&   \nodata&     \nodata&   \nodata&        6\p1&      \nodata \nl
TMC-1C&    \nodata&     \nodata&   \nodata&        3.8\p0.6&  \nodata \nl
TMC-1CS&   \nodata&     \nodata&   \nodata&        11\p3&      \nodata \nl
L1517B&   25\p1&       1.61\p0.07& 1.1\p0.1&       7\p3&      0.5\p0.2 \nl
L1512&    12.5\p0.4&   1.74\p0.06& 2\p1&           9\p2&      1.3\p0.2 \nl
L1544&    23.8\p0.5&   1.87\p0.04& 3.1\p0.7&       7\p1&     0.60\p0.08 \nl
L1582A&   16\p2&       3.5\p0.4&   0.7\p0.1&       23&        8 \nl
B35&      9\p2&        8\p1&       5\p2&           4\p2&      4\p2 \nl
L134A&    6.6\p0.4&    3.3\p0.2&   1.4\p0.2&       23&        12 \nl
L183&      \nodata&     \nodata&   \nodata&        8\p1&      \nodata \nl
L1681B-A& 24\p2&       2.6\p0.2&   1.8\p0.2&       23&        3 \nl
L1681B-B& 18\p1&       2.5\p0.2&   2.2\p0.3&       23&        4 \nl
L1696A&   13.3\p0.8&   2.1\p0.1&   2.0\p0.3&       23&        4 \nl
L43&      11.0\p0.5&   3.3\p0.2&   6\p2&           22\p3&     6.8\p0.9 \nl
L260&     4.6\p0.2&    3.0\p0.1&   2.0\p0.3&       3.0\p0.7&  2.0\p0.5 \nl
L158&     50\p4&       1.00\p0.08& 0.7\p0.2&       23&        0.6 \nl
L234A&    8.7\p0.3&    2.27\p0.08& 1.5\p0.2&       23&        7 \nl
L234E&     \nodata&     \nodata&   \nodata&        1.9\p0.8&  \nodata \nl
L63&      5.2\p0.3&    2.9\p0.2&   7\p4&           9\p3&      5\p2 \nl
B68&      12\p1&       2.3\p0.2&   1.0\p0.2&       23&        5 \nl
L483&     29\p2&       4.5\p0.3&   6\p1&           9.3\p0.5&  1.55\p0.08 \nl
B133&     8\p5&        19\p12&     0.7\p0.6&       23&        65 \nl
L778&     2.8\p0.3&    8.6\p0.8&   6\p3&           12\p5&     39\p18 \nl
B335&     89\p6&       1.27\p0.09& 1.1\p0.5&       8\p2&      0.21\p0.06 \nl
L1152&    2.6\p0.2&    9.5\p0.9&   3.7\p0.6&       23&        93 \nl
L1155C&   3.6\p0.3&    5.7\p0.5&   4\p3&           4\p2&      8\p4 \nl
L1082C&   8.2\p0.8&    4.7\p0.4&   2.0\p0.3&       2.5\p0.7&  1.7\p0.5 \nl
L1082A&   3.2\p0.3&    5.9\p0.5&   1.8\p0.3&       23&        47 \nl
L1228&   16\p1&        6.4\p0.6&   5\p1&           12\p2&     5.4\p0.9 \nl
L1174&    26\p7&       23\p6&      0.9\p0.3&       23&        24 \nl
BERN48&   79\p11 &     2.3\p0.3&   0.8\p0.2&       15\p10&    0.6\p0.4 \nl
L1172A&   4.1\p0.8&    10\p2&      6\p4&           5\p2&      13\p5 \nl
B361&     5\p1&        15\p3&      1.7\p0.4&       3\p2&      10\p5 \nl
L1031B&   9\p2&        95\p18&     1.3\p0.3&       23&        276 \nl
L1221&    36\p4&       5.2\p0.6&   3\p1&           12\p2&     1.8\p0.3 \nl
L1251A&   11\p1&       4.5\p0.5&   1.8\p0.4&       6\p3&      3\p1 \nl
L1251C&   12.4\p0.8&   2.4\p0.2&   4\p2&           8\p4&      1.7\p0.8 \nl
L1251E&   23\p2&       37\p4&      1.6\p0.2&       23&        39 \nl
L1262&    12\p1&       3.5\p0.4&   7\p4&           6\p1&      1.9\p0.5 \nl
\tablenotetext{a}{$n_{\rm vir}$ and $M_{\rm vir}$ is the virial volume density 
and mass, respectively. Data are not reported for those cores 
where the size cannot be determined.}
\tablenotetext{b}{Fractional abundance of N$_2$H$^+$ (X(N$_2$H$^+$) =
$N({\rm N_2H^+})$/$N({\rm H_2})$) calculated from $n_{\rm vir}$ and 
assuming a uniform sphere with $N({\rm H_2})$ = 4/3 $n_{\rm vir}$/1.1 $r$ 
(the factor 1.1 is to convert $n$ to $n({\rm H_2})$).} 
\tablenotetext{c}{Volume density and mass coming from the density to critical 
density ($n_{\rm cr}$ = 2$\times$10$^5$ cm$^{-3}$; \cite{ubg97}, 
ApJ, 482, 245) ratio, calculated by using equation (43) in \cite{g92}.
Values with no associated errors imply an assumed $T_{\rm ex}$ value 
(= 5 K).  The excitation temperature and the optical depth of the 
``averaged'' or the ``peak''  spectrum (see Table 
2) have been used, whichever has the smallest error. In the calculation, 
$T_{\rm kin}$ = 10 K and $T_{\rm bb}$ = 2.7 K.} 
\enddata
\label{tdensity}
\end{deluxetable}
\end{center}

\begin{center}
\begin{deluxetable}{lccccccc}
\scriptsize
\tablewidth{0pc}
\tablecaption{Results of Gradient Fitting}
\tablehead{
\colhead{Core}& \colhead{Number} & \colhead{${\cal G}$}& 
\colhead{$\Theta_{\cal G}$}& 
\colhead{${\cal G}$$\times$ $r$}&
\colhead{$\beta$\tablenotemark{a}} &
\colhead{$<V_{\rm LSR}-V_{\rm fit}>$\tablenotemark{b}} & 
\colhead{$s_{<V_{\rm LSR}-V_{\rm fit}>}$\tablenotemark{b}} \nl
 & \colhead{of points} & \colhead{(km/s/pc)}& \colhead{(deg E of N)}& 
   \colhead{(km/s)}& (10$^{-3}$) & \colhead{(km/s)}& \colhead{(km/s)} }
 \startdata
PER4&   10& 0.64\p0.05& -17\p6& \nodata& \nodata&  -0.01\p0.01& 0.07\p0.01  \nl
B5&     11& 0.86\p0.08& 84\p2& 0.082& 9.2&        0.015\p0.007& 0.07\p0.01 \nl
L1489&  19& 0.7\p0.1& 110\p12&  0.018& 0.74&     0.004\p0.004& 0.02\p0.01  \nl
L1498&  26& 0.5\p0.1& 9\p10&    0.024& 1.4&     -0.002\p0.003& 0.013\p0.003  \nl
L1495&   9& 0.9\p0.2& 87\p8&   0.045& 4.5&       0.004\p0.006& 0.046\p0.008 \nl 
B217&   11& 2.2\p0.2& 36\p4&   0.093& 16&         0.024\p0.008& 0.060\p0.008 \nl
L1524&   9& 1.8\p0.3& 22\p11&    0.11& 29&       0.012\p0.009& 0.03\p0.01 \nl
L1400K& 10& 1.8\p0.1& 62\p4&   0.10& 27&      0.000\p0.005& 0.019\p0.006 \nl 
TMC-2&  24& 0.7\p0.1& 86\p7&   0.067& 6.6&    -0.018\p0.009& 0.093\p0.009  \nl
L1536&  21& 2.11\p0.07& -7\p1&  0.18& 66&       -0.002\p0.004& 0.041\p0.004  \nl 
L1534&  15& 2.5\p0.2& 25\p3&   0.17& 53&        -0.011\p0.007& 0.083\p0.008  \nl
TMC-1NH3& 16& 5.98\p0.08& 35.1\p0.7& \nodata & \nodata& -0.008\p0.005& 0.159\p0.006 \nl
TMC-1C2& 14& 1.25\p0.09& 94\p6&  \nodata & \nodata& 0.004\p0.005& 0.036\p0.007 \nl
TMC-1C& 15& 0.8\p0.1& -136\p10& \nodata & \nodata& 0.004\p0.005& 0.040\p0.005  \nl
TMC-1CS& 17& 2.7\p0.2& 44\p3&  \nodata& \nodata& -0.003\p0.006& 0.030\p0.008  \nl
L1512&  28& 1.41\p0.07& 166\p3&  0.054& 7.7& -0.002\p0.002& 0.017\p0.003  \nl
L1544&  32& 1.0\p0.1& -171\p5& 0.032& 2.0& 0.011\p0.003& 0.033\p0.003  \nl 
L183&   35& 1.19\p0.08& -55\p3&  \nodata & \nodata& -0.006\p0.003& 0.055\p0.005  \nl
L43&    42& 2.72\p0.06& 127\p2&  0.13& 33& -0.022\p0.003& 0.045\p0.004 \nl
L260&   9& 0.1\p0.1& 86\p94&  0.0078& 0.11& -0.022\p0.007& 0.04\p0.01 \nl
L483&   44& 2.38\p0.06& 63\p2&   0.096& 9.5& 0.022\p0.004& 0.063\p0.004 \nl
B335&   10& 0.1\p0.3& -58\p187& 0.0014& 0.005& -0.005\p0.008& 0.021\p0.009  \nl
L1228& 43& 2.2\p0.1& -64\p2&  0.12& 15& 0.048\p0.009& 0.10\p0.01  \nl
L1174&  13& 4.0\p0.2& -158\p4& 0.29& 30& 0.01\p0.02& 0.08\p0.03  \nl
L1221&  22& 3.2\p0.3& -138\p5& 0.13& 14& 0.02\p0.01& 0.07\p0.01  \nl 
L1251E& 90& 1.66\p0.06& -129\p3& 0.15& 5.8& -0.175\p0.007& 0.355\p0.006  \nl
\tablenotetext{a}{$\beta$ is the ratio of rotational kinetic 
energy to gravitational energy (see equation 6 of Goodman et al.
1993). Assuming $\rho_0$ = $m \times n_{\rm vir}$, with $m$ = 2.33 amu, 
and $n_{\rm vir}$ $\equiv$ volume density from Table 4, 
$\beta$ = 4.86$\times$10$^{2}$ $<{\cal G}^2>/n_{\rm vir}$ with 
${\cal G}$ in km/s/pc.}
\tablenotetext{b}{Average value (and relative standard deviation) of the fit 
residuals $V_{\rm LSR}(i)$ - $V_{\rm fit}(i)$ across the core, where 
$V_{\rm fit}(i)$ is the LSR velocity at position ($\alpha(i)$,
$\delta(i)$) determined by the least square fit of a velocity gradient.} 
\enddata
\label{tgradient}
\end{deluxetable}
\end{center}

\clearpage

\begin{center}
\begin{deluxetable}{lcccc}
\footnotesize
\tablewidth{0pc}
\tablecaption{Integrated intensity profiles}
\tablehead{
\colhead{Core}& \colhead{$\alpha$} & \colhead{$\chi^2_1$} & 
\colhead{$b_{\rm break}$} & \colhead{$\chi^2_2$} \nl  
 & & \colhead{($\times 10^2$)} & \colhead{(pc)} & \colhead{($\times 10^2$)}} 
 \startdata
\multicolumn{5}{c}{Cores with stars} \nl
\hline
PER6& 1.9 & 5.8 & 0.051 & 3.7 \nl
L1489& 2.2 & 0.65 & $<$0.007\tablenotemark{a} & 8.7 \nl 
L1495& 1.9 & 3.4 & 0.024 & 1.3 \nl
L43&  1.8 & 31 & 0.019 & 23 \nl
L483& 2.2 & 23 & $<$0.010 & 55 \nl 
L1082C& 2.5 & 0.24 & $<$0.021 & 9.4 \nl
L1228& 2.8 & 23 & $<$0.015 & 220 \nl
L1174&  2.2 & 0.61 & $<$0.021 & 8.8 \nl
L1221&  2.1 & 22 & $<$0.010 & 28 \nl
\hline \nl
\multicolumn{5}{c}{Starless cores} \nl
\hline 
L1498& 1.8 & 2.4 & 0.014 & 3.0 \nl
B217&  1.9 & 1.9 & 0.020 & 0.86 \nl
L1400K& 1.9 & 0.32 & 0.021 & 0.33 \nl
TMC-2&  1.7 & 15 & 0.041 & 6.5 \nl
L1536&  1.6 & 16 & 0.037 & 3.0 \nl
L1512&  1.9 & 6.4 & 0.020 & 3.3 \nl
L1544&  1.9 & 2.1 & 0.014 & 0.96 \nl
L63& 1.8 & 13 & 0.031 & 4.8 \nl 
\enddata
\tablenotetext{a}{The smallest $\chi^2$ is obtained with the 
smallest $b_{\rm break}$ (10$^{\prime\prime}$), indicating
that the profile is consistent with a single power law
(see text for details).}
\label{tprofile}
\end{deluxetable}
\end{center}

\clearpage

\begin{center}
\begin{deluxetable}{lcccccccc}
\scriptsize
\tablewidth{0pt}
\tablecaption{Variation of $\Delta v$ across the cores}
\tablehead{
\colhead{Core}& \colhead{p\tablenotemark{a}}&  \colhead{q\tablenotemark{a}}&  
\colhead{$cc$}&
\colhead{$<\Delta v>$\tablenotemark{b}}& 
\colhead{$s_{<\Delta v>}$\tablenotemark{b}}& 
\colhead{$p^{\prime}$\tablenotemark{c}}& 
\colhead{$q^{\prime}$\tablenotemark{c}}&  \colhead{$cc^{\prime}$}\nl
 & \colhead{(km s$^{-1}$)}& \colhead{(km s$^{-1}$ pc$^{-1}$)}& & 
   \colhead{(km s$^{-1}$)}& \colhead{(km s$^{-1}$)}& 
   \colhead{(km s$^{-1}$)}& \colhead{(km s$^{-1}$ pc$^{-1}$)} & }
 \startdata
PER4& 0.43\p0.04& -0.9\p0.4& -0.43& 0.37\p0.02& 0.10\p0.03& 
      0.2\p0.1& -0.4\p0.9& -0.91 \nl
B5& 0.48\p0.03& -1.8\p0.3& -0.72& 0.33\p0.01& 0.09\p0.02& 
    0.04\p0.05& 0.4\p0.4& 0.91 \nl  
L1489& 0.24\p0.02& 1.4\p0.7& 0.30& 0.298\p0.008& 0.06\p0.02& 
       0.03\p0.01& 0.2\p0.4& 0.23 \nl
L1498& 0.19\p0.01& 1.2\p0.4& 0.41& 0.235\p0.006& 0.038\p0.009& 
       0.008\p0.010& 0.8\p0.3& 0.81 \nl 
L1495& 0.21\p0.03& 2.0\p0.8& 0.53& 0.29\p0.01& 0.05\p0.01& 
      0.05\p0.06& 0\p1& 0.11 \nl 
B217&  0.33\p0.03& -0.5\p0.8& -0.14& 0.33\p0.01& 0.07\p0.02& 
       -0.01\p0.05& 1\p1& 0.68 \nl
L1524& 0.24\p0.04& 2\p1& 0.52& 0.34\p0.02& 0.07\p0.02& 
       0.1\p0.1& -1\p2& -0.95 \nl
L1400K& 0.18\p0.03& 1.3\p0.6& 0.80& 0.24\p0.01& 0.02\p0.01& 
        0.04\p0.04& -0.3\p0.8& -0.92 \nl
TMC-2&  0.35\p0.03& -0.7\p0.5& -0.14& 0.40\p0.01& 0.11\p0.02& 
        0.01\p0.02& 1.4\p0.4& 0.88 \nl
L1536&  0.25\p0.02& -0.2\p0.3& -0.08& 0.248\p0.007& 0.044\p0.008& 
        0.04\p0.01& 0.1\p0.2& 0.20 \nl
L1534&  0.38\p0.03& -0.7\p0.6& -0.25& 0.36\p0.01& 0.05\p0.01& 
        0.00\p0.02& 0.7\p0.4& 0.91 \nl
TMC-1NH3& 0.40\p0.02& -1.3\p0.4& -0.30& 0.39\p0.01& 0.13\p0.02& 
          -0.08\p0.03& 3.2\p0.5& 0.79 \nl
TMC-1C2& 0.34\p0.03& -1.5\p0.6& -0.60& 0.282\p0.009& 0.04\p0.01&
         0.01\p0.03& 0.6\p0.6& 0.54 \nl
TMC-1C&  0.26\p0.03& 0.1\p0.6& 0.03& 0.29\p0.01& 0.05\p0.01& 
         0.03\p0.03& 0.5\p0.5& 0.54 \nl
TMC-1CS& 0.38\p0.03& 0.2\p0.6& 0.06& 0.40\p0.01& 0.07\p0.02&  
         0.03\p0.03& 0.5\p0.6& 0.62 \nl
L1512& 0.17\p0.01& 0.8\p0.4& 0.29& 0.202\p0.005& 0.033\p0.006& 
       0.019\p0.007& 0.3\p0.2& 0.70 \nl 
L1544&  0.30\p0.01& -0.1\p0.4& -0.03& 0.303\p0.006& 0.043\p0.008&
        0.017\p0.007& 0.6\p0.2& 0.82 \nl
L183&   0.23\p0.01& 0.7\p0.3& 0.29& 0.282\p0.008& 0.06\p0.01&
        -0.005\p0.009& 0.9\p0.2& 0.85 \nl
L43&   0.27\p0.01& 1.7\p0.3& 0.30& 0.394\p0.007& 0.11\p0.01& 
       0.031\p0.006& 1.2\p0.1& 0.84 \nl 
L260& 0.18\p0.03& 1.3\p0.8& 0.51& 0.24\p0.01& 0.05\p0.02&  
      -0.01\p0.08& 1\p2& 0.59 \nl
L483&  0.40\p0.01& -0.4\p0.3& -0.07& 0.45\p0.01& 0.15\p0.02& 
       0.067\p0.009& 1.0\p0.2& 0.65 \nl 
B335&  0.43\p0.04& -1\p1& -0.27& 0.39\p0.02& 0.07\p0.02& 
       0.05\p0.08& 0\p2& 0.43 \nl
L1228& 0.65\p0.03& -0.7\p0.4& -0.16& 0.68\p0.02& 0.19\p0.02&
       0.12\p0.02& 0.7\p0.3& 0.61 \nl
L1174& 1.2\p0.1& -9\p2& -0.73& 0.81\p0.04& 0.36\p0.07&
       0.4\p0.3& 0\p3& 0.49 \nl
L1221& 0.70\p0.04& -2\p1& -0.24& 0.69\p0.02& 0.15\p0.04&
       0.00\p0.04& 3\p1& 0.90 \nl
L1251E& 0.90\p0.02& -3.0\p0.3& -0.31& 0.93\p0.02& 0.34\p0.02& 
        0.25\p0.01& 1.0\p0.1& 0.80 \nl
\tablenotetext{a}{$p$ is the intercept and $q$ is the slope of the
best--fit linear $\Delta v$ -- $b$ relation in each core (see text).}
\tablenotetext{b}{Average linewidth and corresponding standard deviation.}
\tablenotetext{c}{$p^{\prime}$ is the intercept and $q^{\prime}$ is the slope
of the $s_{\Delta v}$ -- $b$ relation in each core (see text).}
\enddata
\label{tdeltav}
\end{deluxetable}
\end{center}

\clearpage

\begin{center}
\begin{deluxetable}{lcccccc}
\scriptsize
\tablewidth{0pt}
\tablecaption{Statistics on cores with and without stars}
\tablehead{
 & \multicolumn{3}{c}{Cores with Stars} & \multicolumn{3}{c}{Starless Cores} \nl
  & \colhead{Mean} & \colhead{Standard} & \colhead{Number}  &
 \colhead{Mean} & \colhead{Standard} & \colhead{Number} \nl
 \colhead{Parameter}  & & \colhead{Deviation} & \colhead{of Cores} &  
   & \colhead{Deviation} & \colhead{of Cores} }
 \startdata
$\Delta v_{\rm NT}$ (km s$^{-1}$) \tablenotemark{a} & 0.5 & 0.3 & 35 & 0.3 & 0.1 & 25 \nl
$T_{\rm ex}$ (K) & 5 & 1 & 15 & 4.4 & 0.8 & 14 \nl
$N_{\rm TOT}$(N$_2$H$^+$) (10$^{12}$ cm$^{-2}$) & 8 & 5 & 35 & 6 & 3 & 25 \nl
$r$ (pc) & 0.07 & 0.03 & 22 & 0.05 & 0.02 & 13 \nl
Aspect ratio & 1.9 & 0.7 & 35 & 2 & 1 & 19 \nl
$n_{\rm vir}$ (10$^5$ cm$^{-3}$) & 2 & 2 & 34 & 1.2 & 0.7 & 19 \nl
$M_{\rm vir}$ (M$_{\odot}$) & 9 & 16 & 34 & 3 & 2 & 19 \nl
$n_{\rm ex}$ (10$^5$ cm$^{-3}$) & 0.9 & 0.7 & 21 & 0.8 & 0.4 & 17 \nl
$M_{\rm ex}$ (M$_{\odot}$) & 8 & 11 & 21 & 3 & 6 & 17  \nl
$X$(N$_2$H$^+$) (10$^{-10}$) & 3 & 2 & 34 & 2 & 1 & 18 \nl
Gradient (km/s/pc) & 2 & 1 & 14& 2 & 1 & 12 \nl
$\beta$ & 0.02 & 0.02 & 13 & 0.02 & 0.02 & 7 \nl 
\enddata
\tablenotetext{a}{$\Delta v_{\rm NT}$ is the non--thermal part of the line width: 
$\Delta v_{\rm NT}^2$ = $\Delta v_{\rm obs}^2$ - $8 ln(2) \times 
(k T/m_{\rm obs})$.}   
\label{tsummary}
\end{deluxetable}
\end{center}

\clearpage

\figcaption{Average \NTHP (1--0) spectra of selected cores, obtained by summing 
all the spectra inside the half maximum map contour. The spectra are in 
antenna temperature units.  Three starless cores (L1498, L1544, TMC--2)
and three cores with stars, indicated by the symbol ``*'', are displayed.  
This sub--sample show the variation in the \NTHP \ profile from quiescent
(L1498) to more ``turbulent'' starless cores (L1544, TMC--2), and
from relatively quiescent cores with stars (L1489), to cores with young 
stellar objects driving powerful outflows (L1228), to cores with complex 
internal structure (L1215E). 
\label{fspectra}}

\figcaption{Maps of the N$_2$H$^+$(1--0) intensity integrated over the 
seven hyperfine components.  The angular
scale is the same for each core.  The contours and the grey scale mark
the 20\%, 35\%, 50\%, 65\%, 80\%, and 95\% of the map peak, reported in
Table 1 (see column 7).  The 
thick contour is the half maximum (50\%) level, which defines 
the core size.  Small circles are observed positions and the stars
indicate the location of the associated infrared  source detected 
by IRAS.  The FCRAO half power beam width (HPBW) is shown in the map of 
Per 4.
\label{fmap}}

\figcaption{Correlations between N$_2$H$^+$ and NH$_3$ (from BM89) properties
at the map peaks for the cores with distance $D$ $<$ 200 pc: 
(a) column density; (b) integrated intensity; (c) excitation temperature.
Best--fit lines are referred to starless cores (black dots).  
Cores with stars (grey dots) do not show significant correlations 
between \NTHP \ and \AMM \ properties (see text). 
\label{fcorr}}

\figcaption{Distribution of core radii mapped in N$_2$H$^+$ (shaded histogram)
and NH$_3$ (thin lines), for  the entire sample ({\it top}), 
 cores with stars ({\it center}), and  starless cores ({\it bottom}).  Cores
associated with young stellar objects tend to have smaller sizes in 
N$_2$H$^+$ than in NH$_3$ maps, suggesting higher central densities than 
starless cores.
\label{fhisto}}

\figcaption{The ``excitation'' mass $M_{\rm ex}$, calculated assuming a spherical 
core with constant density $n_{\rm ex}$ (see text), 
as a function of the virial mass $M_{\rm vir}$ 
for starless cores (empty circles) and cores with stars (filled circles).
Cores with $M_{\rm ex}/\sigma_{M_{\rm ex}}$ 
or $M_{\rm vir}/\sigma_{M_{\rm vir}}$ $<$ 2
are not reported in the figure. 
\label{fvir}}

\figcaption{N$_2$H$^+$(1--0) integrated intensity maps of those cores
where ``local'' velocity gradients have been calculated.  The grey--scale
levels represent the 30\%, 50\%, 70\%, and 90\% of the map peak.  
Small dots mark the position of observed spectra where the determination
of $V_{\rm LSR}$ from hfs fit has been possible (see Sect.~3.2). The 
white arrows show the magnitude and the direction of the velocity 
gradient calculated by applying the least square fitting routine to 
the grid of positions centered on the corresponding arrow.
The black arrow in the bottom of each panel represents the total 
velocity gradient listed in Table 5 (the magnitude of the white arrows
is in units of the total gradient).
\label{fgradient}}

\figcaption{N$_2$H$^+$(1--0) integrated intensity as a function of 
impact parameter $b$.  Symbols are different for starless cores
(empty circles) and cores associated with stars (filled circles).
Thin (dashed) curves are Model 1 (Model 2) best--fit profiles (see text). 
Dotted profiles have been
obtained by using the normalized integrated intensity of the ``thin''
hyperfine 
component ($F_1$,F = 1,0 $\rightarrow$ 1,1). Most of the starless cores 
present a ``shallow'' structure at $b$ $<$ $\sim$ 0.03 pc, in agreement with 
results from dust continuum emission maps.
\label{fprofiles}}

\figcaption{Maps of line width (grey scale) overlapped with integrated intensity maps 
(contours; levels are 30, 50, 70, and 90\% of the peak). Grey contours range from 
$\Delta v_{\rm min}$ to $ \leq \Delta v_{\rm max}$ in steps of 
$2 <\sigma_{\Delta v}>$, where $<\sigma_{\Delta v}>$ is the mean line width error 
in the selected positions. Grey areas enclose all the points with $\Delta v$ 
values between two adjacent grey contours. The dots mark the positions which 
have been used in 
the $\Delta v$ maps (i.e. where $\Delta v/\sigma_{\Delta v} \geq 3$ and 
$I/\sigma_I \geq 5$).
Values of $\Delta v_{\rm min}$, $\Delta v_{\rm max}$, and $2 <\sigma_{\Delta v}>$ 
(in km s$^{-1}$) in each core are the following: i) 0.15, 0.29, 0.05 in 
L1498; ii) 0.21, 0.53, 0.11 in TMC--2; iii) 0.13, 0.24, 0.05 in 
L1512; iv)  0.19, 0.39, 0.07 in L1544; v) 0.18, 0.38, 0.08 in L183; vi) 
0.24, 0.70, 0.08 in L43; vii) 0.24, 0.85, 0.11  in L483;
viii) 0.32, 1.28, 0.21 in L1228; ix) 0.37, 1.91, 0.27 in L1251E.
\label{fdvmap}}

\figcaption{Intrinsic N$_2$H$^+$(1--0) line width $\Delta v$ as a function
of impact parameter $b$ for selected cores with positive (L1498, L1495),
null (TMC--2, L1228), and negative (TMC--1C2, B5) correlations 
(see text for details). Cores with stars are marked with filled symbols
whereas starless cores have emtpy symbols.
\label{fdvb}}

\figcaption{The variance of the average nonthermal line 
width of a core as a function of $<\Delta v_{\rm NT}>$.  
Empty circles represent starless cores, whereas filled circles 
are cores with stars.  Lines indicate 
how dispersion is expected to increase with $<\Delta v_{\rm NT}>$ 
in a simple model of ``cells'' along the line of sight (see text).
The number of cells is indicated. In both classes of cores, dispersion is 
increasing with increasing $<\Delta v_{\rm NT}>$, following models
with $N$ $\sim$ 10.
\label{fsigmadv}}

\figcaption{Dispersion of the average gradient fit residual
(Tab.~\ref{tgradient}) in a core vs. the average nonthermal 
line width (Tab.~\ref{tdeltav}). The ``cell'' model is indicated
by thin lines for number of cells equal to 1, 3, 10, and 30, as in 
Fig.~\ref{fsigmadv}.  A linear least square fit to the data gives
$N$ = 13.  These positive correlations
strongly suggest the presence of turbulent motions which cause line width
broadening and contribute to line width and LSR velocity dispersion across
the cores.
\label{fdvvlsr}}

\newpage
\plotone{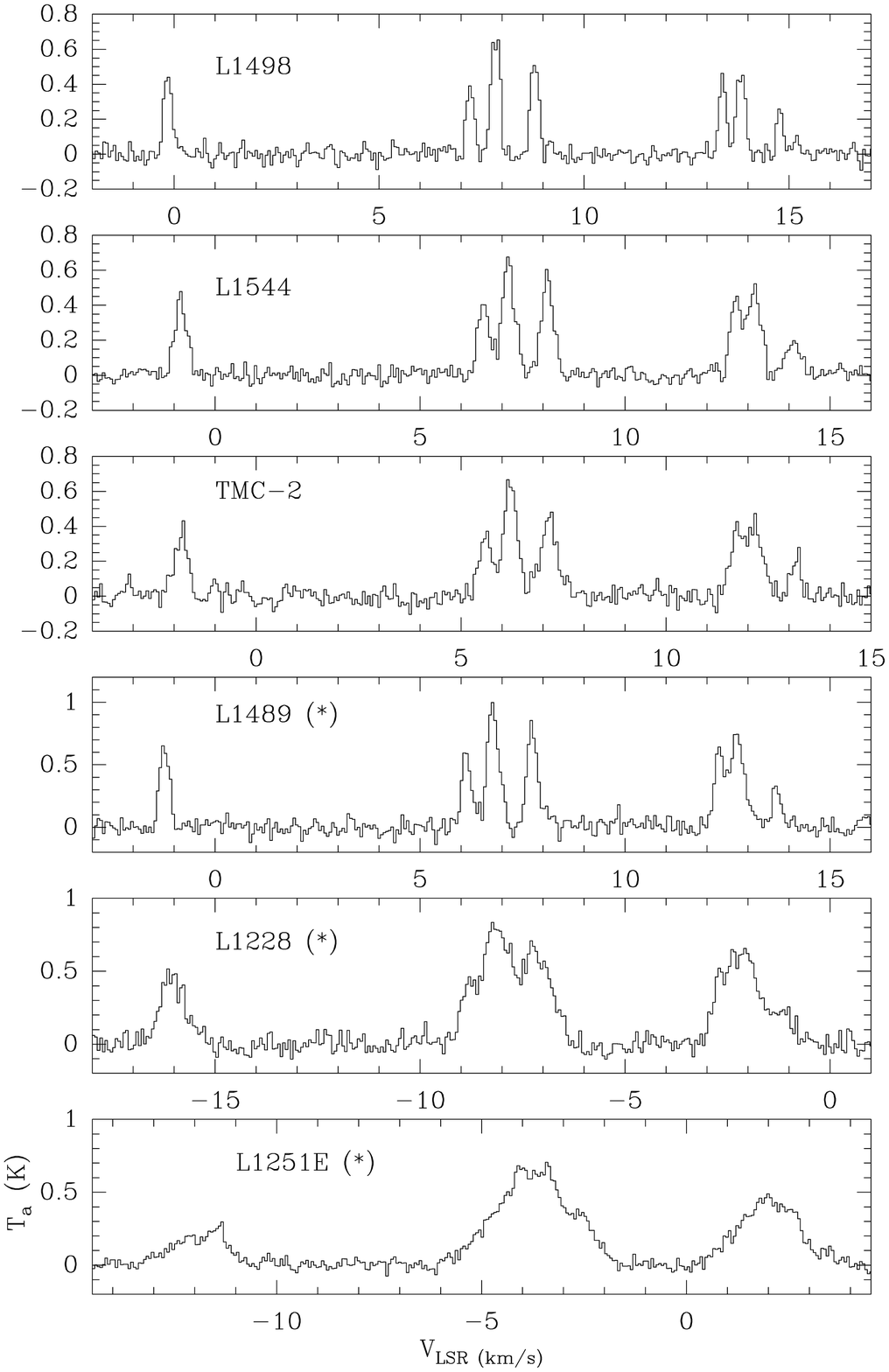}
\newpage
\plotone{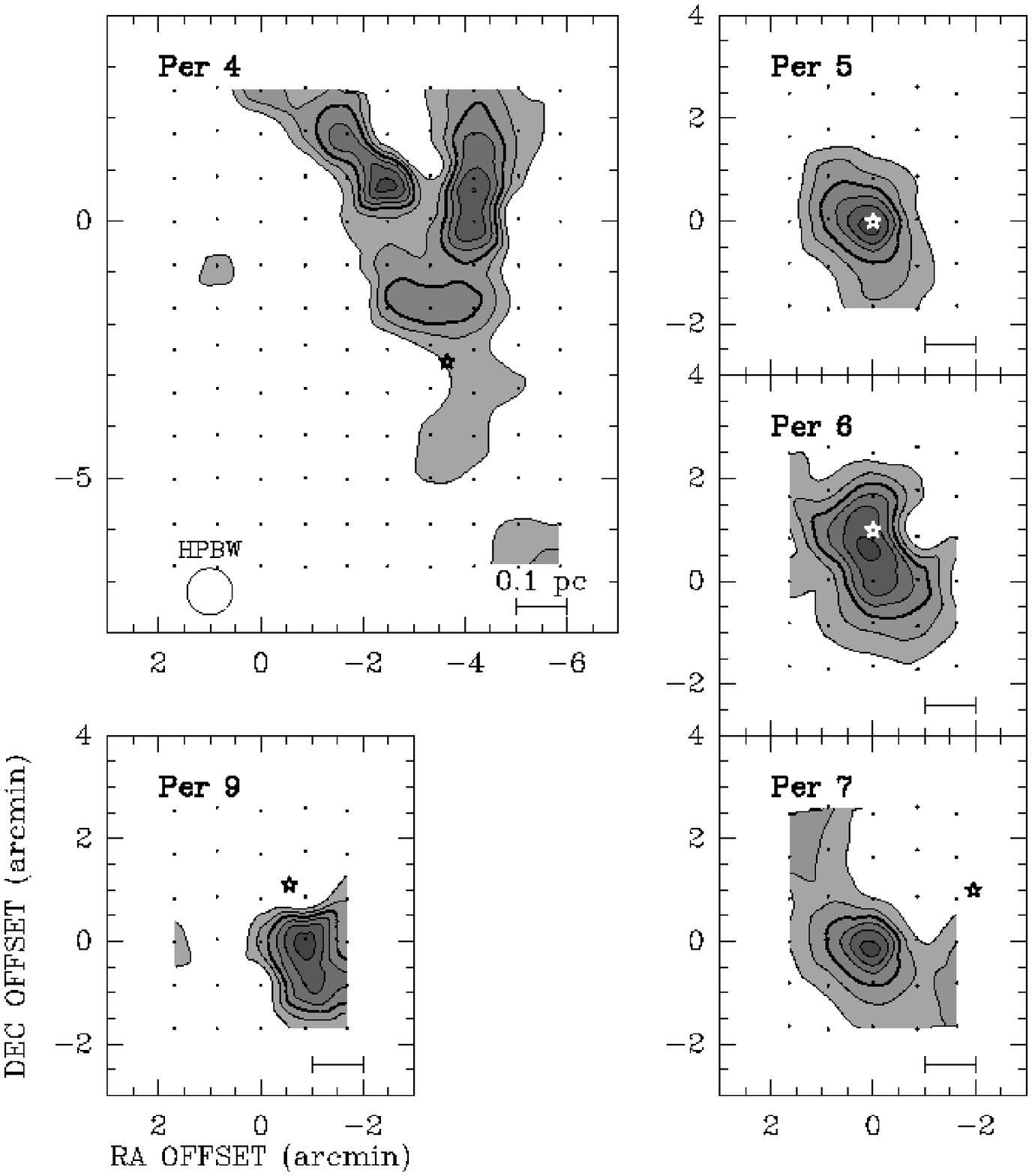}
\newpage
\plotone{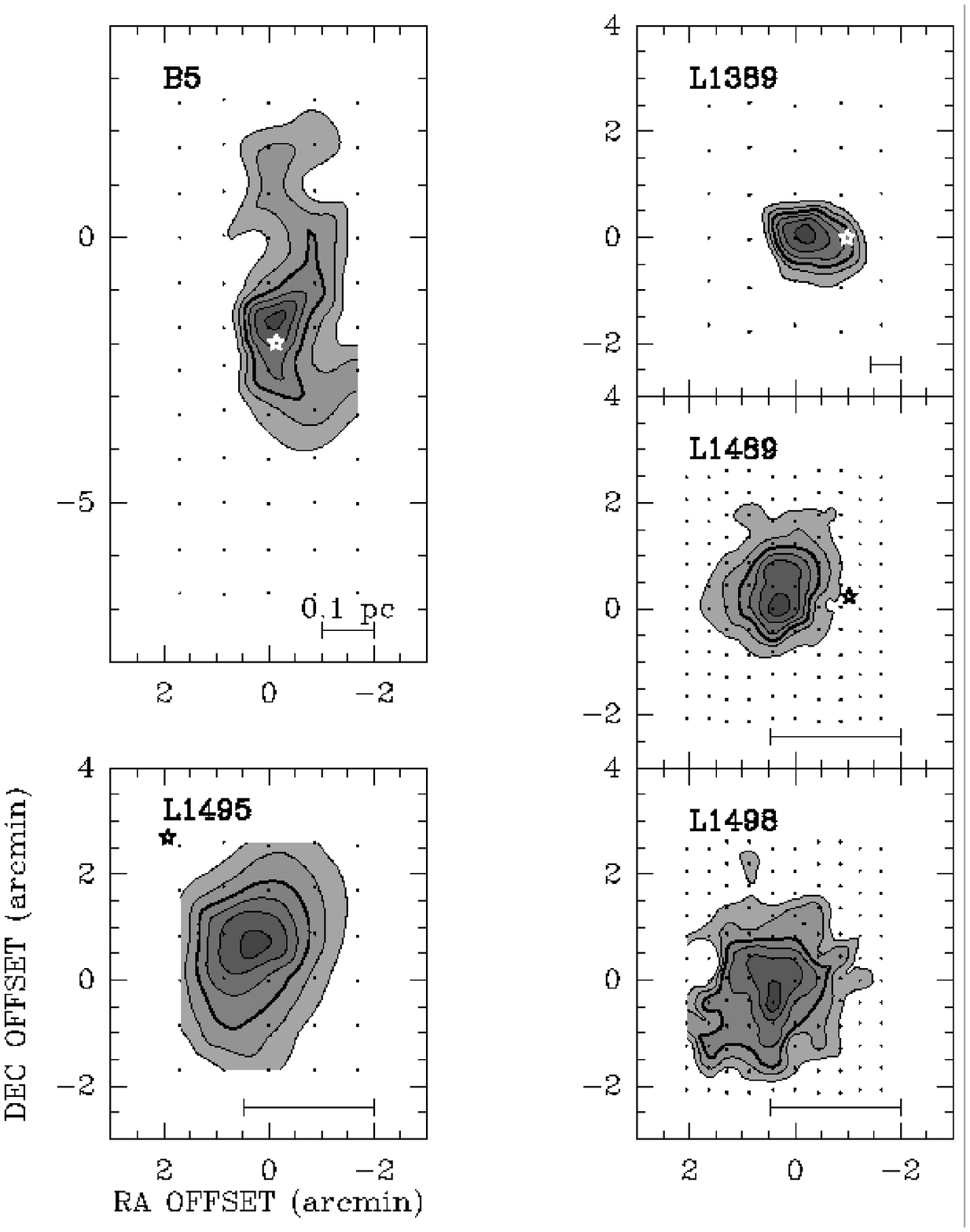}
\newpage
\plotone{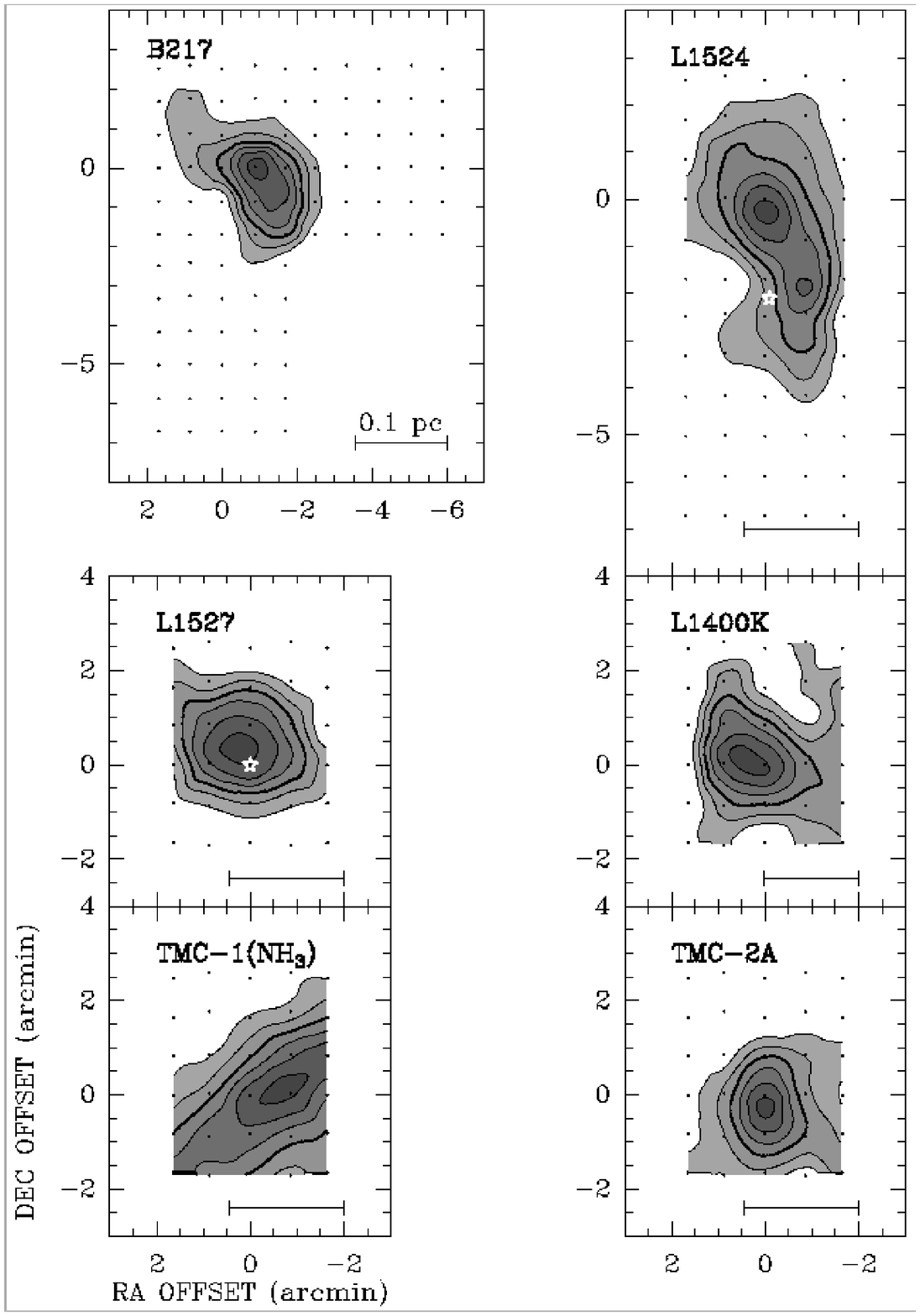}
\newpage
\plotone{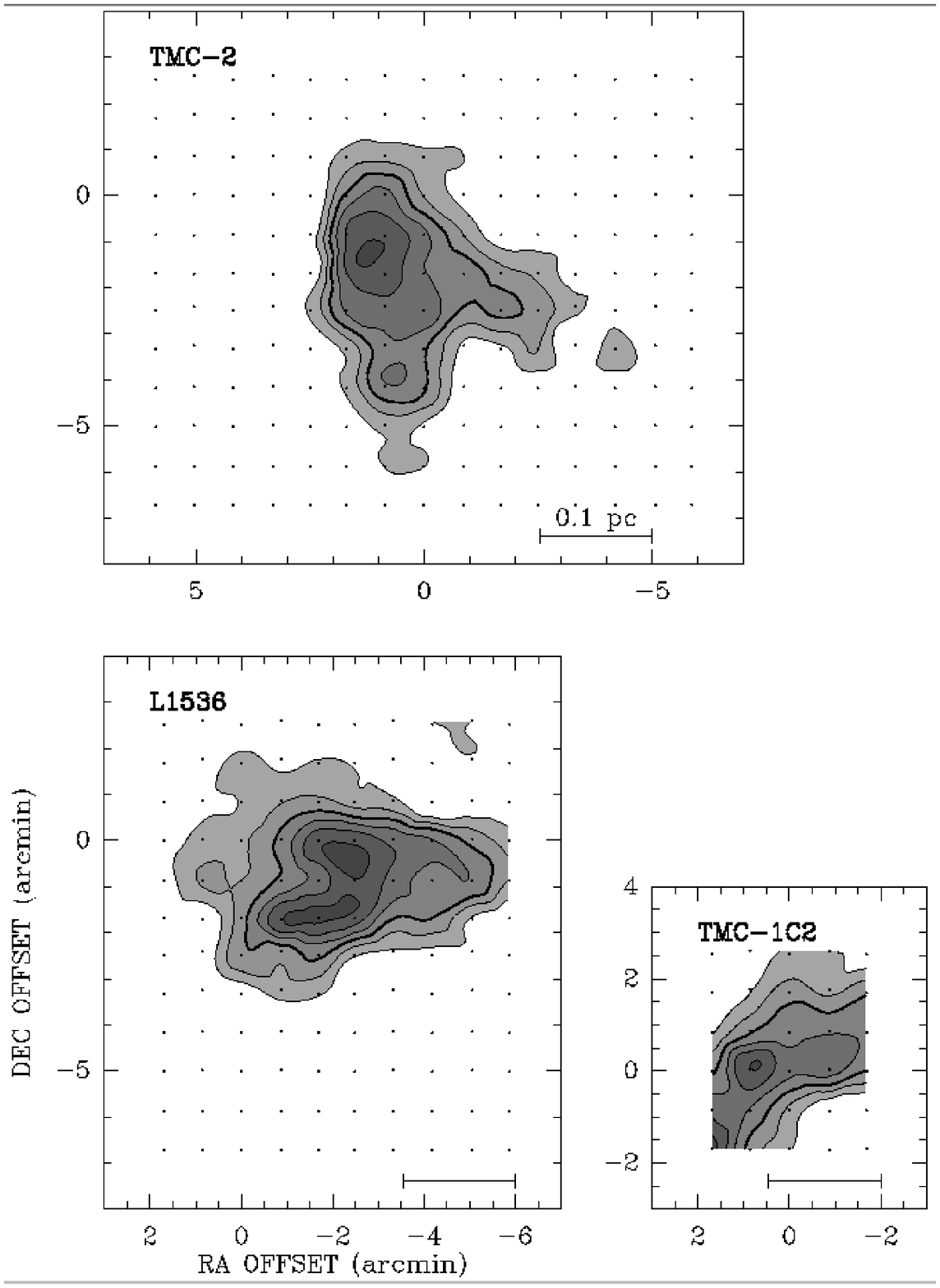}
\newpage
\plotone{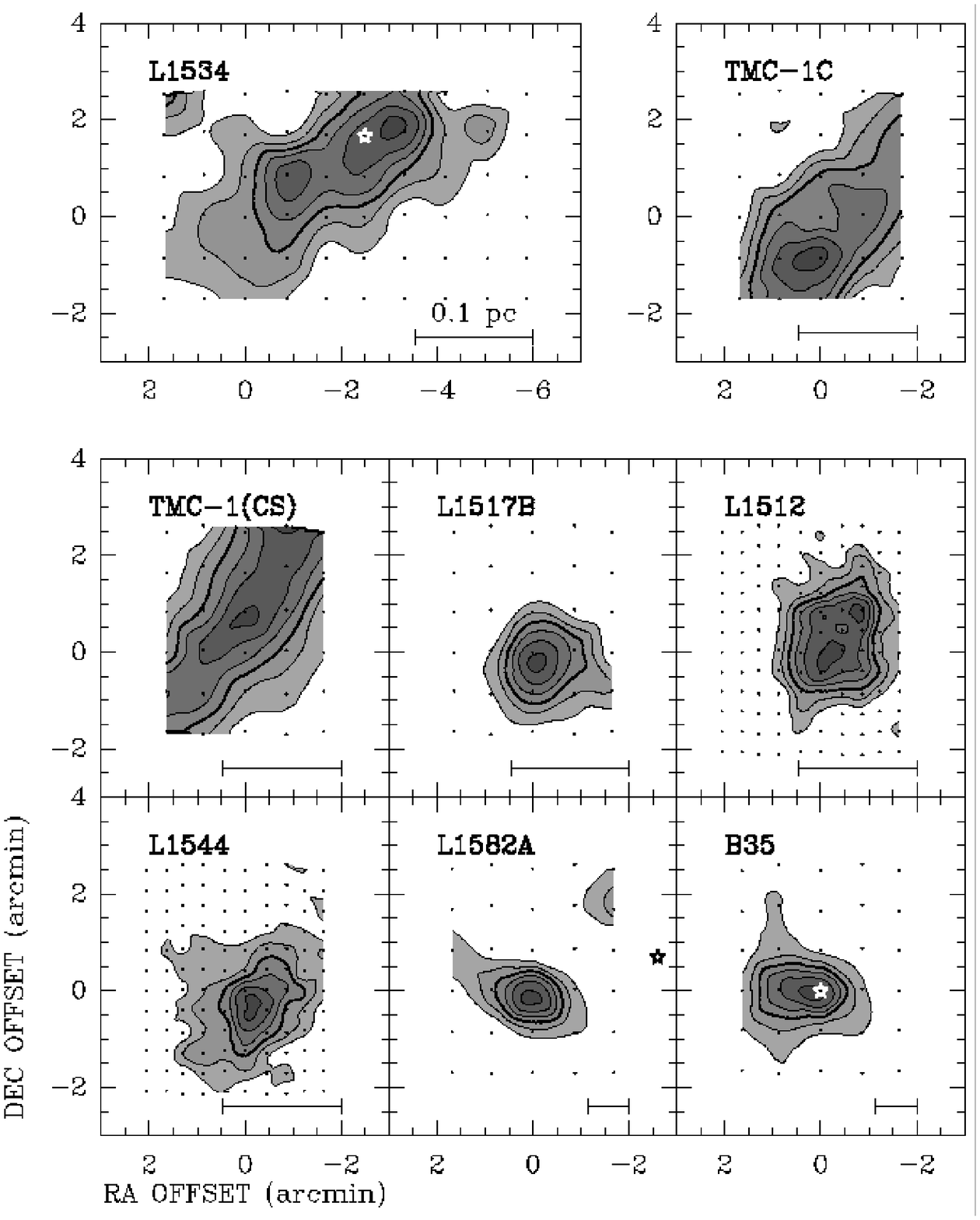}
\newpage
\plotone{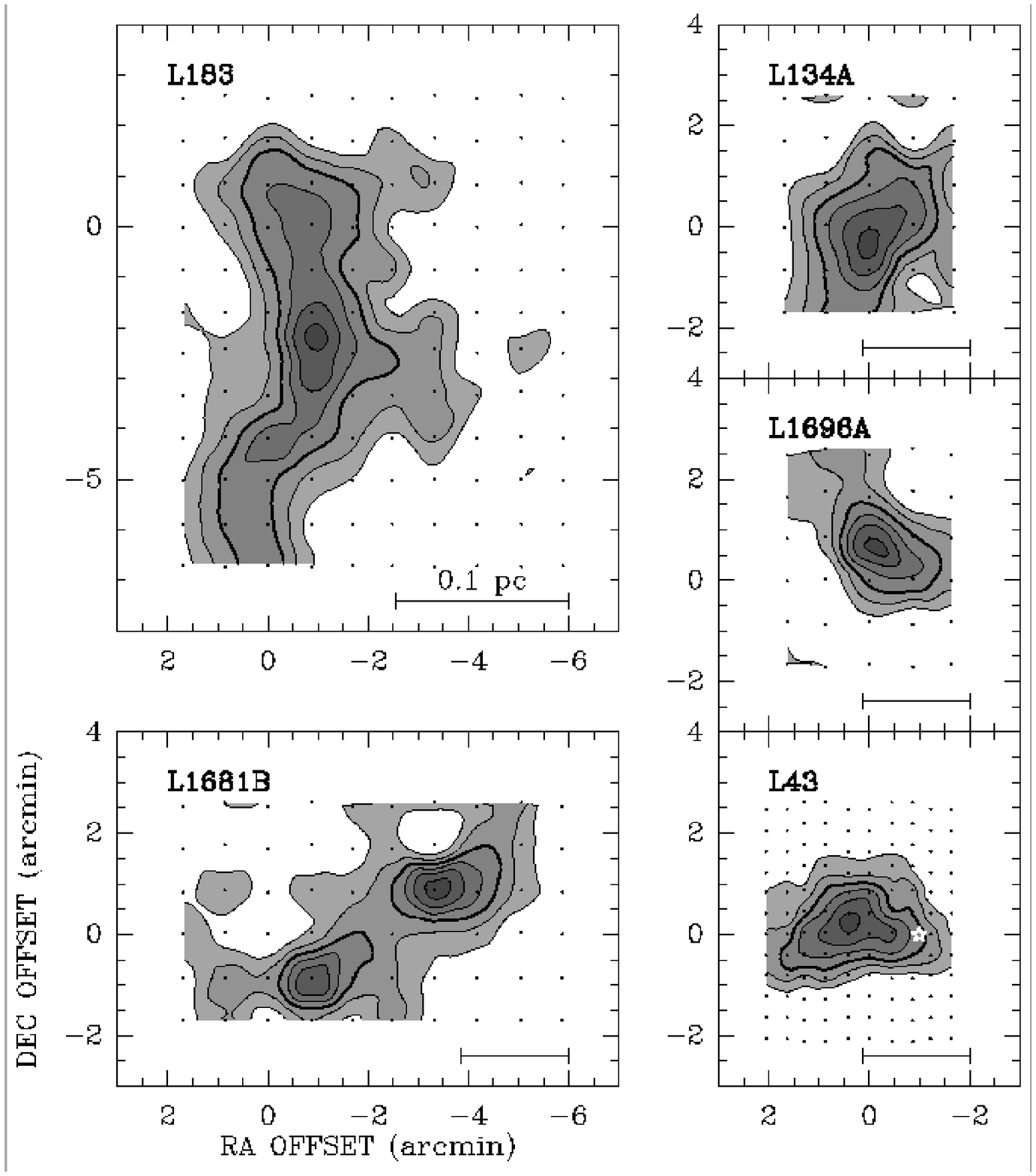}
\newpage
\plotone{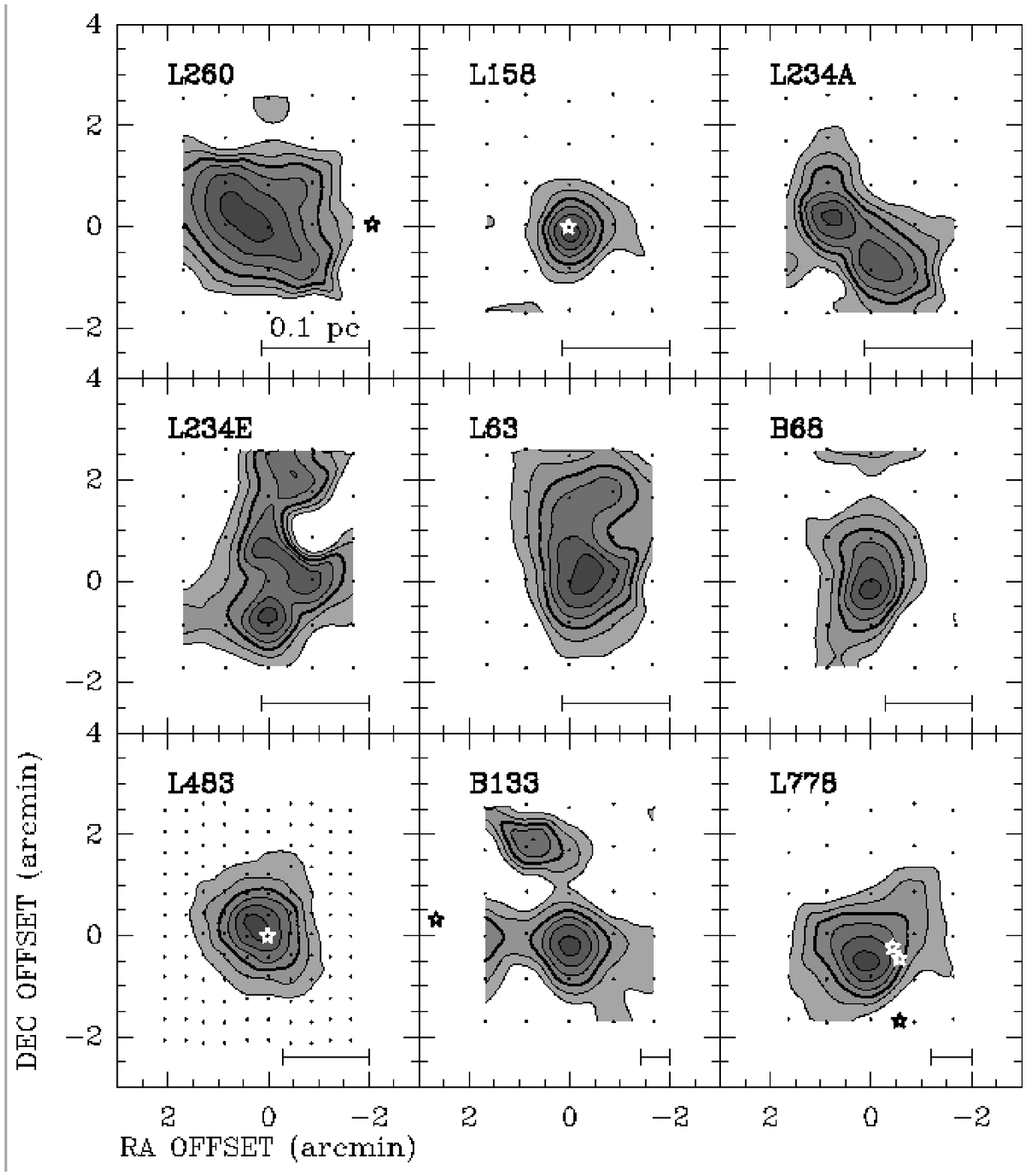}
\newpage
\plotone{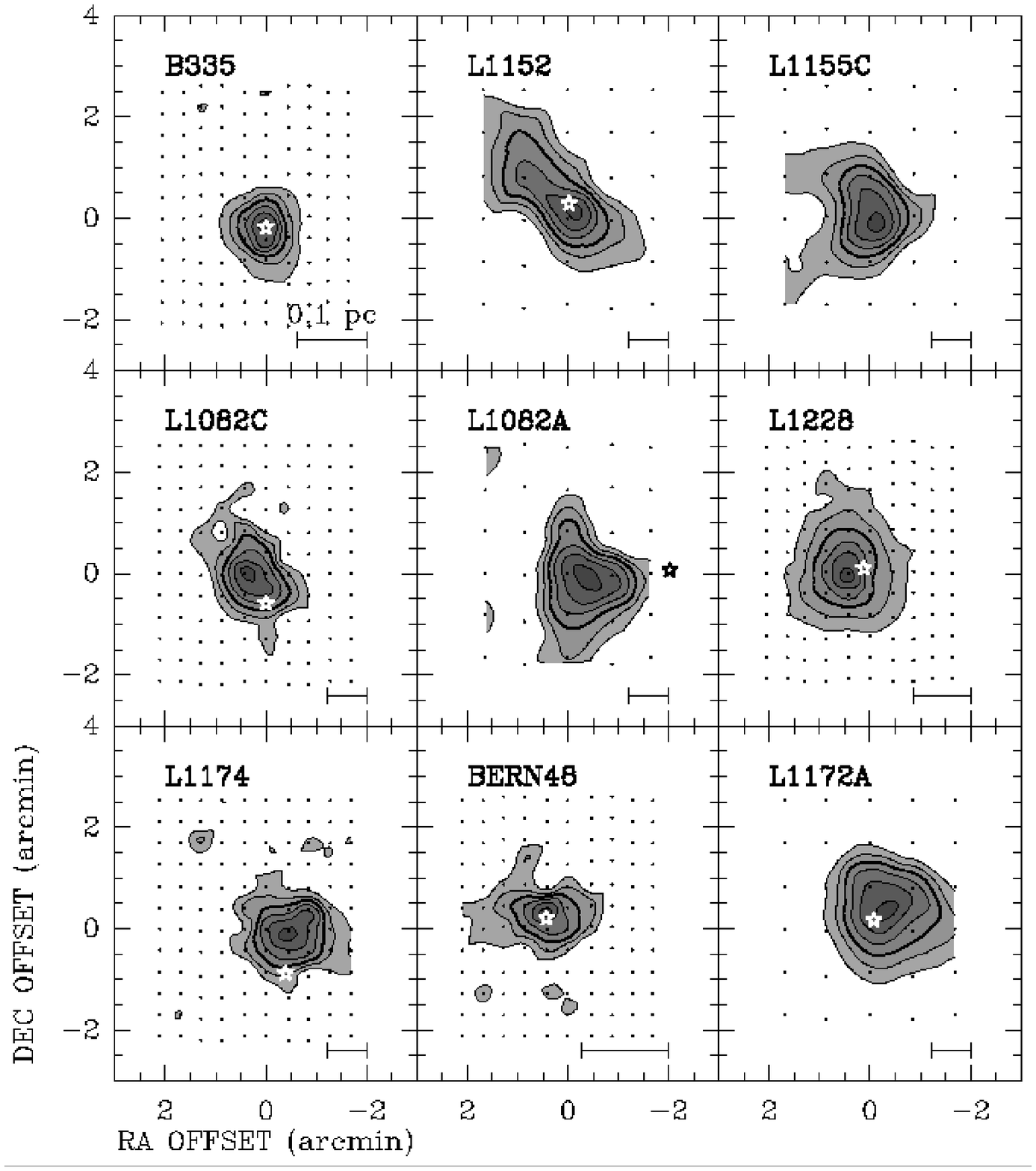}
\newpage
\plotone{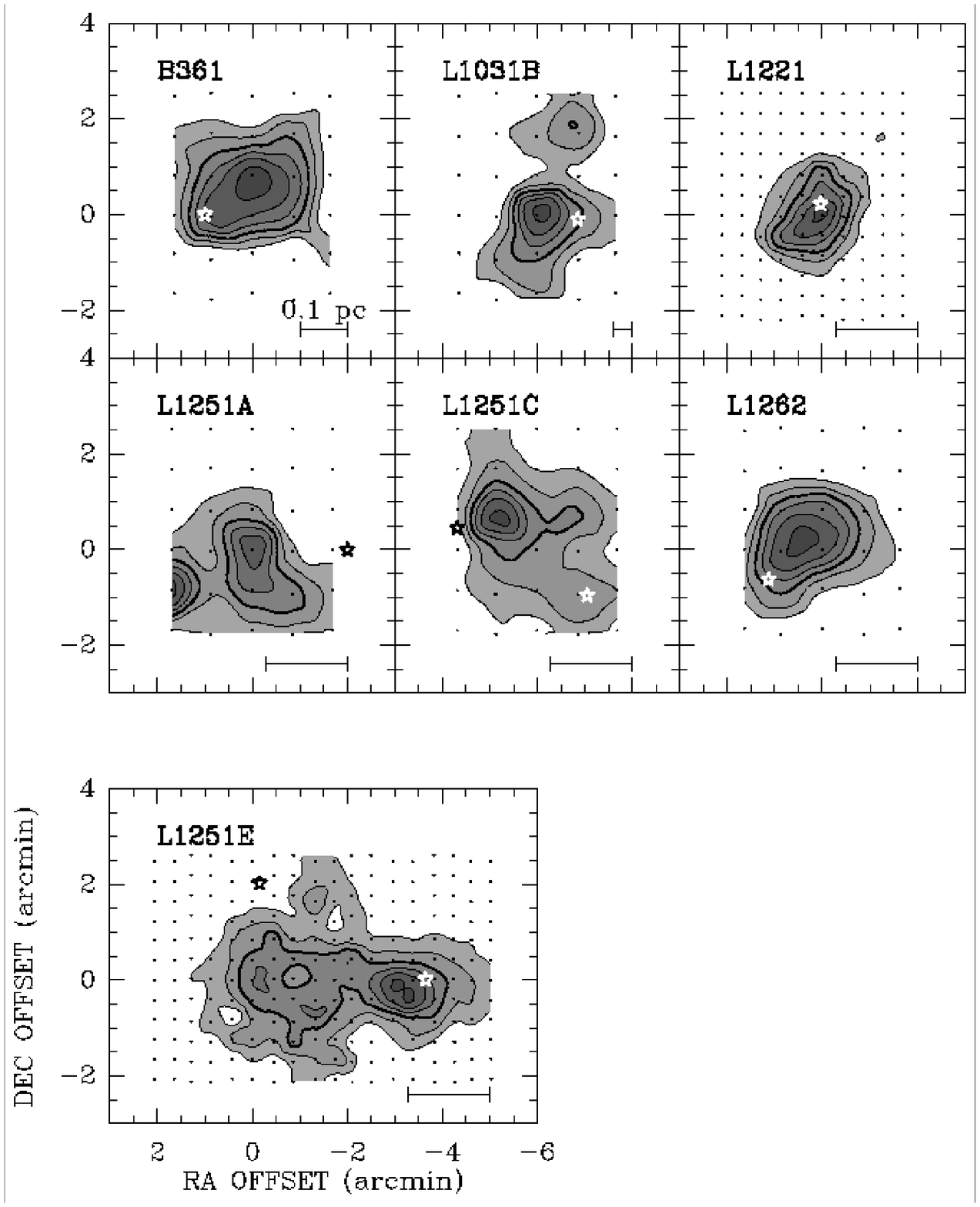}
\newpage
\epsscale{.47}
\plotone{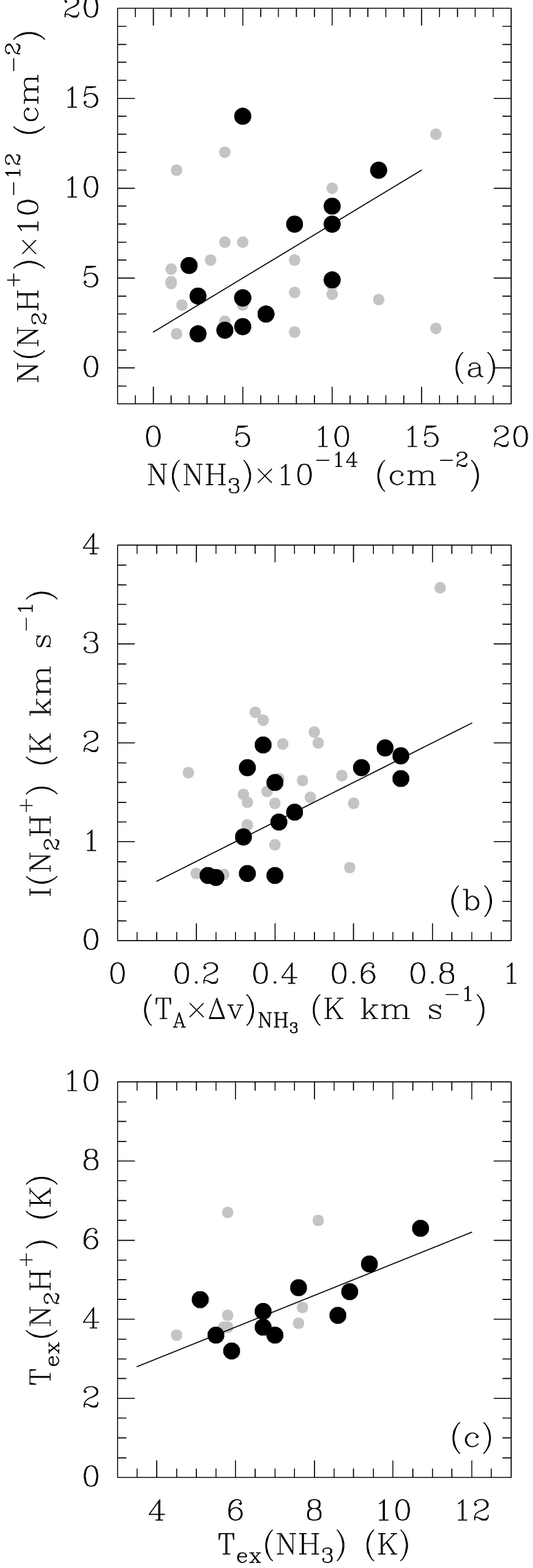}
\newpage
\epsscale{.65}
\plotone{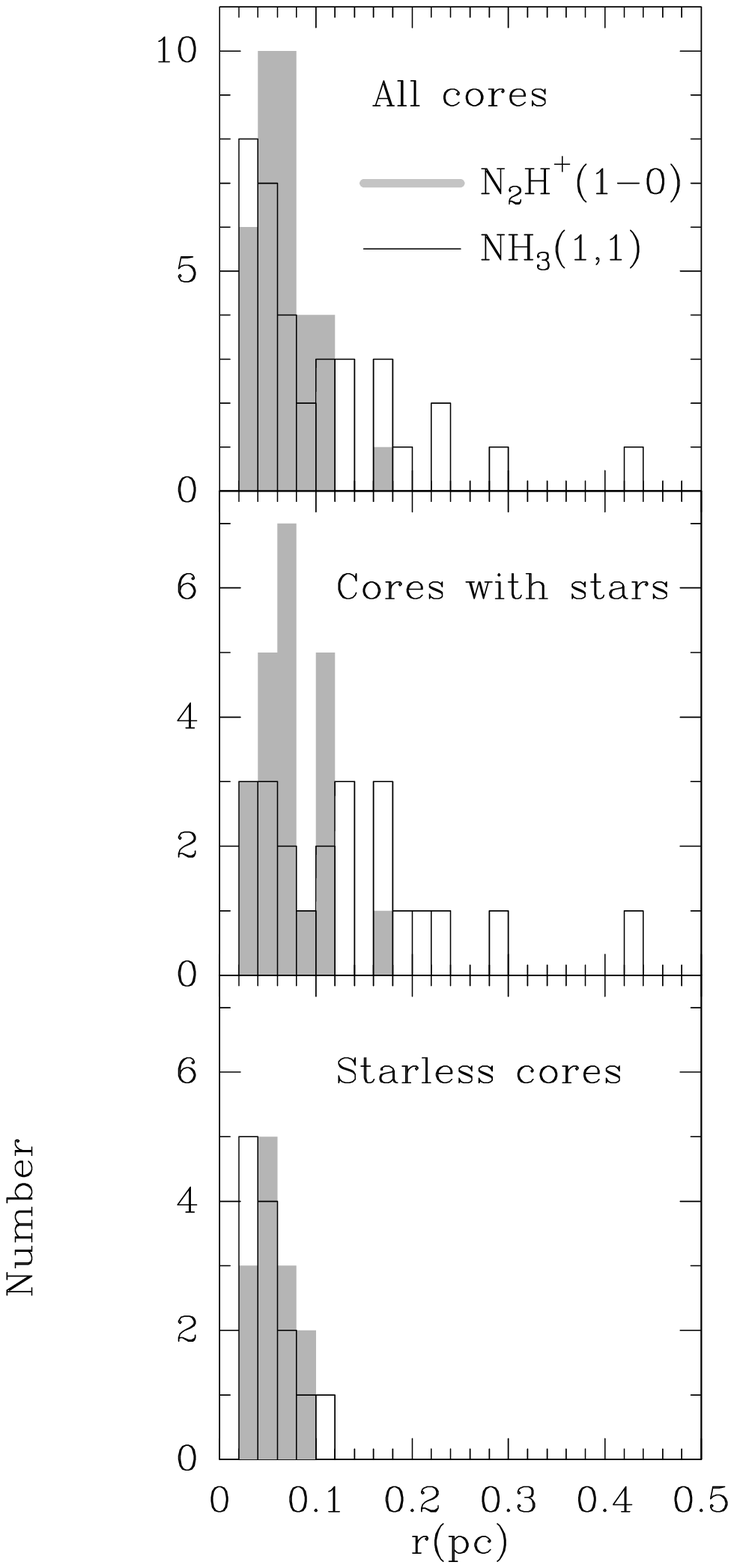}
\newpage
\epsscale{.85}
\plotone{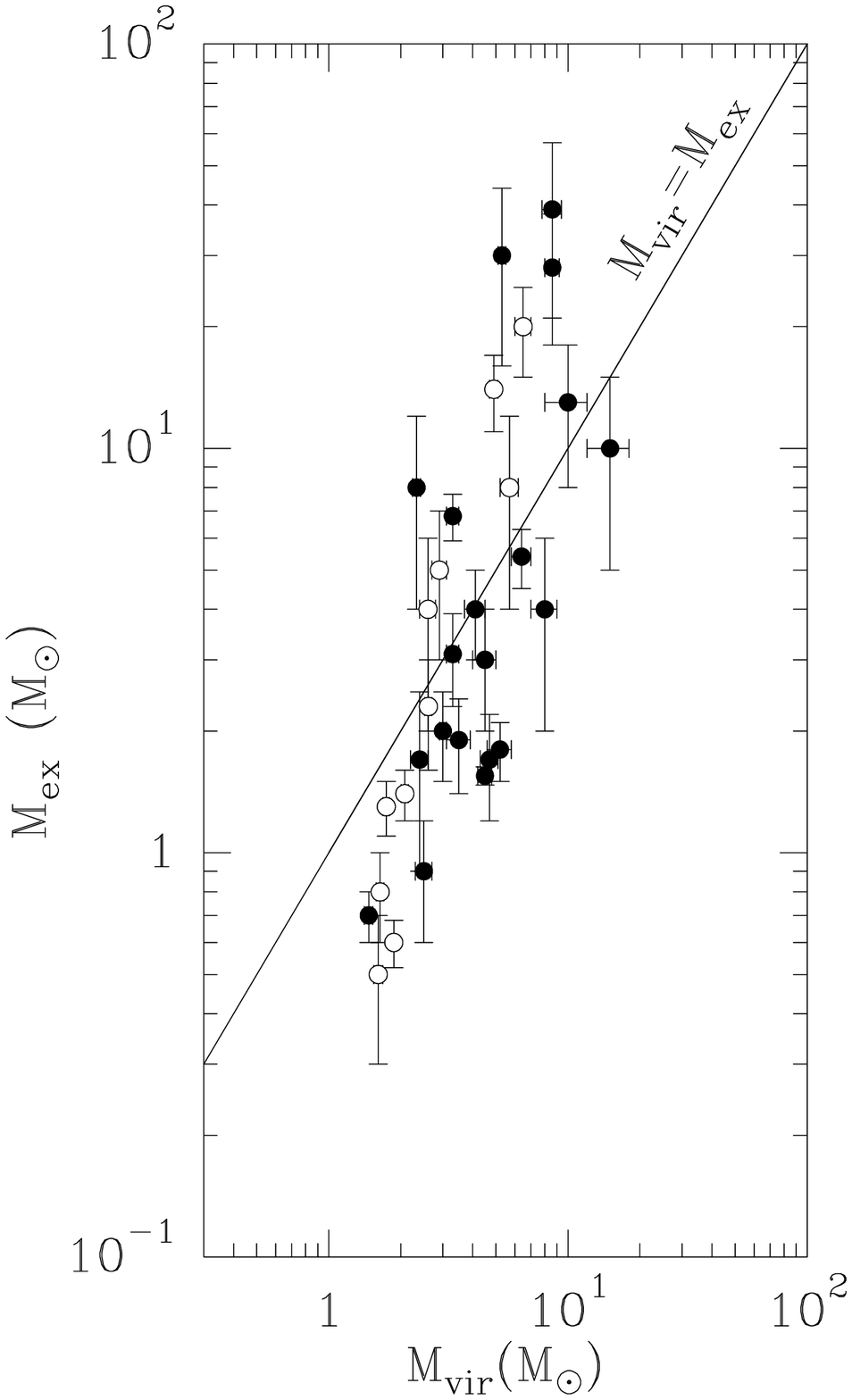}
\newpage
\epsscale{1.0}
\plotone{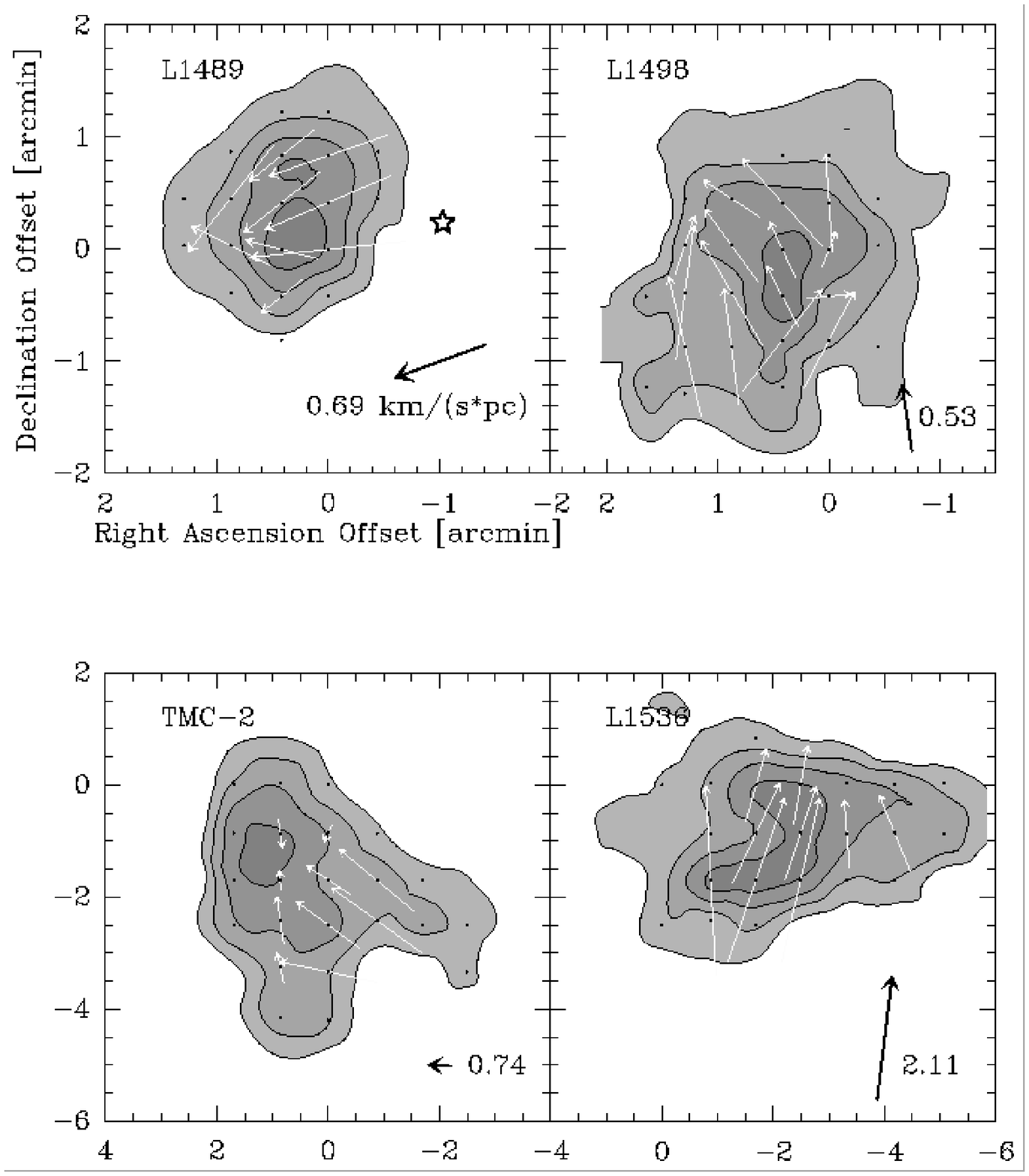}
\newpage
\plotone{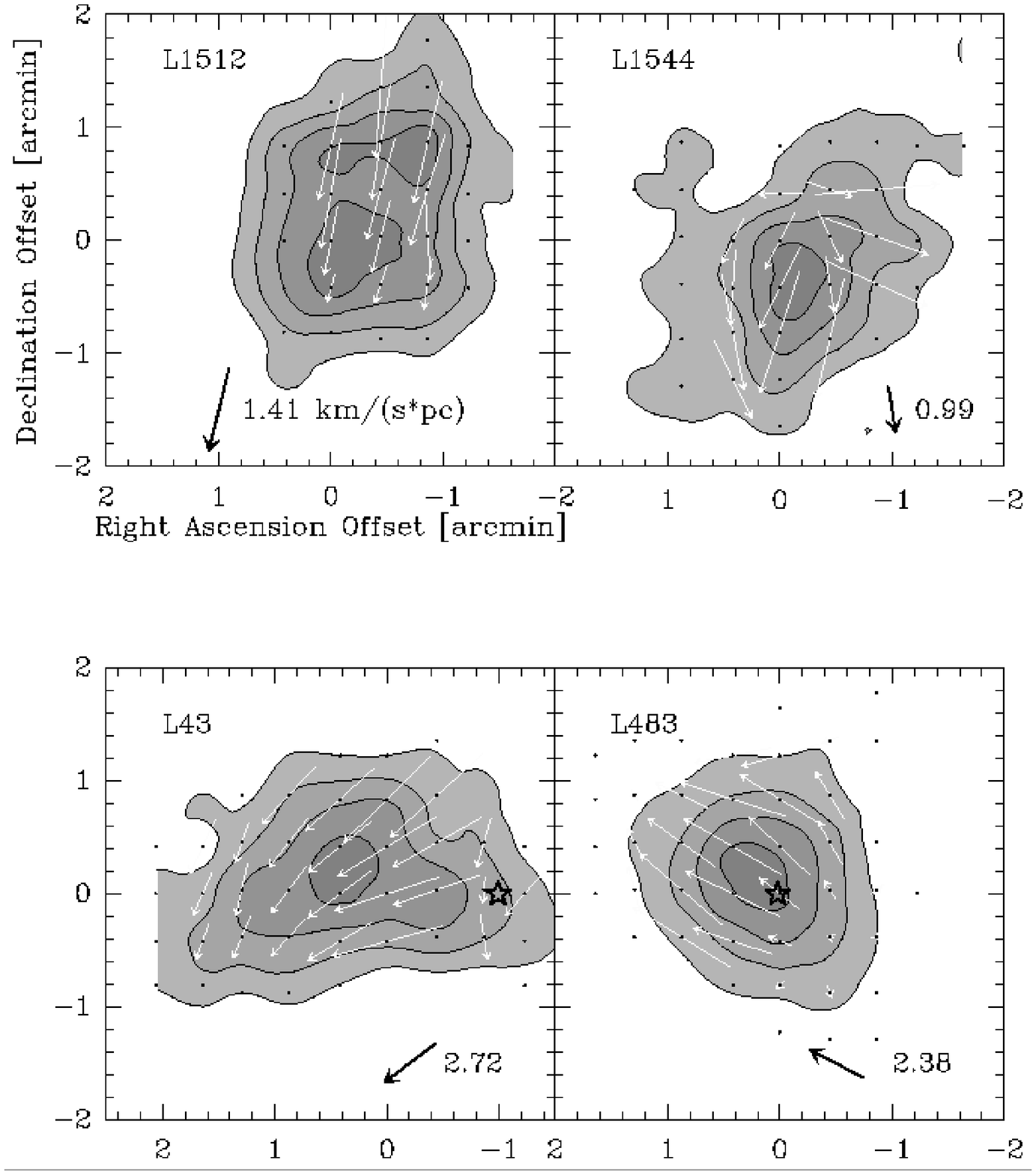}
\newpage
\plotone{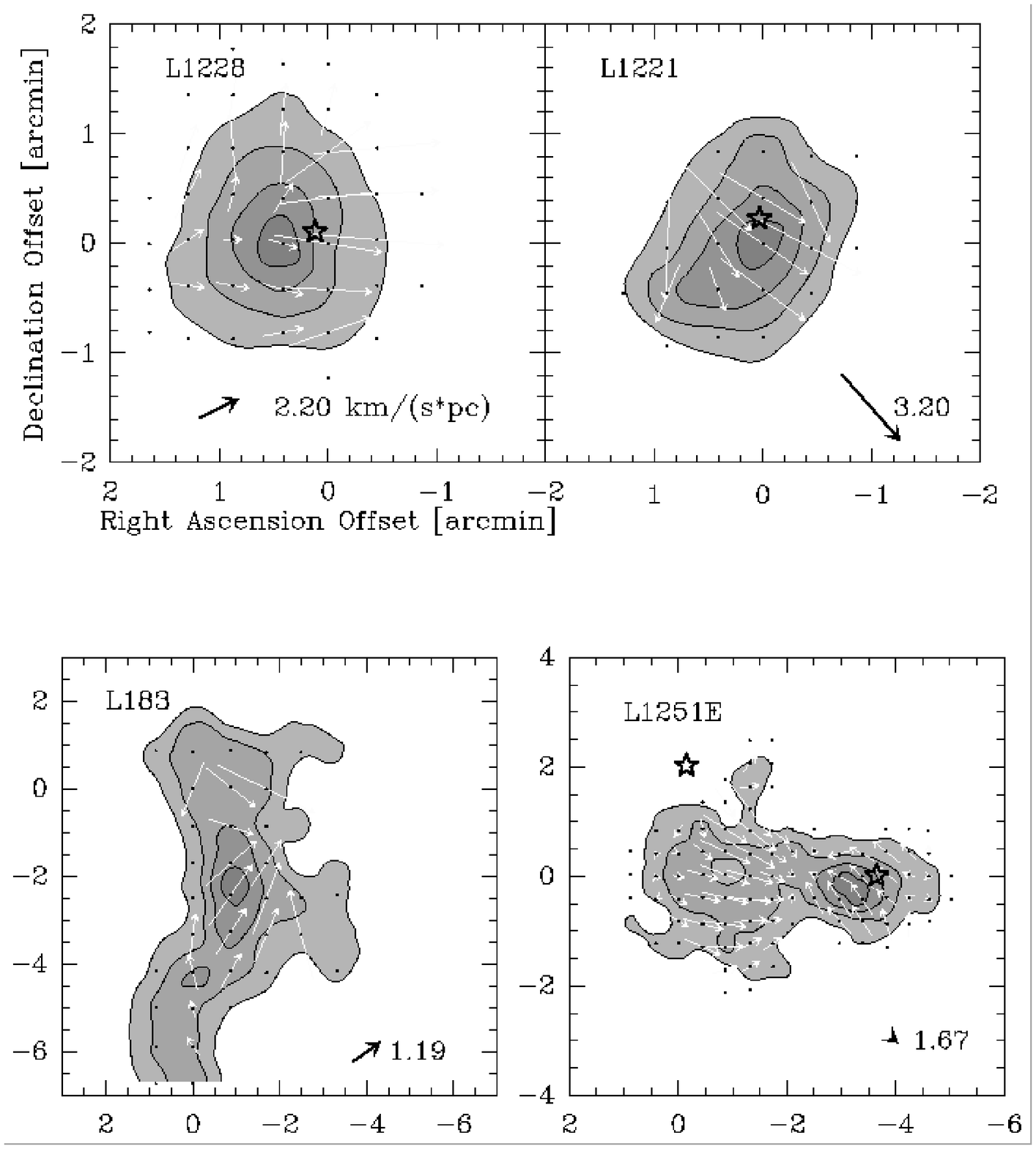}
\newpage
\epsscale{.80}
\plotone{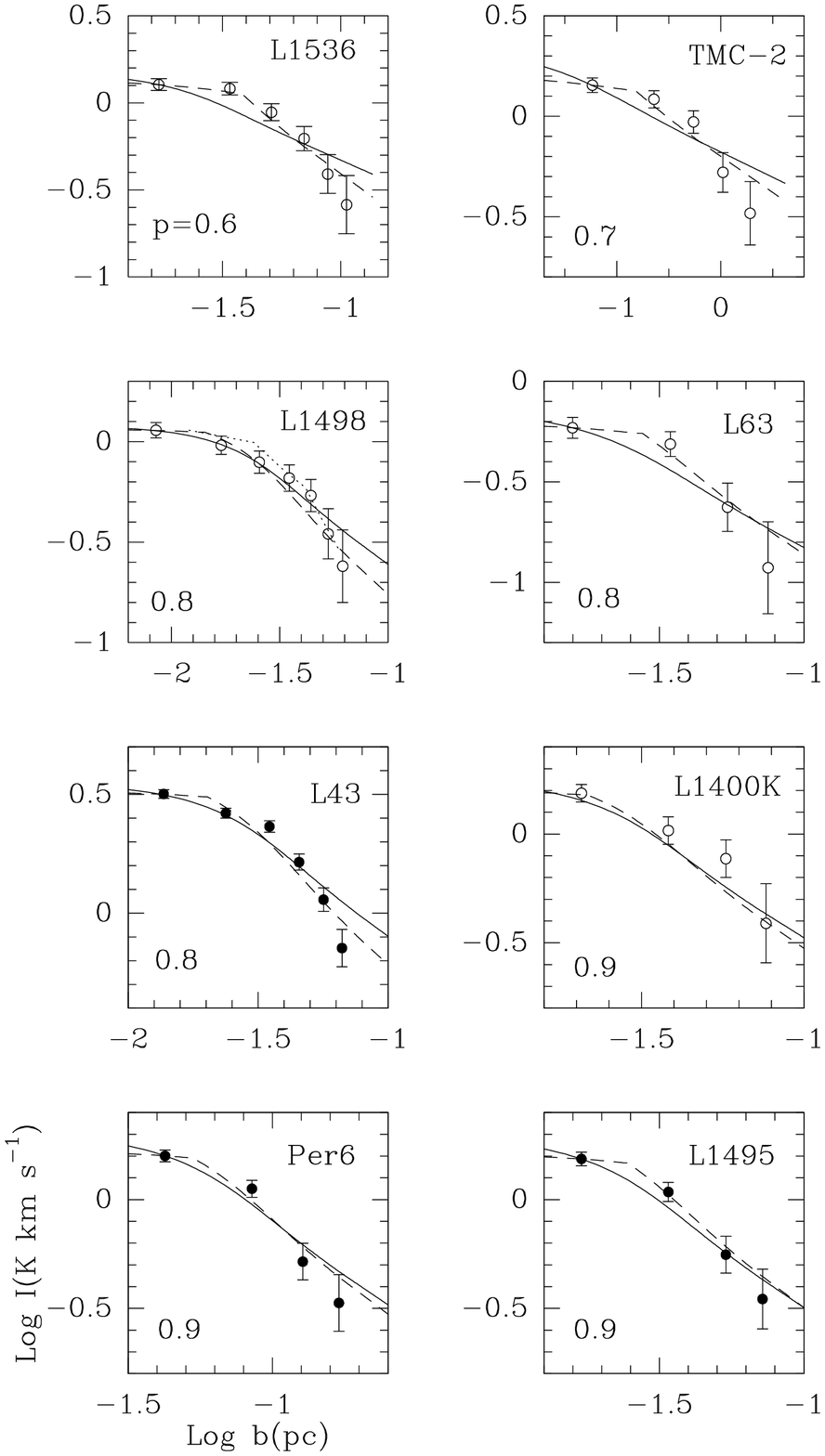}
\newpage
\epsscale{1.0}
\plotone{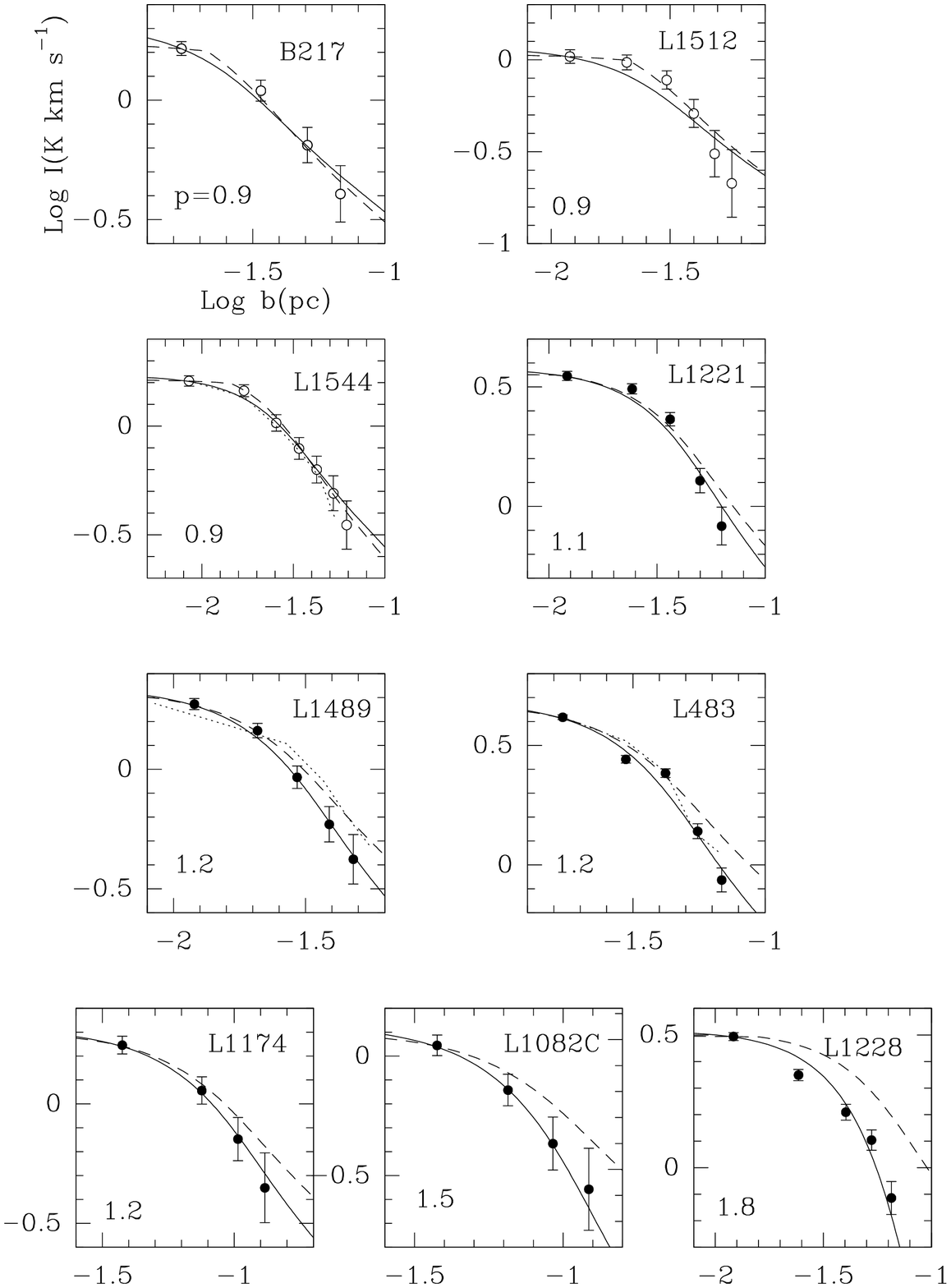}
\newpage
\plotone{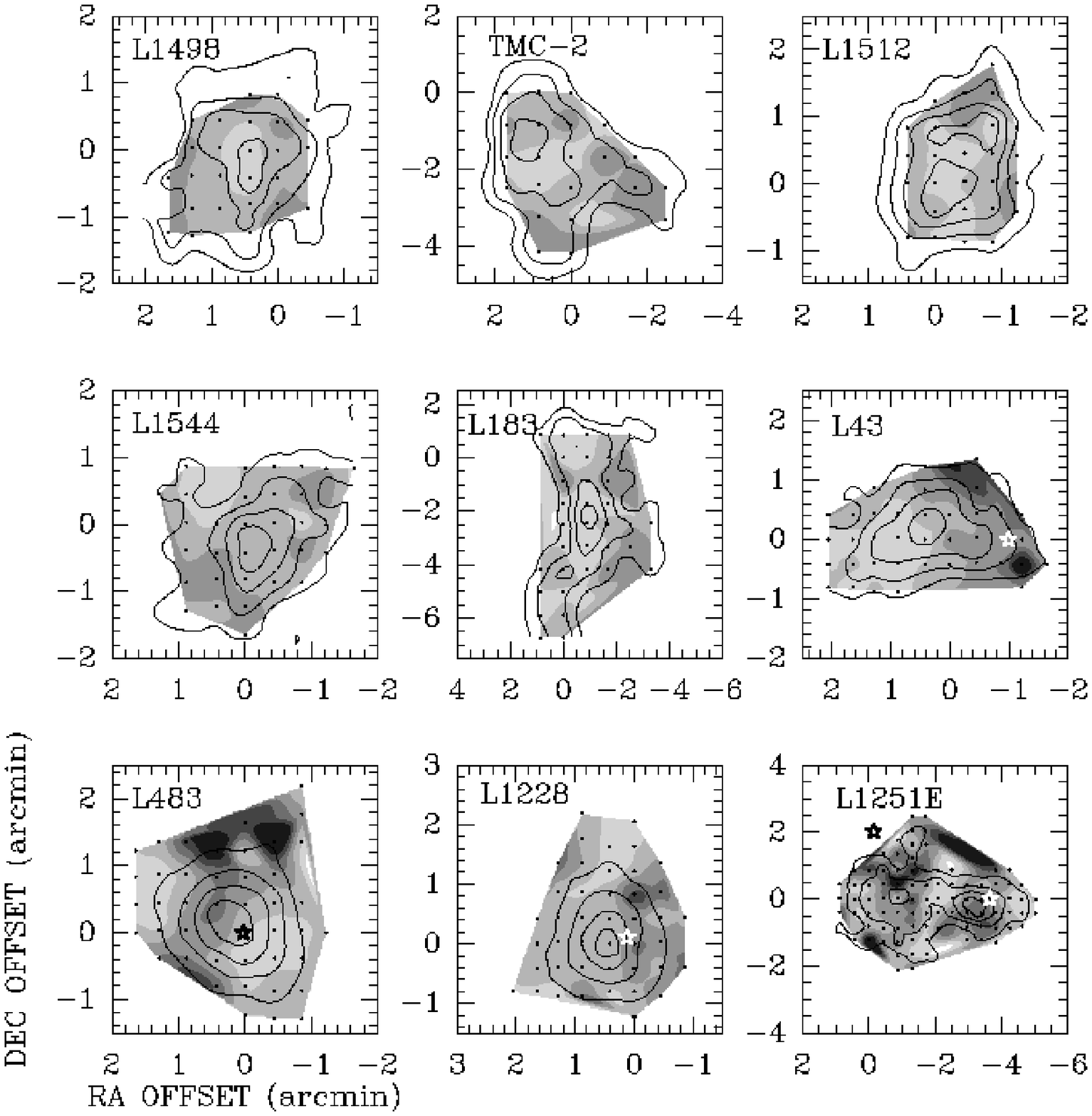}
\newpage
\plotone{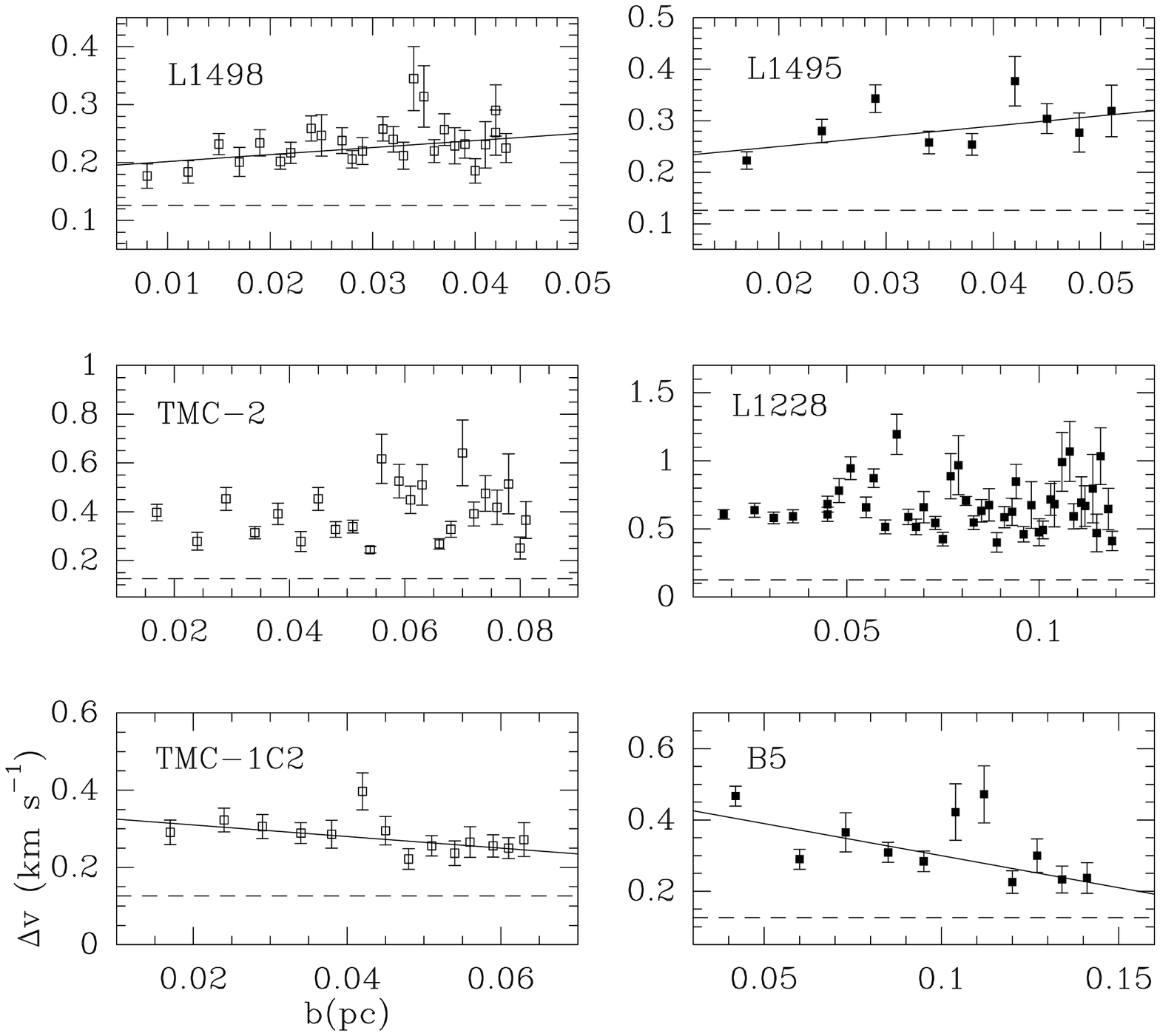}
\newpage
\plotone{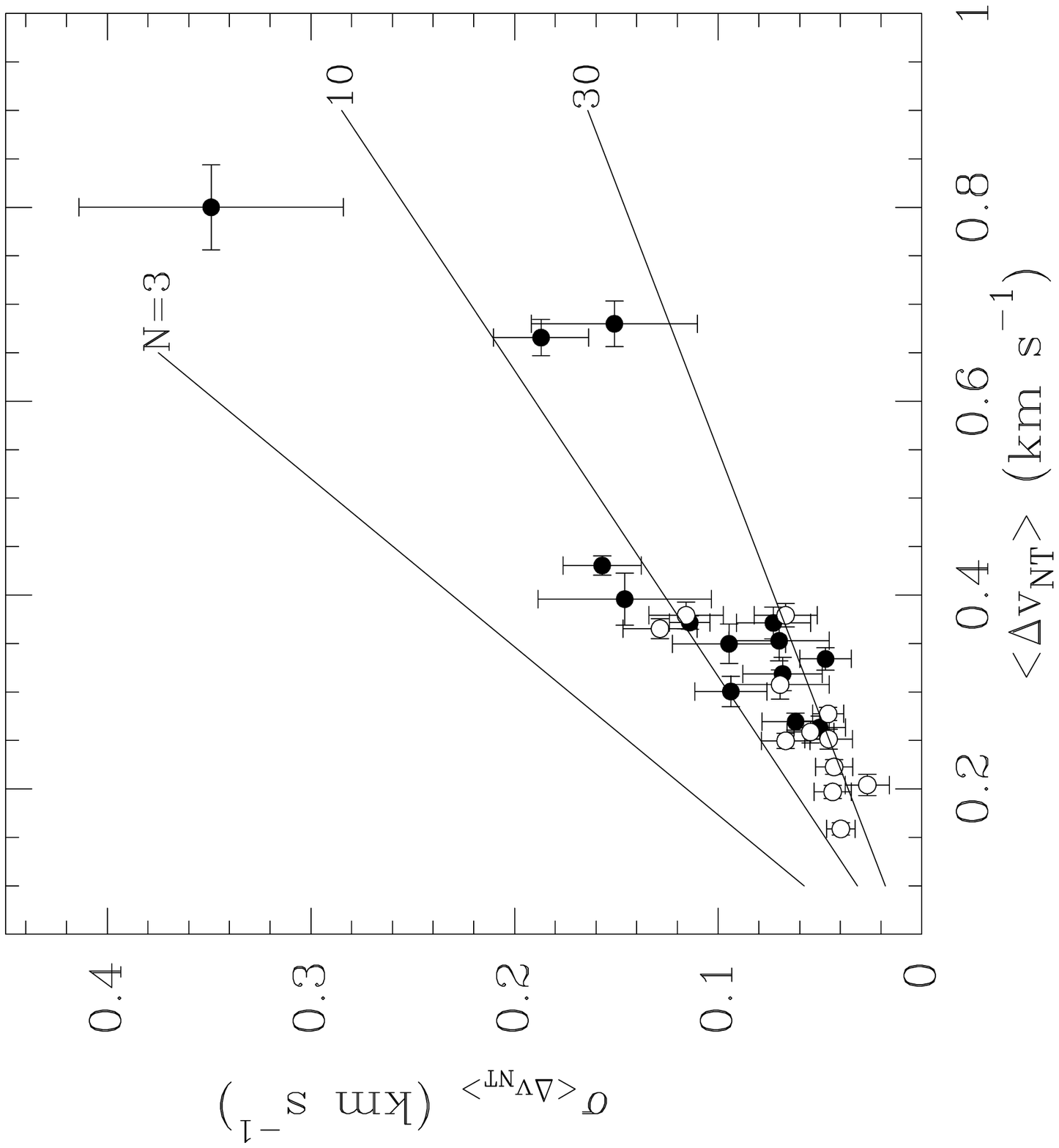}
\newpage
\plotone{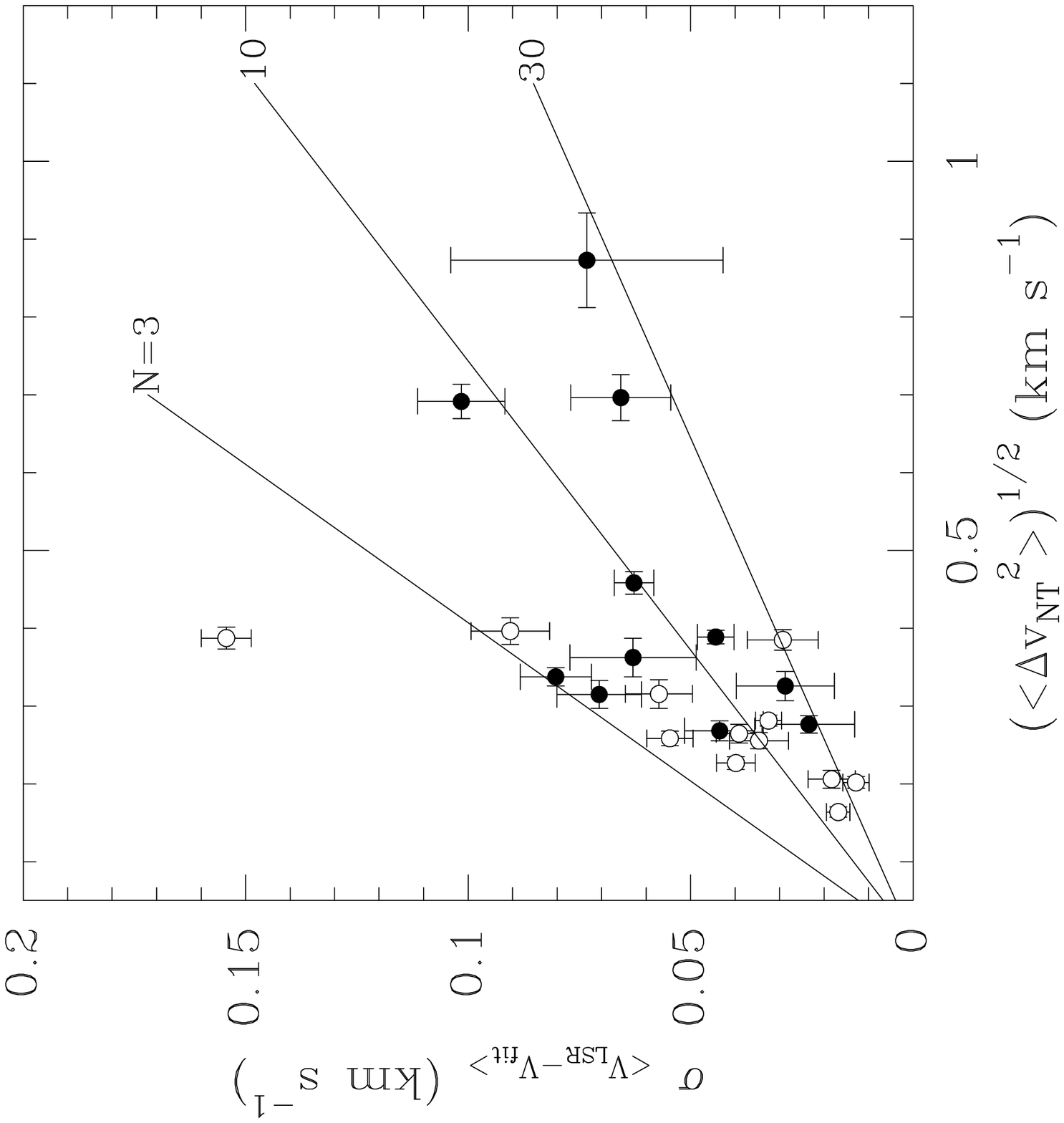}

\end{document}